\documentclass[12pt, oneside, hidelinks]{article}

\usepackage[margin=1.0in]{geometry} 

\usepackage{amsmath,amsthm,amssymb, graphicx, multicol, array,hyperref, bbm, comment, subcaption,  float, booktabs, xpatch, color}
\usepackage{setspace} 
\usepackage{longtable}
\usepackage{lscape}
\usepackage{algorithm}
\usepackage{algorithmicx}
\usepackage{algpseudocode}

\def\*#1{\mathbf{#1}}
\def\+#1{\mathbb{#1}}
\def\wt#1{\widetilde{#1}}

\newcommand{\tr}{\mathrm{trace}}

\newcommand{\diag}{\mathrm{diag}}

\newcommand{\RNum}[1]{\uppercase\expandafter{\romannumeral #1\relax}}

\usepackage[T1]{fontenc}
\usepackage[utf8]{inputenc}
\usepackage{lmodern}
\usepackage[english]{babel}
\usepackage[autostyle]{csquotes}
\usepackage{amsmath}
\usepackage{natbib}
\setcitestyle{open={(},close={)}}

\usepackage{hyperref} 
\usepackage{xr}
\makeatletter
\newcommand*{\addFileDependency}[1]{
	\typeout{(#1)}
	\@addtofilelist{#1}
	\IfFileExists{#1}{}{\typeout{No file #1.}}
}
\makeatother

\usepackage{enumitem}
\setlist{itemsep=.01em}
\setlist{topsep=.5em}

\newtheorem{innercustomgeneric}{\customgenericname}
\providecommand{\customgenericname}{}

\newtheorem{theorem}{Theorem}
\newtheorem{lemma}{Lemma}
\newtheorem{proposition}{Proposition}
\newtheorem{remark}{Remark}
\newtheorem{corollary}{Corollary}

\newtheorem{assumption}{{Assumption}}
\newtheorem{algm}{{Algorithm}}

\def\beq{\begin{equation}}
\def\eeq{\end{equation}}
\def\beqr{\begin{eqnarray}}
\def\eeqr{\end{eqnarray}}
\def\beqrs{\begin{eqnarray*}}
\def\eeqrs{\end{eqnarray*}}
\def\bet{\begin{theorem}}
\def\eet{\end{theorem}}
\def\bel{\begin{lemma}}
\def\eel{\end{lemma}}
\def\bep{\begin{proposition}}
\def\eep{\end{proposition}}
\def\bg{\begin{figure}[tbph]\begin{center}}
\def\eg{\end{center}\end{figure}}

\def\bc{\begin{center}}
\def\ec{\end{center}}

\def\wt{\widetilde}
\def\wh{\widehat}

\def\diag{\mbox{diag}}

\newcommand{\Var}{\textnormal{Var}}

\newcommand{\bA}{{\mathbf A}}
\newcommand{\bB}{{\mathbf B}}
\newcommand{\bF}{{\mathbf F}}
\newcommand{\bE}{{\mathbf E}}
\newcommand{\bG}{{\mathbf G}}
\newcommand{\bH}{{\mathbf H}}
\newcommand{\bI}{{\mathbf I}}

\newcommand{\bL}{{\mathbf L}}
\newcommand{\bM}{{\mathbf M}}
\newcommand{\bN}{{\mathbf N}}
\newcommand{\bQ}{{\mathbf Q}}

\newcommand{\bS}{{\mathbf S}}

\newcommand{\bU}{{\mathbf U}}
\newcommand{\bV}{{\mathbf V}}

\newcommand{\bX}{{\mathbf X}}

\newcommand{\bb}{{\mathbf b}}

\newcommand{\be}{{\mathbf e}}
\newcommand{\bff}{{\mathbf f}}

\newcommand{\bq}{{\mathbf q}}

\newcommand{\bs}{{\mathbf s}}
\newcommand{\bu}{{\mathbf u}}
\newcommand{\bv}{{\mathbf v}}

\newcommand{\bx}{{\mathbf x}}
\newcommand{\by}{{\mathbf y}}

\newcommand{\bfeta}  {\boldsymbol{\eta}}

\newcommand{\blambda}{\boldsymbol{\lambda}}

\newcommand{\bSigma}{\boldsymbol{\Sigma}}

\newcommand{\bve}{\mbox{\boldmath$\varepsilon$}}

\newcommand{\bTheta} {\boldsymbol{\Theta}}
\newcommand{\bPhi} {\boldsymbol{\Phi}}
\newcommand{\bPsi} {\boldsymbol{\Psi}}

\newcommand{\bGamma} {\boldsymbol{\Gamma}}
\newcommand{\bLambda} {\boldsymbol{\Lambda}}

\newcommand{\bD}{{\mathbf D}}

\newcommand{\ve}{{\varepsilon}}
\renewcommand{\epsilon}{{\ve}}
\renewcommand{\hat}{\widehat}
\newcommand{\Trunc}{\text{Truncate}}
\newcommand{\supp}{\text{supp}}
\def\wt{\widetilde}

\newcommand{\Tr}{\mbox{tr}}

\newcolumntype{L}[1]{>{\raggedright\arraybackslash}p{#1}}

\usepackage[margin=10pt,labelfont=bf]{caption}

\begin{document}

{
 
\title{Sparse Asymptotic PCA: Identifying Sparse Latent Factors Across  Time Horizon in High-Dimensional Time Series}
	\author{
		Zhaoxing Gao\thanks{\scriptsize  School of Mathematical Sciences, University of Electronic Science and Technology of China, Chengdu, 611731 PR China. Email: \url{zhaoxing.gao@uestc.edu.cn}. I gratefully acknowledge the
research support from the National Natural Science Foundation of China (NSFC) under the grant numbers 12201558 and U23A2064.}\\School of Mathematical Sciences\\University of Electronic Science and Technology of China
}	
 \date{}

\maketitle
}

\begin{onehalfspacing}
\begin{abstract}{\normalsize
 This paper introduces a novel sparse latent factor modeling framework using sparse asymptotic Principal Component Analysis (APCA) to analyze the co-movements of high-dimensional time series data. Unlike existing methods based on sparse PCA, which assume sparsity in the loading matrices, our approach posits sparsity in the factor processes while allowing non-sparse loadings.  This is motivated by the fact that financial returns typically exhibit universal and non-sparse exposure to market factors. The proposed sparse APCA employs a truncated power method to estimate the leading sparse factor and a sequential deflation method for multi-factor cases under $\ell_0$-constraints. Furthermore, we develop a data-driven approach to identify the sparsity of risk factors over the time horizon using a novel cross-sectional cross-validation method. We establish the consistency of our estimators under mild conditions for dependent data as both the dimension $N$ and the sample size $T$ grow. Monte Carlo simulations demonstrate that the proposed method performs well in finite samples. Empirically, we apply our method to daily S\&P 500 stock returns from 2004 to 2016. Using textual analysis, we investigate specific events linked to the identified sparse factors and uncover nine key risk factors that influence the stock market.}
\end{abstract}

\noindent%
				
\vfill

\newpage


\abovedisplayskip=0.1pt
\belowdisplayskip=0.1pt


		\section{Introduction}

In the realm of big-data analysis, the study of large-dimensional panel data has gained significant prominence in recent years. Large-dimensional panel data refers to datasets where observations are recorded for multiple individuals or entities over time, 
resulting in a wealth of information over the spatial and time horizons. 
Analyzing such datasets is essential for understanding complex economic, financial, and social phenomena. However, as the dimensionality of the data increases, traditional statistical techniques encounter numerous challenges, including multicollinearity, computational complexity, and difficulties in extracting meaningful insights. To address these challenges, latent factor modeling, also referred to as statistical factor modeling in \cite{campbell1997econometrics}, has emerged as a powerful and versatile approach in the field of panel data analysis. Latent factor models offer a structured framework for dimension reduction and capturing common sources of variation in high-dimensional data. For example, asset returns in finance are often modeled as functions of a small number of factors; see \cite{Ross1976} and \cite{connor1986performance,connor1988risk}. Macroeconomic variables of multiple countries are found to have common components; see \cite{stock1989new}, \cite{gregory1999common}, and \cite{forni2000reference}. In demand systems, the Engle curves can be expressed in terms of a finite number of factors; see \cite{lewbel1991rank}. 
As the dimensions of the panel data systems increase, various factor models have been developed to reduce the dimensionalities under different scenarios. See, for example, the approximate factor models in \cite{Chamberlain1983}, \cite{Stock2002a,Stock2002b}, \cite{bai2002determining}, \cite{bai2003inferential}, \cite{fan2013large}, \cite{lettaupelger2018}, \cite{pelger2019}, and \cite{gaotsay2022,gao2023supervised}, and the dynamic factor models in \cite{forni2000generalized}, 
among others. 

One of the most profound challenges in factor modeling is to handle high-dimensional 
time series data efficiently and effectively. Sparse factor modeling represents a groundbreaking approach to this challenge, offering a powerful technique to extract meaningful insights from complex datasets characterized by a multitude of variables. Sparse factor modeling seeks to identify and capture the underlying structure of data while simultaneously promoting simplicity by selecting only a subset of relevant variables (or features) from the original set. The central idea behind sparse factor modeling is to uncover latent factors that drive the observed data's variation while enforcing a degree of sparsity, meaning that only a few variables are deemed essential to explain the data's structure.  This approach is particularly valuable in scenarios where the number of variables far exceeds the number of observations, as it not only reduces computational demands but also enhances the interpretability of the results.
Examples of articles concerning this approach include the sparse PCA in \cite{jolliffe2003modified}, \cite{zou2006}, \cite{shen2008sparse}, \cite{johnstone2009consistency}, \cite{witten2009penalized}, and \cite{ma2013sparse}, and the sparse factor analysis with sparse loadings in \cite{kristensen2017diffusion} and \cite{uematsu2022estimation}, 
among others, where all of the aforementioned works are in line with factor analysis with sparse loadings, implying that each factor process is a linear combination of a small subset of the original panel series only.

\begin{figure}[ht]
\begin{center}
{\includegraphics[width=13cm,height=6cm]{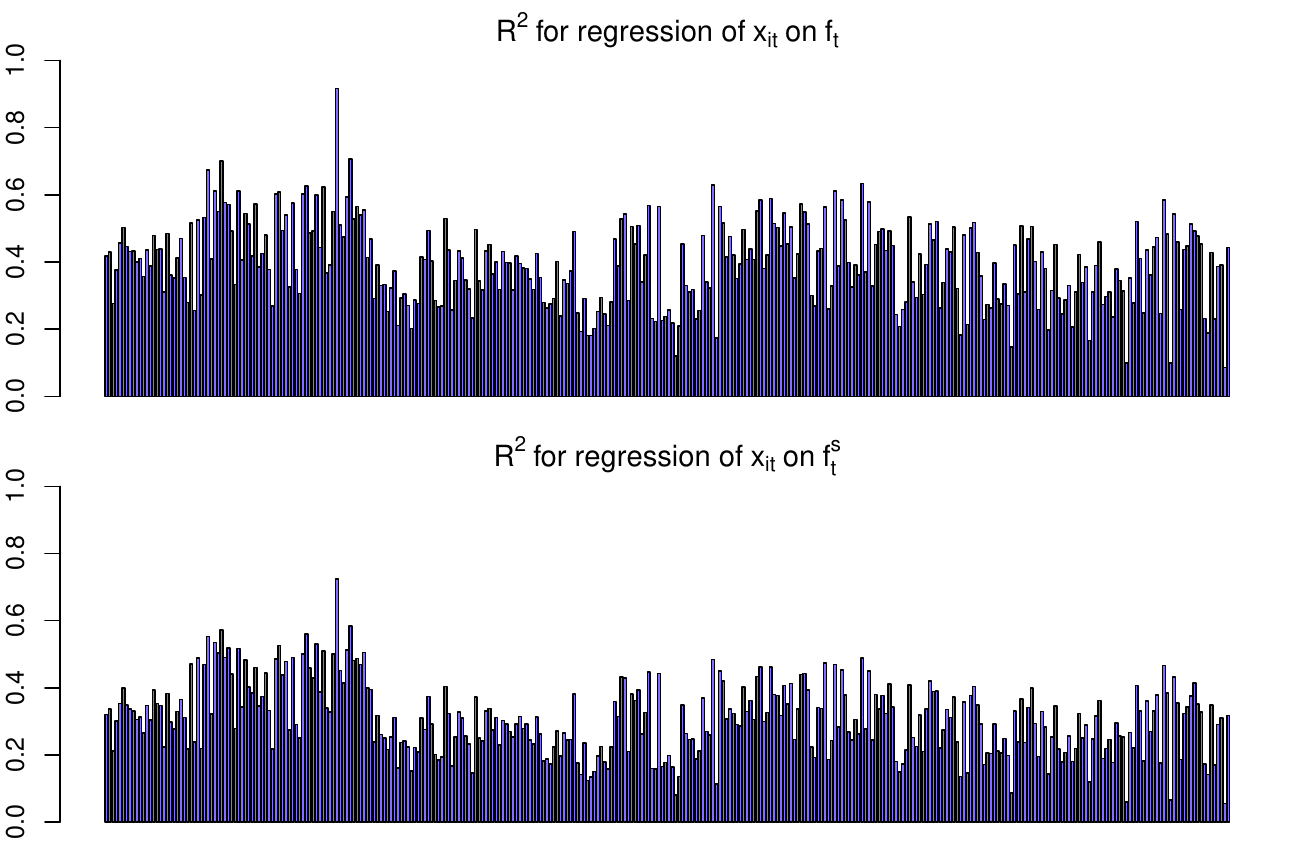}}
\caption{$R^2$ for regression of $x_{i,t}$ on $f_t$ and $f_t^s$, respectively.}\label{fig-int}
\end{center}
\end{figure}

 It is widely known that the key idea in a large-dimensional factor model is that the dimensions of both the timeline and the cross-section of the data are large, and most of the co-movements can be explained by a few factors. These factors and loadings are usually estimated by the conventional PCA method due to its ability to parsimoniously capture much of the information in a large number of variables.
However, the resulting latent PCs or factors are typically linear combinations of all cross-sectional units/variables, which are usually hard to interpret. The sparse PCA restricts the cardinality of the weight vectors for the PCs, or equivalently, the loadings for the factor processes, so the PCs are sparse linear combinations of the underlying variables. Consequently, the PCs or factors are only linear combinations of a small subset of the cross-sectional units, which facilitates the interpretations of the resulting PCs or factors. Nevertheless, there is little literature concerning the justificaton of the sparsity assumption in the loadings.
For example, the assumption may not be appropriate for financial returns, where  the
exposure of the returns to a market factor is universal and non-sparse, as pointed out in \cite{pelger2022interpretable}. 
{To illustrate, we examine the risk factors of the daily return data of the S\&P 500 stocks studied in \cite{pelger2019}, with \( N = 332 \) stocks over \( T = 3273 \) time points. We first apply PCA to the panel, and the estimated  factors \( f_t \) are shown in Figure~S.1 of the Supplement. We then regress each return series \( x_{i,t} \) on \( f_t \), and the resulting \( p \)-values for testing the significance of the regression coefficients are all close to zero, implying that the dependence of the returns on the factors may be non-sparse.
Next, we define a sparse factor \( f_t^s \), where \( f_t^s = f_t \) if \( f_t \) is among the 500 largest values in absolute magnitude, and \( f_t^s = 0 \) otherwise. We find that
${\sum_{t=1}^{3273}(f_t^s)^2}/{\sum_{t=1}^{3273} f_t^2} = 79.4\%$,
indicating that the top 500 factor points account for 79.4\% of the total factor variance over the entire time horizon.
Finally, we calculate the \( R^2 \) values for regressing \( x_{i,t} \) on \( f_t \) and \( f_t^s \), respectively. The results are presented in Figure~\ref{fig-int}. From the figure, we observe that the \( R^2 \) values based on only 500  factor points represent a substantial proportion of those based on the full factor series. The average ratio of \( R^2 \) values from the sparse to the full factor regressions is 77.5\%, suggesting that the explanatory power of the sparse factor over time is substantial and deserves further attention.
}

In view of the above discussion, 
this paper marks a further development in the sparse factor modeling of large-dimensional panel data from a different perspective. 
Motivated by the empirical success of a general approximate factor model, see \cite{pelger2019} and the references therein, we also adopt such a similar framework.
Unlike the sparse factor modeling approach with sparse loadings in existing literature, we assume the latent factor processes are sparse over time, while the loadings can be nonzero in general settings.  {Specifically, over a given time period, cross-sectional economic or financial series may exhibit varying degrees of co-movement. During certain periods, weak co-movement allows idiosyncratic components to dominate, resulting in relatively lower systematic risk—see, for example, the empirical evidence in \cite{campbell2001have} and \cite{goyal2003idiosyncratic}. Since the loading vector for each series reflects a linear combination of the underlying factor series, sparsity—where only the most significant factors are retained—implies that loadings are primarily driven by a subset of factors.
This formulation is intuitive in financial applications, where asset returns often alternate between periods of high volatility and relative calm. Co-movements tend to intensify during major events such as interest rate hikes by the Federal Reserve or financial crises related to debt, currency, or mortgage markets. Our proposed method offers a useful framework for uncovering connections between major economic or policy events and the evolving co-movement structure of economic or financial systems.}

In this paper, 
we present a sparse APCA technique by formulating a sparse factor modeling framework with sparse factors over time horizons.  {Under the $\ell_0$-constraint imposed on the factors, we propose a truncated power method to estimate the sparse factors in the one-factor case, corresponding to the hard-thresholding technique widely used in the statistics and machine learning literature. As noted in \cite{johnstone2009consistency}, a key advantage of hard-thresholding over soft-thresholding is that it preserves the magnitude of the retained signals, whereas soft-thresholding may introduce bias into the estimators.} While hard-thresholding has been employed in studies such as \cite{johnstone2009consistency} and \cite{ma2013sparse}, their focus is primarily on independent and identically distributed (i.i.d.) data with independent/white noises, where the distinction between factors and loadings therein is minimal.  {From a theoretical perspective, the presence of serial correlation in both the data and the factors poses additional challenges in establishing the convergence rate of the sample covariance matrix and conducting inference on the estimated loadings. In particular, large deviation inequalities such as Bernstein’s inequality cannot be directly applied to time series data. Furthermore, the thresholding values proposed in \cite{johnstone2009consistency} and \cite{ma2013sparse}, which rely on the assumption of normality, are not applicable in our setting, as time series data are generally not normally distributed.} 
For multi-factor cases, we develop a sequential deflation estimation procedure that integrates the truncated power method with projection techniques. Additionally, we introduce a data-driven approach to determine the sparsity of risk factors over time using a novel cross-sectional cross-validation method.  
Theoretical properties of the proposed estimators are established under mild conditions for dependent data as both the dimension and sample size tend to infinity. 
Simulated examples are used to illustrate the proposed method. We empirically examine specific events associated with the identified sparse factors that systematically influence the stock market. Our findings reveal nine key time factors that may systematically affect the financial market. 

{The proposed framework is fundamentally different from existing sparse PCA and sparse factor models in both formulation and interpretation, and it also employs a distinct estimation technique. In the sparse PCA literature (e.g., \cite{zou2006}, \cite{shen2008sparse}, \cite{witten2009penalized}), the true sparse eigenvectors are treated as constant vectors, and regularized algorithms are used for their estimation. In sparse factor models (e.g., \cite{kristensen2017diffusion}, \cite{uematsu2022estimation}), the factor loadings are assumed to be sparse and non-random, with estimation based on $\ell_1$-regularization. The sparse proximate factor model of \cite{pelger2022interpretable} assumes sparsity only in factor weights, while the resulting factor and loading processes remain dense. Time-varying factor models, such as \cite{su2017time}, allow for temporal dynamics of the loadings but do not impose sparsity.
In contrast, our sparse APCA framework assumes that the factor processes themselves exhibit sparsity—i.e., each factor entry is either an active random variable or zero at a given time. A zero factor value suggests weaker co-movement  at that  time point. Our method imposes $\ell_0$-constraints directly on the factors and is thus theoretically distinct from prior approaches, which focus on cross-sectional sparsity. This temporal sparsity structure enables us to identify specific time points where systematic risk factors significantly influence the panel.}


The contributions of this paper are multi-fold. First, to the best of our knowledge, this is the first study to consider sparse factors over the time horizon, providing economists with a novel approach to bridging the gap between time and risk in economic and financial systems. Second, unlike the $\ell_1$-penalized methods in most sparse PCA or factor modeling literature, we propose a truncated power method based on an 
$\ell_0$-constraint for the one-factor case, along with a sequential optimization procedure for estimating multiple factors.    Our approach differs from the hard-thresholding methods for i.i.d. data in \cite{johnstone2009consistency} and \cite{ma2013sparse} by explicitly accounting for time series data with both cross-sectional and dynamic dependence. Consequently, our method extends the asymptotic PCA technique developed by \cite{connor1986performance,connor1988risk} to high-dimensional factor analysis.
 Third,  instead of relying on traditional cross-validation to determine sparsity structures, we introduce a novel cross-validation approach that partitions the spatial dimensions. Fourth, most thresholding techniques require specifying a threshold value, as seen in \cite{johnstone2009consistency} and \cite{ma2013sparse}. In contrast, our method introduces a selection criterion that determines the number of nonzero factor points without requiring pre-specified threshold values. Theoretically, we establish the consistency of the proposed estimators as the dimension and sample size approach infinity. Importantly, we provide statistical guarantees for the proposed cross-sectional cross-validation method, demonstrating its ability to consistently estimate the sparsity structure over the time horizon of factor models. 

The rest of the paper is organized as follows. Section \ref{sec2} introduces the sparse factor modeling framework and its estimation procedure. Section \ref{sec3} presents asymptotic properties of the estimators obtained in Section \ref{sec2}. Section \ref{sec4} studies the finite-sample performance 
of the proposed approach via simulation, and \ref{sec5} illustrates the proposed procedure with an empirical application. Section \ref{sec6} concludes. Some illustrative examples of sparse factors over time horizons, additional tables and figures used in the numerical analyses, all theoretical proofs of the main theorems, and a description of the text data extracted from {\it CNN} are relegated to an online Supplement.


{\bf Notation:}  We use the following notation. For a $p\times 1$ vector
$\bu=(u_1,..., u_p)'$, $\|\bu\|_1=\sum_{i=1}^p|u_i|$ is the $\ell_1$-norm and $\|\bu\|_\infty=\max_{1\leq i\leq p}|u_i|$ is the $\ell_\infty$-norm. $\bI_p$ denotes the $p\times p$ identity matrix. For a matrix $\bH$, its Frobenius norm is $\|\bH\|=[\tr(\bH'\bH)]^{1/2}$ and its operator norm is $\|\bH
\|_2=\sqrt{\lambda_{\max} (\bH' \bH ) }$, where
$\lambda_{\max} (\cdot) $ denotes the largest eigenvalue of a matrix, and $\|\bH\|_{\min}$ is the square root of the minimum non-zero eigenvalue of $\bH\bH'$. $|\mathbf{H}|$ denotes the element-wise absolute value of $\mathbf{H}$. The superscript ${'}$ denotes the 
transpose of a vector or matrix. We also use the notation $a\asymp b$ to denote $a=O(b)$ and $b=O(a)$ or $a$ and $b$ have the same order of stochastic bound when they are random variables.

		\section{Model and Methodology} \label{sec2}
		
		\subsection{Model Setup}\label{model_overview}
		Let $x_{i,t}$ be the $i$-th unit of the cross-sectional panel of time series $\bx_t=(x_{1,t},...,x_{N,t})'\in R^N$ at time $t$, for example, $x_{i,t}$ can be the stock return of the $i$-th asset at time $t$, we consider the following approximate factor model:
		\begin{equation}\label{ftm}
		x_{i,t}=\blambda_i'\bff_t+e_{i,t},
		\end{equation}
		where $x_{i,t}$ is the only observed datum for the $i$-th cross-section at time $t$ ($i=1,...,N$; $t=1,...,T$), $\bff_t$ is an $r$-dimensional vector of common or systematic risk factors, $\blambda_i$ is an $r$-dimensional vector of factor loadings, and $e_{i,t}$ is the idiosyncratic component of $x_{i,t}$ that cannot be explained by the common risk factors.  We may combine all the cross-sectional units together and write the above equation as		
\begin{equation}\label{comF}
\bx_t=\bLambda\bff_t+\be_t,
\end{equation}
where $\bLambda=(\blambda_1,...,\blambda_N)'$ and $\be_t=(e_{1,t},...,e_{N,t})'$. 
		Assume a panel data set of $T$ time-series observations and $N$ cross-sectional observations, denoted as $\bX \in R^{T\times N}$, has a factor structure with $r$ common factors.  Let $\bF=(\bff_1,...,\bff_T)'$ and $\be=(\be_1,...,\be_T)'$, then Models (\ref{ftm}) and (\ref{comF}) can be written  in the following matrix form
		\begin{equation}\label{MF}
		\bX=\bF\bLambda'+\be.
		\end{equation}
		For ease of notation, we also denote $\bX=(\underline{\bx}_1,...,\underline{\bx}_N)$, $\bF=(\underline{\bff}_1,...,\underline{\bff}_r)$, and $\bLambda=(\underline{\blambda}_1,...,\underline{\blambda}_r)$ when referring to their columns.
		In the econometrics/statistical and finance literature, Model (\ref{MF}) is usually estimated using the Principal Components or Asymptotic Principal Components estimation method (see \cite{connor1986performance,connor1988risk}, \cite{bai2002determining}, \cite{bai2003inferential},  and \cite{fan2013large}, among others). As a result, all the elements in the estimators $\wh\bF$ and $\wh\bLambda$ are usually nonzero, making it difficult to interpret the dynamic and cross-sectional relationships of the data. Moreover, the role of the idiosyncratic terms is rarely considered or even ignored in characterizing the individual dynamics of the series.
		
		{
In Model~(\ref{MF}), only the panel $\bX$ is observed, while the factor $\bF$ and loading $\bLambda$ are latent and unidentifiable up to a rotation: for any invertible $r \times r$ matrix $\bH$, the pair $(\bF\bH', \bLambda\bH^{-1})$ yields the same model. To ensure identification, two commonly used sets of conditions are adopted in the PCA framework:
\begin{align}
\label{Con1}
\frac{1}{T}\sum_{t=1}^T \bff_t\bff_t' = \bI_r, \quad \text{and} \quad \frac{1}{N} \bLambda'\bLambda \text{ is diagonal}, \quad \text{or} \quad \frac{1}{N} \bLambda'\bLambda = \bI_r, \quad \text{and} \quad \frac{1}{T} \sum_{t=1}^T \bff_t\bff_t' \text{ is diagonal}.
\end{align}
See \cite{bai2002determining}, \cite{fan2013large}, and \cite{jiang2023revisiting}, among others. These identification restrictions help resolve the rotational indeterminacy when the leading eigenvalues of the covariance matrix are distinct; see \cite{bai2013principal}. Under either set of conditions, PCA provides consistent estimates of the common component $\wh\bF \wh\bLambda'$.

In this paper, we adopt the first set of conditions in~(\ref{Con1}) and apply the Asymptotic Principal Component Analysis (APCA) approach to uniquely identify $\bF$ and $\bLambda$. Suppose the true model is $\bX = \bF^*\bLambda^{*'} + \be$, \cite{jiang2023revisiting} shows that there exists a rotation matrix $\bH$ such that the estimated APCA factors $\wh\bF$ satisfy $\wh\bF'\wh\bF/T = \bI_r$ and are consistent for $\bF^*\bH'$. Redefining $\bF = \bF^*\bH'$ and $\bLambda = \bLambda^*\bH^{-1}$ transforms the model to $\bX = \bF\bLambda' + \be$, which allows us to impose sparsity assumptions directly on $\bF$ without encountering identification issues.

Our first goal is to extend the APCA framework of \cite{connor1986performance, connor1988risk} to high-dimensional settings using modern machine learning techniques. Moreover, since factors and loadings in approximate factor models are typically difficult to interpret, our second goal, building on the framework of \cite{bai2013principal}, is to impose additional sparse structure on the factors. This enables interpretation by linking the estimated factors to specific time events in the data.

		}

		\subsection{Sparse APCA: Formulation and Estimation}\label{est}
		In this paper, unlike the existing approach in sparse factor modeling where the loadings are sparse,  we assume the systematic risk factor process $\bff_t$ is sparse along the time horizon as it characterizes the co-movement of the panel dynamically. When all components of $\bff_t$ are zero at some certain time point, it implies that the panel is driven dominantly by individual factors or the corresponding idiosyncratic terms rather than a systematic one.  On the other hand, for the economic and financial panel series,  the nonzero factor $\bff_t$ at certain data points can be treated as a systematic response of the panel to some important and influential economic or financial events or outcomes.
		
		We first introduce some further notation used in the estimation procedure. For an $N\times 1$ vector $\bv=(v_1,...,v_N)'$, $\|\bv\|_2=\sqrt{\sum_{i=1}^2v_i^2}$ is the Euclidean norm and $\|\bv\|_0=\text{card}\{\text{support}(\bv)\}$ is the cardinality or number of non-zero elements in $\bv$. Let $\mathbb{V}$ be the set of $N\times r$ semi-orthogonal matrices and $\mathbb{V}^{\perp}$ be the set consisting of the orthogonal complements of the ones in $\mathbb{V}$.  {For a matrix $\bV = (\bv_1, \ldots, \bv_r) \in \mathbb{V}$ and a vector $\bs = (s_1, \ldots, s_r)'$,  $\|\bV\|_0 \leq \bs$ means that each column $\bv_i$ satisfies $\|\bv_i\|_0 \leq s_i$ for $1 \leq i \leq r$.}

In the following subsection, we first formulate the estimation procedure in the one-factor case, i.e. the number of factors $r=1$ in Model (\ref{MF}), and the multi-factor case can be carried out based on the one-factor estimation procedure. 
  
		\subsubsection{One-Factor Case}

		We consider the case when $r=1$ and the  first factor process over the timeline is $\underline{\bff}_1=(f_{1,1},...,f_{1,T})'$.
		Let $\bS={\bX\bX'}/{(NT)}$,  it follows that $\underline{\bff}_1/\sqrt{T}$ can be approximated by the  first normalized eigenvector of the $T\times T$ positive semi-definite matrix $\bS$ according to Model (\ref{MF}). We assume $\|\bv\|_0\leq s_1$, it suffices to solve the following optimization problem:
		\begin{equation}\label{opt1}
		\wh\bv=\arg\max_{\bv'\bv=1}\bv'\bS\bv,\,\,\text{subject to}\,\, \|\bv\|_0\leq s_1,
		\end{equation}
  and consequently, the estimator for $\underline{\bff}_1$ is denoted as $\underline{\wh\bff}_1=\sqrt{T}\wh\bv$ such that $\underline{\wh\bff}_1'\underline{\wh\bff}_1/T=1$, which satisfies the first condition in (\ref{Con1}). 
Note that the solutions in (\ref{opt1}) are the same as those obtained by PCA in \cite{bai2002determining} and \cite{gaotsay2022}, among others, if the constraint $\|\bv\|_0\leq s_1$ is removed. Furthermore, the problem in (\ref{opt1}) is also equivalent to finding the largest eigenvalue associated with an $s_1$-sparse eigenvector of the matrix $\bS$:
\begin{equation}\label{opt:eigenvalue}
\lambda_{\max}(\bS,s_1)=\max_{\bv'\bv=1} \bv'\bS\bv,   \,\,\text{subject to}\,\, \|\bv\|_0\leq s_1,
\end{equation}
which is a non-convex problem in general. In fact, it is  NP-hard because it can be reduced to the subset selection problem for ordinary least-squares problem; see \cite{moghaddam2006generalized}.

The optimization  in (\ref{opt1}) is still a sparse PCA problem symbolically, and numerous methods have been developed to obtain sparse eigenvectors during the past decades.  See the regularization method with elastic net in \cite{zou2006} and the penalized matrix decomposition (PMD) algorithm using $\ell_1$-penalty in \cite{witten2009penalized}, among others.

 In this paper, we extend the APCA method of \cite{connor1986performance,connor1988risk} to high dimensions by adopting an $\ell_0$-constraint on the sparse factors. We propose to use the truncated power method introduced in \cite{yuan2013} to solve the first normalized sparse eigenvector in problem (\ref{opt1}). The method is similar to the classical power method but includes an additional truncation operation to ensure sparsity. The rationale for this is as follows. First, the conventional APCA is conducted based on the following matrix perturbation formulation:
 \begin{equation}\label{mt:pert}
     \bS=\bSigma+\bE,
 \end{equation}
where $\bS$ is the matrix obtained by the noisy observation $\bx_t$, $\bSigma$ is a symmetric matrix whose eigenvectors are the true ones, and $\bE$ is a random perturbation. The decomposition of (\ref{mt:pert}) is formulated in (S.1) of the online Supplement. If the true largest eigenvector $\bv_1$ of $\bSigma$ is sparse, then it is natural to recover $\bv_1$ from the noisy matrix $\bS$. This recovery is guaranteed when the error $\bE$ is of a smaller order using the well-known $\sin \theta$ theorem in \cite{davis1970rotation} under the approximate factor model. Second, for any given vector $\bu_0\in R^{N}$, the power method estimates the first eigenvector by 
\begin{equation}\label{power}
\bu_{k}=\bS\bu_{k-1}=...=\bS^k\bu_{0}, \,\, k=1,2,...,
\end{equation}
and a normalized $\bu_{k}$ converges to the first eigenvector of $\bS$. To see this, note that there exist $c_1$,..., $c_N$ such that
\[\bu_0=c_1\wh\bv_1+c_2\wh\bv_2+...+c_N\wh\bv_N,\]
 where $\wh\bv_1,....,\wh\bv_N$ are the $N$ eigenvectors associated with the eigenvalues $\{\wh\lambda_i,i=1,...,N\}$ of $\bS$. Then, 
 \[\bu_k=\bS^k\bu_0=\sum_{i=1}^Nc_i\bS^k\wh\bv_i=\wh\lambda_1^k\left[c_1\wh\bv_1+\sum_{i=2}^N c_i(\frac{\wh\lambda_i}{\wh\lambda_1})^k\wh\bv_i\right].\]
Assuming that $|\wh\lambda_1|>|\wh\lambda_i|$ for $i\geq 2$, it follows that the direction of $\bu_k$ converges to that of $\wh\bv_1$.

 Motivated by the above discussion, we modify the power method by adding a truncation in each iteration, and the procedure is given in Algorithm \ref{alg:framework} below.   
		
\begin{algm}
[Estimation procedure of the first sparse risk factor process]
\label{alg:framework}\hfill
\begin{algorithmic}[1]
\State \textbf{Input}: The scaled covariance matrix $\bS=\bX\bX'/(NT)$, an initial vector $\bu_0\in R^{T}$, a cardinality integer $s_1\in\{1,...,T\}$;

\State Let $t=1$;
\Repeat
\State Compute $\bu_{t}^* = \bS\bu_{t-1}/\|\bS \bu_{t-1}\|$;

\State Let $l_t = \supp(\bu_t^*,s_1)$ be the indices of $\bu_t^*$ with the largest $s_1$
absolute values;

\State Compute $\hat{\bu}_t = \Trunc( \bu_t^*, l_t )$, where it only keeps the elements in $\bu_t^*$ with indices $l_t$;

\State Normalize $\bu_t = \hat{\bu}_t/\|\hat{\bu}_t\|$;

\State $t \leftarrow t + 1$;
\Until {Convergence};
\State \textbf{Output}: $\wh\bv_1=\bu_t$.
\end{algorithmic}
\end{algm}
The procedure, presented in Algorithm \ref{alg:framework}, generates a sequence of intermediate $s_1$-sparse eigenvectors $\{\bu_1, \bu_2,...\}$  from an initial sparse approximation $\bu_0$. At each time stamp $t$, the intermediate vector $\bu_{t-1}$ is multiplied by $\bS$, and then the entries are truncated to zeros except for the largest $s_1$ entries. The resulting vector is then normalized to unit length. Finally, the estimated first factor process is $\underline{\wh\bff}_1=\sqrt{T}\wh\bv_1$, where $\wh\bv_1$ is the output one in Algorithm \ref{alg:framework}. 
Instead of pre-specifying threshold values as in \cite{johnstone2009consistency} and \cite{ma2013sparse}, we rely on the sparsity parameter 
$s_1$, which is unknown in practice. To address this, we will propose a cross-validation procedure to determine the optimal value of 
$s_1$ later.

Algorithm~\ref{alg:framework} can consistently estimate sparse eigenvectors under a factor structure when the corresponding eigenvalues diverge. However, this property may not extend to general covariance matrices. For instance, it is possible to construct a diagonal covariance matrix with sparse eigenvectors, where the algorithm may fail to correctly identify the leading sparse eigenvectors for some initial vectors. For instance, consider 
$\bS=\diag(2,1)$. In this case, the leading eigenvector is 
$\bv_1=(1,0)'$, but the algorithm will incorrectly output 
$\bv_1=(0,1)'$ if the initial vector is 
$\bu_0=(1,3)'$. This occurs because the top eigenvalue is not sufficiently large. However, if 
$\bS=\diag(4,1)$, where the top eigenvalue is significantly larger—as is common in a factor model, the algorithm correctly identifies the leading eigenvector.

In practice, the convergence criterion in Algorithm~\ref{alg:framework} needs to be given first.  A common criterion for the convergence of Algorithm \ref{alg:framework} is that the two eigenvector iterates $\bu_{t}$ and $\bu_{t-1}$ for $t\geq 1$ satisfy
\begin{equation}\label{cri}
 \|\bu_t-\bu_{t-1}\|_\infty\leq \epsilon,
\end{equation}
for a small $\epsilon>0$. 
Simulation results suggest that the convergence is not sensitive to a sufficiently small $\epsilon$, and the numerical results in Section~\ref{sec4} indicate that the algorithm works well when we take $\epsilon=10^{-3}$ in (\ref{cri}).

		\subsubsection{Multi-Factor Case}
{In this section, we consider the general case where multiple factor processes are present in Model~(\ref{MF}), i.e., when $r > 1$. We extend the truncated power method to accommodate the multi-factor setting.}
	In the presence of more than one common factor process,  we need to apply Algorithm \ref{alg:framework} multiple times and extract all the common factors in a sequential way.  Specifically,  when $\wh\bv_1$ is given via Algorithm \ref{alg:framework}, 	we subtract the projection on first factor component of the panel and the residual would be $\wt\bX=\bX-\wh\bv_1\wh\bv_1'\bX=(\bI_T-\wh\bv_1\wh\bv_1')\bX$. The resulting scaled covariance $\wt\bS_1=(\bI_T-\wh\bv_1\wh\bv_1')\bS(\bI_T-\wh\bv_1\wh\bv_1')$. Then the second eigenvector $\wh\bv_2$ can be obtained by solving the following optimization problem (\ref{opt2}) below:
	\begin{equation}\label{opt2}
		\wt\bv_2=\arg\max \bv'\wt\bS_1\bv,\,\,\text{subject to}\,\, \bv'(\bI_T-\wh\bv_1\wh\bv_1')\bv=1\,\,\text{and}\,\,\|\bv\|_0\leq s_2,
		\end{equation}
		and the second eigenvector is $\wh\bv_2=\wt\bv_2/\|\wt\bv_2\|_2$. Consequently,  the second estimated factor process is $\underline{\wh\bff}_2=\sqrt{T}\wh\bv_2$.  We use the normalization condition $\bv'(\bI_T-\wh\bv_1\wh\bv_1')\bv=1$ to maximize the additional variance of the original matrix $\bS$, which is the same as the deflation framework in \cite{mackey2008}.  In other words, we need to eliminate the effect of the previous eigenvector directions by a projection method. Consequently, we formulate the procedure in Algorithm \ref{alg:2} below.

  \begin{algm}[A sequential estimation procedure of the  sparse risk factors]\hfill
\label{alg:2}
\begin{algorithmic}[1]
\State \textbf{Input}: The scaled covariance matrix $\bS$, the cardinalities of $r$ columns $\{s_1,...,s_r\}$ where $s_i\in\{1,2,...,T\}$;

\State Let $i=1$, $\bB_1=\bI_T$;
\Repeat
\State Given $s=s_i$, solve \[\wh\bv_i=\arg\min_{\bv'\bB_i\bv=1,\|\bv\|_0\leq s_i}\bv'\bS\bv;\]

\State Compute $\bq_i=\bB_i\wh\bv_i$;

\State Update $\bS$ by $\bS\leftarrow(\bI_T-\bq_i\bq_i')\bS(\bI_T-\bq_i\bq_i')$;

\State Update $\bB_{i+1}=\bB_i(\bI_T-\bq_i\bq_i')$;
\State Return $\wh\bv_i\leftarrow\wh\bv_i/\|\wh\bv_i\|_2$;
\State $i \leftarrow i + 1$;
\Until {$i=r+1$};
\State \textbf{Output}: $\{\wh\bv_1,...,\wh\bv_r\}$.
\end{algorithmic}
\end{algm}

From the optimization problem in (\ref{opt2}) and the procedure in Algorithm~\ref{alg:2}, we cannot enforce the orthogonality and sparsity at the same time, which is similar to the case in the sparse PCA framework. When there is no $\ell_0$-constraint in (\ref{opt1}) and (\ref{opt2}), we can easily obtain that $\wh\bv_1$ and $\wh\bv_2$ are the two normalized and orthogonal eigenvectors associated with the top two eigenvalues of $\bS$,  and  the approach reduces to the traditional APCA method.  {One of the differences between the one-factor case in (\ref{opt1}) and the multi-factor case in (\ref{opt2}) is that the estimated factors and loadings in the one-factor case satisfy the conditions in (\ref{Con1}), whereas those obtained in the multi-factor case may not. It is important to note that this discrepancy is only a finite sample outcome. Asymptotically, we can demonstrate that the multi-factor case still satisfies the identification conditions in (\ref{Con1}). Similar outcomes are also observed in the sparse PCA literature, including \cite{zou2006} and \cite{witten2009penalized}, as well as in the sparse factor modeling frameworks discussed in \cite{kristensen2017diffusion}.}

For each sparse vector $\wh\bv_i$ obtained in Algorithm~\ref{alg:2}, the estimated factor process is obtained as $\underline{\wh\bff}_i=\sqrt{T}\wh\bv_i$.	It is not hard to see that the number of iterations is $r$, and the $i$th iteration outputs a sparse estimator (multiplied by $\sqrt{T}$) of the $i$th column of $\bF$. For the optimization problem in Step 4 of Algorithm \ref{alg:2},  by the argument in Lemma~1 of the Supplement,  the matrix $\bB_i$ in each step is a symmetric one.  Therefore, we modify the truncated power method in Algorithm~\ref{alg:framework} and propose the following algorithm to obtain a sparse eigenvector for Step 4 of Algorithm~\ref{alg:2}.

\begin{algm}[Estimation of the $i$-th eigenvector in Algorithm \ref{alg:2}]\hfill
\label{alg:3}
\begin{algorithmic}[1]
\State For each $\bB=\bB_i$, perform a singular-value-decomposition (SVD) $\bB=\bU\bD\bU'$ and hence $\bB^{1/2}=\bU\bD^{1/2}\bU'$.  Let $\bA=\bB^{-1/2}\bS\bB^{-1/2}$;
\State  Initialize $j=1$ and choose an initial vector $\bx_0\in R^{N}$;
\Repeat
\State Compute $\wt\bx_j=\frac{\bB^{-1/2}\bA\bx_{j-1}}{\|\bA\bx_{j-1}\|_2}$;

\State Let $l_j=\text{supp}(\wt\bx_j,s)$ be the indices of $\wh\bx_t$ with the largest $s$ absolute values;
\State  Compute $\bx_j^*=\text{Truncate}(\wt\bx_j, l_j)$;
\State Let $\bx_j=\frac{\bB^{1/2}\bx_j^*}{\|\bB^{1/2}\bx_j^*\|_2}$;
\State $j \leftarrow j + 1$;
\Until {$\bx_j$ is convergent};
\State \textbf{Output:} $\wh\bv_i=\bx_j^*$, which is the last $\bx_j^*$ in Step 6.
\end{algorithmic}
\end{algm}

Note that the symmetric matrix $\bB_i$ in Algorithm~\ref{alg:3} is not strictly positive definite, the half-inverse $\bB_i^{-1/2}$ should be taken as a generalized one in line with the Moore-Penrose generalized inverse matrix.

\subsection{Determination of the Number of Factors}	\label{sec:ft}
The estimation method in Section \ref{est} depends on a known number of factors $r$. In practice, $r$ is unknown and we need to develop a data-driven method to estimate it. There are several useful methods developed in the past decades to estimate the number of factors including the information criterion in \cite{bai2002determining},
the random
matrix theory method in \cite{onatski2010determining}, the ratio-based method in \cite{lam2012factor} and \cite{Ahn2013}, 
and the white noise testing approach
in \cite{gao2022modeling}, among others.

We introduce two commonly used methods. The first is an information criterion (IC) approach adapted from \cite{bai2002determining}, with an added $\log(T)$ penalty to account for the scale of $s$. The number of factors $r$ is estimated by
\begin{equation}\label{rt}
    \wh r = \arg\min_{1 \leq k \leq K} \log\left( \frac{1}{NT} \|\bX - \wh\bF_k \wh\bLambda_k' \|_F^2 \right) + k \frac{N+T}{NT} \log(T) \log\left( \frac{NT}{N+T} \right),
\end{equation}
where $K$ is a pre-specified upper bound, and $\wh\bF_k$, $\wh\bLambda_k$ are the estimated factors and loadings for $k$ factors.
The other method is based on the eigenvalue ratios introduced by \cite{lam2012factor} and \cite{Ahn2013}. Specifically,
let $\wh\lambda_1\geq\wh\lambda_2\geq...\geq\wh\lambda_T$ be the $T$ sample eigenvalues of $\bX\bX'$, we adopt the ratio-based method  to estimate $r$ by
\begin{equation}\label{rhat}
            \wh r=\arg\min_{1\leq k\leq K}\wh\lambda_{k+1}/\wh\lambda_k,
\end{equation}
where $K$ is a prescribed upper bound as in (\ref{rt}) to control the stability of the ratios. In practice, we may choose $K=\min(T,N)/3$ under the assumption that the number of factors $r$ is usually not large.

	\subsection{Cross-Sectional Cross-Validation}\label{sec27}
 
		For a given number of factors $r$, which can be estimated by the methods in Section~\ref{sec:ft} above, if the cardinality of each column in $\bF$ is known, we can obtain the estimated factors $\wh \bF$ by solving the optimization problem using the proposed algorithms in Section~\ref{est}. The resulting estimated loading matrix is $\wh\bLambda'=(\wh\bF'\wh\bF)^{-1}\wh\bF'\bX$, which can be obtained by the Ordinary Least-Squares (OLS) method. When the true factors are sparse and orthogonal, the theoretical results in Section~\ref{sec3} below suggest that $\wh\bLambda\approx \bX'\wh\bF$, implying that each column loading vector is a sparse linear combination of the panel data over the timeline.

{For theoretical purposes, we may assume \( s_i / T = \delta_i \in (0,1) \), analogous to the dimensional asymptotics commonly considered in random matrix theory.} In practice, the cardinality of each column in $\bF$ is unknown, and we may choose them by the cross-validation method which is commonly used in machine learning literature. In fact, even though the sparsity parameters $s_1, ..., s_r$ of $r$ factor processes may not be the same, it is still a convenient way to assume {that $s_1=...=s_r=s_0\asymp T$} which can simplify the cross-validation procedure significantly. According to the theoretical results in Section~\ref{sec3} and the proofs in the Supplement, when the $r$ sparsity parameters are distinct, the convergence of the estimator for $s^*$ can be achieved, where $s^*=\max(s_1,...,s_r)$. Once $s^*$ is estimated in this scenario, it can be fixed, allowing the estimation of the second-largest sparsity parameter using the proposed method. For simplicity, in this section, we assume that the sparsity parameters for each column are identical. The cross-validation procedure is outlined as follows.
  
For each fixed $s\in \mathbb{S}$ where $\mathbb{S}$ is a chosen candidate set for the sparsity parameter $s$,   we first randomly divide the panel into two segments consisting of one training sample with $N_1$ components and the other a testing one with $N_2$ components, where $N_1\asymp N_2\asymp N$ and $N_1+N_2=N$. Equivalently, we partition the data matrix $\bX\in R^{T\times N}$ into $\wt\bX_1$ and $\wt\bX_2$, where $\wt\bX_1\in R^{T\times N_1}$ and $\wt\bX_2\in R^{T\times N_2}$. Let $\wt\bF_1^{s}$ be the estimated factors based on the training sample $\wt\bX_1$ and $\wt\bLambda_1$ be the estimated loading matrix. Define the  testing error as
		\begin{equation}\label{error}
		    R(s,\wt\bF_1^{s})=\|\wt\bX_2-\wt\bF_1^s(\wt\bF_1^{s}{'}\wt\bF_1^{s})^{-1}\wt\bF_1^{s}{'}\wt\bX_2\|_F^2/(N_2T).
		\end{equation}
  When we range the parameter $s$ over the candidate set $\mathbb{S}$, the optimal number of nonzero factors, denoted by $\wh s$, is the one that produces the smallest errors in (\ref{error}). On the other hand, we do not expect $s$ to be a large one to avoid the overfitting issue. Therefore, we define $g(N_1,T)$ as a penalty function, and let
  \[PC(s)=R(s,\wt\bF_1^{s})+rC_T\frac{s}{T}g(N_1,T),\]
 for some   $C_T>0$ which will be specified later. Then, $\wh s$ is defined as
  \begin{equation}\label{ICN}
   		\wh s=\arg\min_{s\in\mathbb{S}}PC(s)=\arg\min_{s\in\mathbb{S}}\{R(s,\wt\bF_1^{s})+rC_T\frac{s}{T}g(N_1,T)\},
  \end{equation}
implying that we choose the optimal $s$ which results in minimal testing errors in the cross-validations. 
There are many choices for the penalty function $g(N_1,T)$ as shown in Theorem~\ref{thm3} of Section~\ref{sec3} below. We will discuss the required properties of $g(N_1,T)$ such that we can establish the consistency of $\wh s$ in theory.

{In practice, we may restrict the candidate set $\mathbb{S}$ to the range $c_1T \leq s \leq c_2T$, where $c_1, c_2 \in (0,1)$ are prescribed constants.  We may choose $C_T=\log(T)$ in order to balance the magnitude of the scale of $s$.}  Note that the cross-validation method is different from the classical one in the machine learning literature where the samples are partitioned along the time horizon, whereas we partition the samples over the cross-section in this paper. In fact, we can even perform more cross-validation experiments in order to obtain an optimal parameter $s$. For example, we can choose a sufficiently large integer $J>0$ and  obtain $J$ testing errors as that in (\ref{error}) for each fixed $s$ by partitioning the samples for $J$ times. We may denote the testing error by $R_j(s,\wt\bF_{1,j}^{s})$ for the $j$-th random partition and the average testing error for $s$ is defined as
  \begin{equation}\label{cr:rs}
      R^J(s)=\frac{1}{J}\sum_{j=1}^J R_j(s,\wt\bF_{1,j}^{s}),
  \end{equation}
  where $\wt\bF_{1,j}^{s}$ is the estimated factors based on the training sample in the $j$-th cross-validation.
Then, the optimal number of nonzero factors $\wh s$ is the one that produces the smallest error of the information criterion in (\ref{ICN}) by replacing the $R(s,\wt\bF_1^{s})$ therein by $R^J(s)$ in Equation (\ref{cr:rs}) for $s\in \mathbb{S}$.

 

\section{Theoretical Properties}\label{sec3}
In this section, 
we present the consistency and asymptotic bounds of the estimators from Section~\ref{sec2}  as  $N,T\rightarrow\infty$.  We first derive the asymptotic results assuming known $r$ and $s_i$, and then establish the consistency of their estimators. Mathematical proofs are provided in the Supplement.

To derive the asymptotic properties of the proposed method and estimators from Section~\ref{sec2}, we need the following high-level assumptions. These assumptions can be readily verified by some standard and more primitive assumptions, which will be discussed in detail later. Most of the assumptions below are commonly used in the PCA or approximate-factor modeling literature, and some of which are stronger than those in \cite{bai2002determining} in order to show that the estimated sparse factor $\wh\bF$ is close to $\bF$. We will use $C$ or $c$ to denote a generic positive constant the value of which may change at different places.

\begin{assumption}\label{asm01}
The process $\{\bff_t\}$ is $\alpha$-mixing with the mixing coefficients satisfying the condition $\alpha_N(k)<\exp(-k)$, where $\alpha_N(k)$ is defined as
\begin{equation}\label{amix}
\alpha_N(k)=\sup_{i}\sup_{A\in\mathcal{F}_{-\infty}^i,B\in \mathcal{F}_{i+k}^\infty}|P(A\cap B)-P(A)P(B)|,
\end{equation}
where $\mathcal{F}_i^j$ is the $\sigma$-field generated by $\{\bff_t:i\leq t\leq j\}$. 
\end{assumption}
\begin{assumption}\label{asm02}
Assume 
$\frac{1}{T}\sum_{t=1}^T\bff_t\bff_t'\rightarrow_p\bI_r$.
\end{assumption}
\begin{assumption}\label{asm03}{
There exists a vector $\bs=(s_1,...,s_r)'$ with $s_i/T=\delta_i\in(0,1)$ for $1\leq i\leq r$ such that $\|\bF\|_0\leq \bs$, where $\bF$ is defined as that in (\ref{MF}).}
\end{assumption}
\begin{assumption}\label{asm04}
The eigenvalues of $\frac{1}{N}\bLambda'\bLambda$ are distinct and bounded away from $0$ and $\infty$ as $N\rightarrow\infty$.
\end{assumption}
\begin{assumption}\label{asm05}
$e_{i,t}=\sigma_i\epsilon_{i,t}$ for some $0<\underline{\sigma}\leq \sigma_i\leq \bar{\sigma}<\infty$, where $\epsilon_{i,t}$ is independent and identically distributed over $i$ and $t$.
\end{assumption}
\begin{assumption}\label{asm06}
$\{\blambda_i\}$, $\{\bff_t\}$, and $\{e_{i,t}\}$ are mutually independent groups.
\end{assumption}
\begin{assumption}\label{asm07}
    $\underline{\bff}_i$ and $\underline{\be}_j$ are sub-Gaussian variables for $i=1,...,r$ and $j=1,...,N$ in the sense that 
    \[P(|\bv'\underline{\bff_i}|>x)\leq c\exp(-cx^2)\quad\text{and}\quad P(|\bv'\underline{\be_j}|>x)\leq c\exp(-cx^2),\]
    for any $\|\bv\|_2=1$.
\end{assumption}
Assumption~\ref{asm01} is standard to characterize the dynamic dependence of 
the factor processes. See, for example, the Lemma 1 in the appendix of \cite{gao2019banded}. Assumptions~\ref{asm02}-\ref{asm03} establish an identification condition and a sparsity structure for the true factor process 
$\bF$. {While the factors are sparse, Assumption~\ref{asm02} is not restrictive, as it is reasonable to assume that \( s_i \) diverges. Moreover, the factors can be normalized such that Assumption~\ref{asm02} holds under the condition \( s_i / T = \delta_i \in (0,1) \), as specified in Assumption~\ref{asm03}, which resembles the setting commonly adopted in random matrix theory such as those in \cite{bai2010spectral}. Alternatively, one could adopt the framework of \cite{uematsu2022estimation}, which models factors as weak with varying strengths. However, specifying the strength levels in practice can be challenging. Instead, we adopt the random matrix theory perspective and assume that the sparsity parameter grows at a fractional rate relative to the time dimension \( T \). This is purely a theoretical device and does not impact our primary goal of identifying nonzero factors over time.} Assumption \ref{asm04} specifies that the top eigenvalues of the covariance of the panel data are distinct in order to avoid multiplicity of eigenvalues.  Assumptions~\ref{asm02} and \ref{asm04} provide identifiability for the Model (\ref{MF}) so that the factors and the loading matrix can be uniquely determined asymptotically. See \cite{bai2013principal} for a detailed argument. {Assumption~\ref{asm05} is primarily imposed to simplify theoretical derivations. It can be relaxed to weaker conditions, such as Conditions (c.i)–(c.iii) in \cite{bai2013principal}, or replaced with strong mixing assumptions in both spatial and temporal dimensions, allowing the use of the Bernstein-type inequality in \cite{merlevede2011bernstein} for the analysis. Similar independence assumptions are also found in \cite{onatski2012asymptotics} and Assumption 3 of \cite{huang2022scaled}. Assumption~\ref{asm07} facilitates the derivation of a Bernstein-type concentration bound for establishing convergence rates of the estimated factors, and can likewise be weakened following \cite{merlevede2011bernstein}.}

The following theorem establishes  the consistency of the first estimated factor process.
		
		\begin{theorem}\label{thm1}
		Let Assumptions \ref{asm01}--\ref{asm07}  hold and $\wh\bv_1$ be the first solution of (\ref{opt1}). As $N$ and $T$ increase, it holds that
		\[\sqrt{1-(\wh\bv_1'\bv_1)^2}=O_p\left(\sqrt{\frac{s_1\log(T)}{NT}}+\frac{s_1\log(T)}{NT}+\frac{1}{T}\right),\]
		where $\wh\bv_1=\underline{\wh\bff}_{1}/\sqrt{T}$ and $\bv_1=\underline{\bff}_{1}/\sqrt{T}$.
        \end{theorem}
{To maintain clear interpretation, we continue to use 
$s_1$
  instead of 
$s_1\asymp T$ in the theorem.  From Theorem~\ref{thm1}, we see that the angle between the estimated direction $\wh\bv_1$ and the true one $\bv_1$ is asymptotically equal to zero or $\pi$ if  $s_1\log(T)/(NT)\asymp \log(T)/N\rightarrow 0$. In fact, we can rewrite $\wh\bv_1'\bv_1$ as $\cos(\theta)$ where $\theta$ is the angle between $\wh\bv_1$ and $\bv_1$. Some remarks for Theorem~\ref{thm1} are as follows.}

  \begin{remark}
      (i) It follows immediately from Theorem~\ref{thm1} that
      \[\sqrt{1-\cos^2(\theta)}=|\sin(\theta)|=O_p\left(\sqrt{\frac{s_1\log(T)}{NT}}+\frac{s_1\log(T)}{NT}+\frac{1}{T}\right).\]
      If $s_1\log(T)/(NT)\rightarrow 0$, we have that $\sin(\theta)\rightarrow 0$ asymptotically, implying that $\theta\rightarrow 0$ or $\theta\rightarrow \pi$. Therefore, $\wh\bv_1$ and $\bv_1$ coincide on the same line.\\
      (ii) {When the two vectors $\wh\bv_1$ and $\bv_1$ point to the same direction in the sense that the angle $\theta$ is acute, then $\wh\bv_1'\bv_1\geq 0$ and $1\leq 1+\wh\bv_1'\bv_1\leq 2$. Theorem~\ref{thm1} implies that
      \[\|\wh\bv_1-\bv_1\|_2=\sqrt{2(1-\wh\bv_1'\bv_1)}\leq \sqrt{2(1-(\wh\bv_1'\bv_1)^2)}=O_p\left(\sqrt{\frac{s_1\log(T)}{NT}}+\frac{s_1\log(T)}{NT}+\frac{1}{T}\right).\]
    Consequently,
    \[\frac{1}{\sqrt{T}}\|\underline{\wh\bff}_{1}-\underline{\bff}_{1}\|_2=O_p\left(\sqrt{\frac{\log(T)}{N}}+\frac{1}{T}\right),\]
which is in line with the conventional result in factor modeling if we ignore the $\log(T)$ term. See \cite{gao2023supervised} for details. }   
      
  \end{remark}

  Next, we present the theorem concerning the consistency of all the estimated factors along the time horizon. For the distance between two matrices $\bH_1$ and $\bH_2$, there are several measures that are  used in the literature. For example, we adopt the discrepancy measure used by
\cite{pan2008modelling}: for two $T\times r$ semi-orthogonal 
matrices ${\bf H}_1$ and ${\bf H}_2$ satisfying the condition ${\bf
H}_1'{\bf H}_1={\bf H}_2'{\bf H}_2=\bI_{r}$, the difference
between the two linear spaces $\mathcal{M}({\bf H}_1)$ and
$\mathcal{M}({\bf H}_2)$ is measured by
\begin{equation}
D(\mathcal{M}({\bf H}_1),\mathcal{M}({\bf
H}_2))=\sqrt{1-\frac{1}{r}\textrm{tr}({\bf H}_1{\bf H}_1'{\bf
H}_2{\bf H}_2')}.\label{eq:D}
\end{equation}
Note that $D(\mathcal{M}({\bf H}_1),\mathcal{M}({\bf
H}_2)) \in [0,1].$ 
It is equal to $0$ if and only if
$\mathcal{M}({\bf H}_1)=\mathcal{M}({\bf H}_2)$, and to $1$ if and
only if $\mathcal{M}({\bf H}_1)\perp \mathcal{M}({\bf H}_2)$. By Lemma A1(i) in \cite{pan2008modelling}, $D(\cdot,\cdot)$ is a well-defined distance measure on some quotient space of matrices. Alternatively, we may also adopt the measure 
\begin{equation}\label{rho:d}
\rho(\bH_1,\bH_2)=\|\bH_1\bH_1'-\bH_2\bH_2'\|_F,
\end{equation}
which is the Frobenius norm of the difference between the projection matrices of two spaces and is also a well-defined distance between linear subspaces. In addition, if we denote the singular values of $\bH_1'\bH_2$ by $\{\sigma_i\}_{i=1}^r$, in descending order, then the principal angles between $\mathcal{M}(\bH_1)$ and $\mathcal{M}(\bH_2)$, $\bTheta(\bH_1,\bH_2)=\diag(\theta_1,...,\theta_r)$, are defined as $\diag\{\cos^{-1}(\sigma_1),...,\cos^{-1}(\sigma_r)\}$; see, for example,  Theorem I.5.5 of \cite{stewart1990matrix}. The  squared Frobenius norm of the so-called $\sin\bTheta$ distance, defined as 
\begin{equation}\label{sin:t}
\|\sin\bTheta(\bH_1,\bH_2)\|_F^2:=\sum_{i=1}^r\sin^2(\theta_i), 
\end{equation}
can also be used to measure the distance between two linear spaces. In fact, if $r$ is finite, the distances in (\ref{eq:D})--(\ref{sin:t}) are equivalent, because
\begin{align}\label{equi}
\rho^2(\bH_1,\bH_2)=&\|\bH_1\bH_1'\|_F^2+\|\bH_2\bH_2'\|_F^2-2\tr(\bH_1\bH_1'\bH_2\bH_2')=2r\{D(\mathcal{M}({\bf H}_1),\mathcal{M}({\bf
H}_2))\}^2\notag\\
=&2r-2\|\bH_1'\bH_2\|_F^2=2\sum_{i=1}^r(1-\sigma_i^2)=2\sum_{i=1}^r\sin^2(\theta_i)=2\|\sin\bTheta(\bH_1,\bH_2)\|_F^2.
\end{align}
Therefore, we shall only use the distance in (\ref{rho:d}) to present our theoretical results in the main article.

  
				\begin{theorem}\label{thm2}
		Let Assumptions \ref{asm01}--\ref{asm07}  hold and $\wh\bV$ be the matrix consisting of the estimated sparse eigenvectors obtained by Algorithm~\ref{alg:2}. As $N$ and $T$ increase, it holds that
		\[\rho(\wh\bV,\bV):=\|\wh\bV\wh\bV'-\bV\bV'\|_F=O_p\left(\sqrt{\frac{s^*\log(T)}{NT}}+\frac{s^*\log(T)}{NT}+\frac{1}{T}\right),\]
		where $s^*=\max\{s_i\}_{i=1}^{r}$, $\wh\bV=\wh\bF/\sqrt{T}$, and $\bV=\bF/\sqrt{T}$.
		\end{theorem}
		
\begin{remark}
    (i) From Theorem~\ref{thm2}, we may plug the estimated factor and obtain
    \[\sqrt{\sum_{i=1}^r\sin^2(\theta_i)}=O_p\left(\sqrt{\frac{s^*\log(T)}{NT}}+\frac{s^*\log(T)}{NT}+\frac{1}{T}\right),\]
    implying that all the principal angles will be asymptotically zero or $\pi$ if $s^*\log(T)=o(NT)$. \\
    (ii) Suppose the singular values of $\wh\bV'\bV$ are $\{\sigma_i,i=1,...,r\}$ where $1\geq \sigma_i\geq 0$, then $\Tr(\wh\bV'\bV)=\sum_{i=1}^r \sigma_i$. If all the directions of the estimated eigenvectors and the true ones coincide on the same line, by an elementary argument, we can show that
    \[\|\wh\bV-\bV\|_F^2=\Tr[(\wh\bV-\bV)'(\wh\bV-\bV)]=2\sum_{i=1}^r(1-\sigma_i)\leq2\sum_{i=1}^r(1-\sigma_i^2),\]
    where we used the inequality $(1-\sigma_i)\leq (1-\sigma_i)(1+\sigma_i)=(1-\sigma_i^2)$. 
    {Consequently, by (\ref{equi}) and Assumption~\ref{asm03}, we have that 
     \[\frac{1}{\sqrt{T}}\|\wh\bF-\bF\|_F\asymp \rho(\wh\bV,\bV)=O_p\left(\sqrt{\frac{\log(T)}{N}}+\frac{1}{T}\right),\]
     which is similar to the one in Remark 1(ii) for each single factor process.}
\end{remark}

Next, we establish the theoretical results for the estimated loading matrix $\wh\bLambda$ in the following theorem.

\begin{theorem}\label{thm20}{
    Let Assumptions~\ref{asm01}--\ref{asm07} hold.\\
    (i) There exists a rotation matrix $\bH_s$ such that
    \begin{equation}
        \max_{1\leq i\leq N} \|\wh\blambda_i-\bH_s\blambda_i\|_2=O_p(\frac{\sqrt{\log N}}{N}+\sqrt{\frac{\log N}{T}}),
    \end{equation}
    where $\bH_s=(\wh\bF'\wh\bF)^{-1}\wh\bF'\bF$. Furthermore, if ${\sqrt{T}}/{N}=o(1)$, then
    \[    \max_{1\leq i\leq N} \|\wh\blambda_i-\bH_s\blambda_i\|_2=O_p(\sqrt{\frac{\log(N)}{T}}).\]
    (ii) For $1\leq i \leq N$, if ${\sqrt{T}}/{N}=o(1)$, then there exists a rotation matrix $\bH_s$ as above such that
    \begin{equation}\label{lim:dis}
       \sqrt{T}(\wh\blambda_i-\bH_s\blambda_i)\longrightarrow_d N(0,\bQ^{-1}\bGamma_i\bQ^{-1}),
    \end{equation}
    where $\bQ$ is the limit of $\wh\bF'\wh\bF/T$, and $\bGamma_i=\lim_{T\rightarrow\infty}\Var(\frac{1}{\sqrt{T}}\sum_{t=1}^T\bff_te_{i,t})$ is the long-run covariance matrix. }
\end{theorem}
\begin{remark}\label{rm30}
    By Assumptions~\ref{asm02} and the results in Theorem~\ref{thm1}, it is not hard to show that $\bQ=\bI_r$, but we can use the sample version $\wh\bF'\wh\bF/T$ in empirical applications. If Assumption~\ref{asm05} holds, then the variance term in (\ref{lim:dis}) reduces to $\sigma_i^2\bI_r$, where $\sigma_i$ can be estimated from the residuals. Under the general setting that the noise are not $i.i.d.$, we can use the well-known heteroskedasticity and autocorrelation consistent (HAC) estimator such as that in \cite{andrews1991heteroskedasticity} to approximate the long-run covariance term $\bGamma_i$. 
\end{remark}

The following corollary establishes the consistency of the estimated number of factors obtained via (\ref{rt}) or (\ref{rhat}).

\begin{corollary}
Let Assumptions~\ref{asm01}--\ref{asm07} hold. Suppose $N \asymp T$, $\|\be\be'/N\|_2 \leq \bar{c} < \infty$, and the $K$-th largest eigenvalue satisfies $\|\be\be'/N\|_K \geq \underline{c} > 0$, where $K = \min(N, T)/3$ as in (\ref{rt})–(\ref{rhat}). Then both the information criterion in (\ref{rt}) and the ratio-based method in (\ref{rhat}) consistently estimate $r$.
\end{corollary}

Finally, we provide the consistency of the cross-validation method in estimating the sparsity parameters $s_1,...,s_r$. We consider the case where the sparsity parameters of each column of $\bF$ are the same, i.e., $s_1=...=s_r=s_0$, as discussed in Section~\ref{sec27}. Otherwise, the following theorem states the consistency for estimating the largest sparsity parameter among the 
$r$ columns. {Since $s_0 \asymp T$, it is generally challenging to derive consistency of $\widehat{s}$ when both $N$ and $T$ grow. Therefore, we restrict attention to the asymptotic regime where $N \to \infty$ and $T$ is large but fixed.}


\begin{theorem}\label{thm3}
    Let Assumptions~\ref{asm01}--\ref{asm07} hold. Suppose $s_0=c_0T$ and $\wh s$ is the solution in (\ref{ICN}) with $\mathbb{S}=[c_1T,c_2T]$ where $0<c_1\leq c_0\leq c_2<1$.  If $C_Tg(N_1,T)\rightarrow 0$ and $C_{N_1T}^{-2}C_Tg(N_1,T)\rightarrow\infty$ as $N\rightarrow\infty$ with large $T$,  where $C_{NT}=\sqrt{{\log(T)}/{N}}+{1}/{T}$, we have
    \[\lim_{N\rightarrow\infty} P(\wh s=s_0)=1,\]
    where $s_0$ is the sparsity of each column of $\bF$.
\end{theorem}
	\begin{remark}\label{rm3}
	(i) Theorem~\ref{thm3} shows consistency of $\widehat{s}$ as $N \to \infty$ with fixed large $T$. Its proof also implies that $\widehat{s}/T$ consistently estimates $s_0/T$ as both $N, T \to \infty$.\\
	    (ii) When the columns of $\bF$ have distinct sparsity parameters, it is an important step to estimate the largest one among $\{s_1,...,s_r\}$ first, and 
     $\wh s$ in Theorem~\ref{thm3} is an estimator for $s^*=\max\{s_1,...,s_r\}$. In other words, for large $T$, we can show that
     \begin{equation}\label{shat:c}
     \lim_{N\rightarrow\infty} P(\wh s=s^*)=1.    
     \end{equation}
     On the other hand, estimating the largest sparsity parameter is often adequate since all the important factors over the timeline can be recovered. Furthermore, we can subsequently estimate the sparsity of each factor sequence using a coordinate descent approach.\\
     (ii) In practice, there are many choices for $g(N_1,T)$ in (\ref{ICN}). For example, we may take $C_T=\log(T)$,
     $g(N_1,T)=\frac{\sqrt{N_1}+T}{\sqrt{N_1}T}\log(\frac{\sqrt{N_1}T}{\sqrt{N_1}+T})$.
    By an argument similar to that in Corollary 1 of \cite{bai2002determining}, 
     \begin{equation}\label{Infor}
           IC(s)=\ln( R^J(s))+r\frac{s}{T}\frac{\sqrt{N_1}+T}{\sqrt{N_1}T}\log(T)\log(\frac{\sqrt{N_1}T}{\sqrt{N_1}+T})
     \end{equation}
     can also consistently estimate $s_0$, where $r$ can be replaced by $\wh r$ obtained in (\ref{rhat}).
	\end{remark}

\section{Simulation Evidence}\label{sec4}

In this section, we illustrate the finite-sample properties of the proposed methodology under different choices of $N$ and $T$. To ensure our simulation results are reproducible, we set the seed to \texttt{1234} in \texttt{R} programming.
		
\subsection{One-factor Case}\label{sec41}
First, we consider the one-factor case in Model (\ref{comF}), i.e., $r=1$ is used in this section. The factor process is generated by
\[f_t=\phi f_{t-1}+\eta_t,\quad \eta_t\overset{i.i.d.}{\sim} N(0,1),\quad t=1,...,T,\]
where we set $\phi=0.5$. We consider the dimensions $N=50,100,150,300$ and $500$, with sample sizes  $T=200,500,800,1000$, and $1200$ for each $N$. Each loading value is generated independently from the uniform distribution $U(-2,2)$, and the $N$-dimensional loading vector is then re-normalized to have an $l_2$-norm strength of $\sqrt{N}$. For each configuration of $(N,T)$, {we choose the sparsity $s=T/10$},  and we randomly generate $s$ integers from $\{1,...,T\}$ and keep those $s$ elements in $\underline{\bff}_1=(f_1,...,f_T)'$ to be nonzero, re-normalizing the sparse vector $\underline{\bff}_1=(f_1,...,f_T)'$ to have unit variance. We consider two scenarios for the idiosyncratic terms $\be_t$'s in each experiment:
\[(1)\,\, \be_t\overset{i.i.d.}{\sim} N({\bf 0},\bI_N)\quad\text{and}\quad (2)\,\, \be_t=\bPhi\be_{t-1}+\bve_t,\bve_t\overset{i.i.d.}{\sim}N({\bf 0},\bI_N), t=1,...,T,\]
where $\bPhi$ is a diagonal matrix with the diagonal elements generated from $U(0.5,0.9)\cup U(-0.9,-0.5)$. A total of 500 replications are used throughout the experiments.

We first study the accuracy of the estimated factors by defining the estimation error in each replication as
\begin{equation}\label{f:measure}
    d(\underline{\wh\bff}_1,\underline{\bff}_1)=\sqrt{1-(\underline{\wh \bff}_1'\underline{\bff}_1/T)^2}.
\end{equation}
{Table~\ref{Table-a1} reports the average estimation errors over 500 replications under two scenarios for the idiosyncratic components. {\it Proposal} refers to our proposed method, while {\it Lasso} denotes the Lasso-regularized approach applied to the factor process, following a similar procedure to Algorithm 1 in \cite{kristensen2017diffusion}, with the tuning parameter set to $\psi_T = 0.6f_0$, where $f_0$ is the minimum nonzero component (in absolute value) of $\underline{\bf f}_1$. As shown in Table~\ref{Table-a1}, for each scenario, the estimation error decreases as the dimension $N$ increases for a fixed sample size $T$. In contrast, for fixed $N$, the error remains relatively stable or slightly increases as $T$ grows, which is consistent with the theoretical rate $\sqrt{\log(T)/N}$—a term that dominates $1/T$ for large $T$—as established in Theorem~\ref{thm1}. These results are consistent with our asymptotic theory.  Moreover, the results clearly indicate that the Lasso procedure tends to incur additional estimation errors across all configurations of $(N, T)$. This highlights the advantage of the proposed hard-thresholding technique.
}

\begin{table}[ht]
\caption{The estimation accuracy of factors when the number of factors $r=1$. The results are based on the average of the measures  defined in  (\ref{f:measure}) through 500 replications. $N$ and $T$ denote the dimension and the sample size, respectively.  The Proposal refers to our proposed method, while Lasso denotes the Lasso-regularized approach applied to the factor process, which is similar to Algorithm 1 in \cite{kristensen2017diffusion}.} 
          \label{Table-a1}
{\begin{center}
\begin{tabular}{ccccccccccccc}
\toprule
&&\multicolumn{5}{c}{$\be_t\overset{i.i.d.}{\sim} N({\bf 0},\bI_N)$}&&\multicolumn{5}{c}{$\be_t=\Phi\be_{t-1}+\bve_t,\bve_t\overset{i.i.d.}{\sim} N({\bf 0},\bI_N)$}\\
\cline{3-7}\cline{9-13}
&&\multicolumn{5}{c}{$T$}&&\multicolumn{5}{c}{$T$}\\
\cline{3-7}\cline{9-13}
Method&$N$&200&500&800&1000&1200&&200&500&800&1000&1200\\
\hline
&50&0.060&0.062&0.062&0.062&0.062&&0.095&0.102&0.097&0.097&0.096\\
&100&0.040&0.041&0.041&0.041&0.041&&0.064&0.068&0.068&0.071&0.065\\
Proposal&150&0.031&0.033&0.033&0.033&0.033&&0.053&0.054&0.055&0.054&0.054\\
&300&0.021&0.022&0.022&0.022&0.022&&0.035&0.036&0.035&0.035&0.035\\
&500&0.016&0.016&0.012&0.017&0.017&&0.026&0.027&0.027&0.027&0.027\\
\midrule
&50&0.264&0.200&0.185&0.176&0.172&&0.312&0.261&0.240&0.225&0.229\\
&100&0.245&0.176&0.152&0.144&0.137&&0.273&0.214&0.193&0.193&0.177\\
Lasso&150&0.230&0.163&0.139&0.131&0.123&&0.256&0.192&0.174&0.164&0.157\\
&300&0.229&0.156&0.123&0.116&0.108&&0.242&0.172&0.141&0.135&0.127\\
&500&0.218&0.144&0.116&0.111&0.101&&0.226&0.155&0.129&0.126&0.116\\
\bottomrule
\end{tabular}
  \end{center}}
\end{table}

Next, we study the accuracy of the empirical recovery (ER) of the sparse factors. Let $S$ be the set of true indexes of the $s$ nonzero elements in $\underline{\bff}_1$, and $\wh S$ the set of indices of the $\wh s$ nonzero elements in $\underline{\wh \bff}_1$. For each replication, define the empirical recovery rate of the non-sparse indices as
\begin{equation}\label{er:sps}
    ER(s):=\frac{\#\{S\cap\wh S\}}{s},
\end{equation}
where $\#\{S\cap\wh S\}$ is the cardinality of the intersection between the estimated indices and the true ones. Table~\ref{Table-a2} presents the empirical recovery rate of the sparsity in the factor process using the measure defined in (\ref{er:sps}).  {Similar to the comparison in Table~\ref{Table-a1}, we also compare the proposed method with the Lasso approach.  From Table~\ref{Table-a2}, we observe that empirical accuracy improves as the dimension $N$ increases for each fixed sample size $T$, which is consistent with the asymptotic results presented in Section~\ref{sec3}. In addition, the Lasso procedure performs comparably to the proposed method in recovering the sparsity structure of the factor process, with the proposed method only slightly outperforming Lasso in certain cases. Taken together with the results in Table~\ref{Table-a1}, this suggests that the hard-thresholding technique provides an improvement in accuracy while remaining easy to implement.
}

\begin{table}[ht]
\caption{Empirical recovery rate of the sparsity in the factor process when the number of factors $r=1$. The empirical recovery rate in each replication is calculated as in (\ref{er:sps}). $N$ and $T$ denote the dimension and the sample size, respectively. 500 replications are used in the experiments. The Proposal refers to our proposed method, while Lasso denotes the Lasso-regularized approach applied to the factor process, which is similar to Algorithm 1 in \cite{kristensen2017diffusion}.} 
          \label{Table-a2}
{\begin{center}
\begin{tabular}{ccccccccccccc}
\toprule
&&\multicolumn{5}{c}{$\be_t\overset{i.i.d.}{\sim} N({\bf 0},\bI_N)$}&&\multicolumn{5}{c}{$\be_t=\Phi\be_{t-1}+\bve_t,\bve_t\overset{i.i.d.}{\sim} N({\bf 0},\bI_N)$}\\
\cline{3-7}\cline{9-13}
&&\multicolumn{5}{c}{$T$}&&\multicolumn{5}{c}{$T$}\\
\cline{3-7}\cline{9-13}
Method&$N$&200&500&800&1000&1200&&200&500&800&1000&1200\\
\hline
&50&0.916&0.909&0.909&0.909&0.909&&0.880&0.868&0.873&0.880&0.874\\
&100&0.937&0.934&0.933&0.934&0.934&&0.908&0.902&0.904&0.901&0.906\\
Proposal&150&0.947&0.944&0.944&0.944&0.944&&0.919&0.919&0.918&0.918&0.917\\
&300&0.962&0.961&0.959&0.959&0.958&&0.943&0.941&0.941&0.940&0.941\\
&500&0.969&0.969&0.968&0.967&0.968&&0.956&0.952&0.952&0.955&0.953\\
\midrule
&50&0.916&0.909&0.909&0.909&0.909&&0.878&0.867&0.872&0.879&0.873\\
&100&0.937&0.934&0.933&0.934&0.934&&0.908&0.901&0.903&0.901&0.906\\
Lasso&150&0.947&0.944&0.944&0.944&0.944&&0.918&0.918&0.918&0.917&0.917\\
&300&0.962&0.961&0.959&0.959&0.958&&0.943&0.941&0.940&0.940&0.941\\
&500&0.969&0.969&0.968&0.967&0.968&&0.955&0.952&0.952&0.955&0.953\\
\bottomrule
\end{tabular}
  \end{center}}
\end{table}
Finally, we evaluate the distribution of the estimated loadings as described in Theorem~\ref{thm20}. For simplicity, we consider the case when the noises are $i.i.d.$ and plot the empirical histogram of the estimated loading associated with the first series in Figure~S.2 of the Supplement. The variance of the normal curve is based on the average of the estimated variance across 2000 replications. From Figure~S.2, we see that the estimators behave closely to normal, which is in agreement with our asymptotic theory.


\subsection{Multi-factor Case}

In this section, we further verify the efficacy of the proposed algorithms when there are multiple factors. The number of factors is set to be $r=3$, and the factors are generated by
\[\bff_t=\bPsi\bff_{t-1}+\bfeta_t,\bfeta_t\overset{i.i.d.}{\sim}N({\bf 0},\bI_r),t=1,...,T,\]
where $\Psi=\diag(0.5,-0.6,0.7)$ is a diagonal matrix. We consider the dimensions $N=50,100,150,200$, and $300$, with the sample size $T=100,200,300,500$, and $800$ for each $N$ in this experiment. For each configuration of $(N,T)$, we set the sparsity parameters to $s_1=s_2=s_3=T/10$.  We first randomly generate $s_1$ indexes from $\{1,...,T\}$ to form an index set $S_1$, such that we only keep the corresponding $s_1$ time points of $\underline{\bff}_1$ and the remaining ones are set to zero. The index set $S_2$ is formed by generating  $s_2$ indexes  $\{1,...,T\} \setminus S_1$, keeping  only those $s_2$ locations of $\underline{\bff}_2$. the sparse $\underline{\bff}_3$ is obtained by repeating the above procedure. Then each $\underline{\bff}_j$ is normalized to have unit variance. For the generation of the loading matrix $\bLambda\in R^{N\times r}$, we first generate an $N\times r$ matrix $\bM$ with elements independently generated from $U(-2,2)$, we then perform a singular-value decomposition on $\bM$, and the left singular matrix is $\bU$. The loading $\bLambda$ is taken as $\bU$ multiplied by $\sqrt{N}\diag(3,2,1)$ on its right. We also consider the two scenarios for the idiosyncratic terms as in Section~\ref{sec41}. A total of 500 replications are used throughout the experiments.

Now, we first study the estimation accuracy of the factor processes. We define the measure of the errors of the estimated factors as
\begin{equation}\label{F:error}
    d(\wh\bF,\bF)=\|\frac{\wh\bF\wh\bF'}{T}-\frac{\bF\bF'}{T}\|_F.
\end{equation}
The average estimation errors of the factors over 500 replications are reported in Table~S.I of the Supplement. From Table~S.I, we observe a similar pattern to that in Section~\ref{sec41}. For each fixed sample size 
$T$,  the error decreases as the dimension 
$N$ increases, which is in agreement with our asymptotic theory in Theorem~\ref{thm3}, regardless of whether the idiosyncratic terms are \textit{i.i.d.} or dynamically dependent.



Furthermore, we study the empirical recovery (ER) of the sparsity in the factor processes. Similar to the measure in (\ref{er:sps}), we define
\begin{equation}\label{er:sp:3}
    ER(s):=\frac{\#\{\wh S_1\cap S_1 \}+\#\{\wh S_2\cap S_2 \}+\#\{\wh S_3\cap S_3 \}}{3s},
\end{equation}
{where $s=T/10$} and $\wh S_j$ is the the estimated nonzero locations in $\underline{\wh \bff}_j$ for $j=1,2$, and $3$. Table~S.II of the Supplement reports the empirical accuracy of the sparsity locations in the factor processes when $r=3$. From Table~S.II, we see that the pattern is also similar to the case when $r=1$ in Section~\ref{sec41}. The empirical results are in line with our asymptotic ones in the sense that the estimation accuracy will increase as the dimension  increases for each fixed $T$.

 Although the ratio-based method of (\ref{rhat}) for determining the number of factors has been shown to be valid in many previous studies, we conduct an auxiliary experiment to verify its efficacy under our setting. Table~S.III of the Supplement reports the empirical probabilities of $P(\wh r=r)$ in 500 replications under the aforementioned setting. From Table~S.III, we see that the ratio-based method performs well, with most of the empirical probabilities being close to one. This is understandable since all the factors are strong ones. Similar results can be found in \cite{gaotsay2022}.

\subsection{Determining the Sparsity with Cross-Validation}
In this section, we study the estimation accuracy of the information criterion discussed in Section \ref{sec27} and Remark \ref{rm3} in estimating the sparsity of the factors. For simplicity, we only consider the case when $r=1$, but similar results can be obtained for $r>1$. The generation of the factors and the data is similar to those in Section~\ref{sec41}, but we only keep the largest $s=T/10$ elements of the factor process in absolute value for each sample size $T$. We consider the dimension $N=50,150,200$, and $300$, and the sample size $T=100,200,300,500$, and $800$ for each $N$ in this section. For each configuration of $(N,T)$, we set the number of partitions $J=1$ for simplicity and set $N_1=N_2=N/2$ in the partition. Since $s=T/10$ is diverging with the sample size $T$, we will use the information criterion defined in (\ref{Infor}) to estimate the sparsity. 500 replications are used throughout the experiment. Table~S.IV of the Supplement reports the Empirical probabilities (EP) of determining the sparsity parameter using the information criterion in (\ref{Infor})  when the number of factors $r=1$.  The empirical probabilities are calculated based on the 500 experiments.  From Table~S.IV we see that  that the proposed information criterion estimates the sparsity parameter effectively. While dynamic dependence in idiosyncratic terms reduces accuracy,  its performance generally improves with increasing $N$ for each fixed $T$,  which is consistent with Theorem~\ref{thm3}.


\section{An Empirical Application to Stock Returns} \label{sec5}

	\subsection{Data}	
		
In this section, we estimate the sparse latent risk factors across the time horizons in daily returns of individual stocks. The nonzero factors over certain time period can provide us one way to bridge time and certain type of risks in the financial market. The daily returns are downloaded from the CRSP daily security database and adjusted for dividend and stock splits. 
The data set used is the same as that in \cite{pelger2019} and consists of the daily stock returns for the balanced panel of S\&P 500 stocks from January 1st 2004 to December 31st 2016. The daily interest rates from Prof. Kenneth French's website (\url{https://mba.tuck.dartmouth.edu/pages/faculty/ken.french/data_library.html}) are used to adjust the daily returns of individual stocks. The full data set is also available at \url{https://mpelger.people.stanford.edu/data-and-code}, where  only the stocks with returns available for the full-time horizon are included, leaving us with a panel of $N=332$ and $T=3273$. 

Similar to \cite{pelger2019}, we group these 332 stocks into 14 categories by industry sector, as shown in Table~S.V of the Supplement. The data encompasses a wide range of industry sectors within the stock market, including Oil, Finance, Electricity, Technology, Food, Manufacturing, Pharma \& Chemicals, Primary Manufacturing, Machinery, Health, Transportation, Trade, Services, and Mining.

\subsection{Sparse Factor Estimation}
\begin{figure}[htp]
\begin{center}
{\includegraphics[width=14cm,height=5cm]{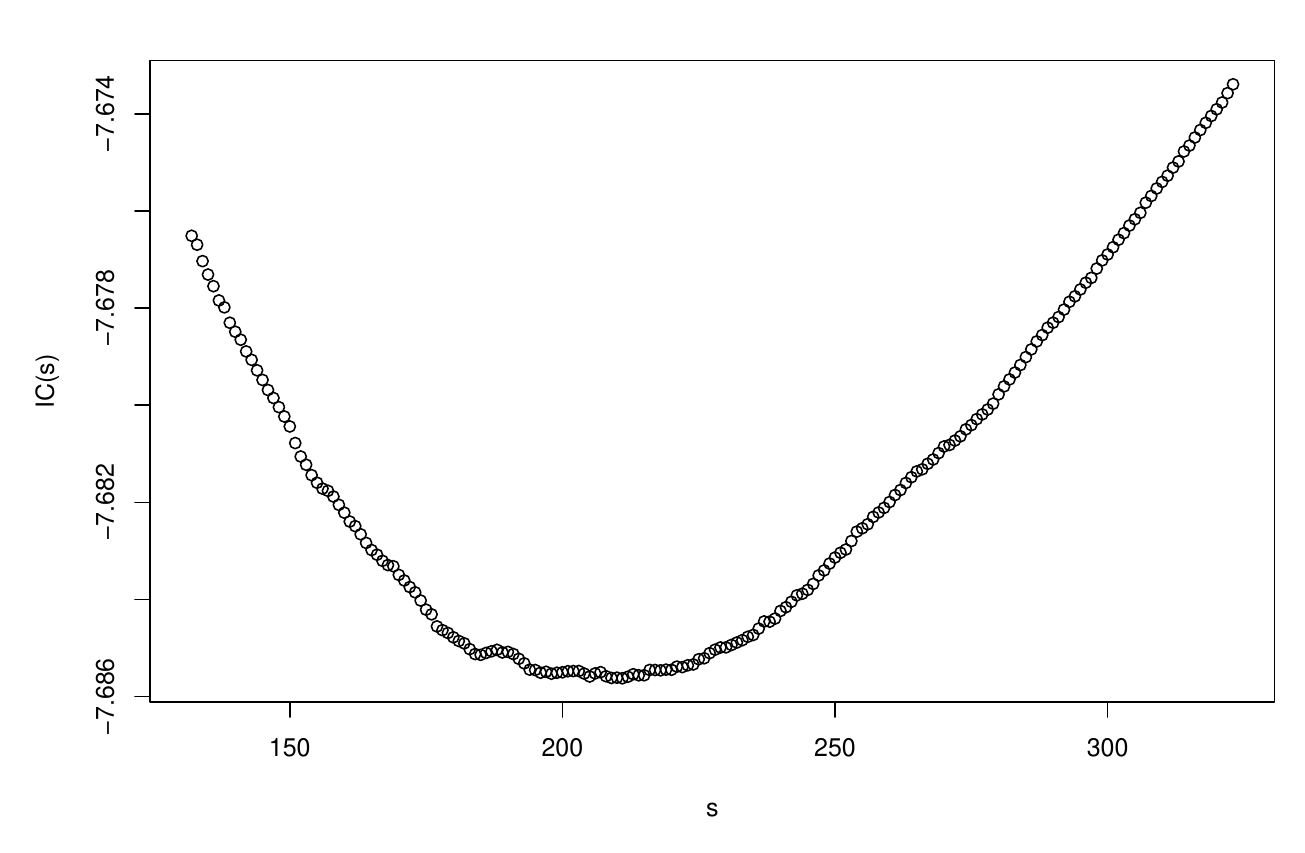}}
\caption{Plots of the information criterion $IC(s)$, defined in equation (\ref{Infor}) of the main text, over the range $s\in[\lceil \frac{T}{18} \rceil-50,\lceil \frac{T}{12} \rceil+50]=[132,232]$. We use $J=10$ cross-validations for each $s$. }\label{fig-222}
\end{center}
\end{figure}

We first determine the number of factors for the centered data $\bX$ using the eigenvalue-ratio method in (\ref{rhat}). Figure~S.3 of the Supplement plots the ratios of eigenvalues of $\bX\bX'$ and shows
that the largest gap between the eigenvalues occurs between $\wh\lambda_1$ and $\wh\lambda_2$, implying that the number of factors is  $\wh r=1$. Next, we apply Algorithm~\ref{alg:framework} and the cross-validation method in Section~\ref{sec27} with the number of random partitions $J=10$, where the seed number is set to be \texttt{1234} in \texttt{R}. {For the sparsity of the factors over the time horizon, we conduct  a grid search over the time interval $s\in[\lceil \frac{T}{18} \rceil-50,\lceil \frac{T}{12} \rceil+50]=[132,232]$, where $\lceil x \rceil$ denotes the smallest integer that exceeds $x$.   Figure~\ref{fig-222} plots the information criterion $IC(s)$ defined in (\ref{Infor}) and 
the estimated  $\wh s=211$.  This suggests that there are 211 days during which stock returns were more strongly influenced by significant systematic risks. We also searched over the neighborhoods of \( T/10 \) and \( T/6 \), and found that the minimum of \( IC(s) \) in these regions is greater than \( IC(211) \). Moreover, the entire path of \( IC(s) \) increases after \( s = 211 \), indicating that \( \wh s = 211 \) is a suitable estimate.
  It is important to note that this does not imply the absence of systematic risk on the remaining days; rather, the method is designed to highlight periods with stronger and more dominant risk signals. Compared to the full time span of $T=3273$, this represents a substantial reduction in temporal dimensionality, thereby simplifying subsequent analysis.}

\begin{figure}[htp]
\begin{center}
{\includegraphics[width=14cm,height=6cm]{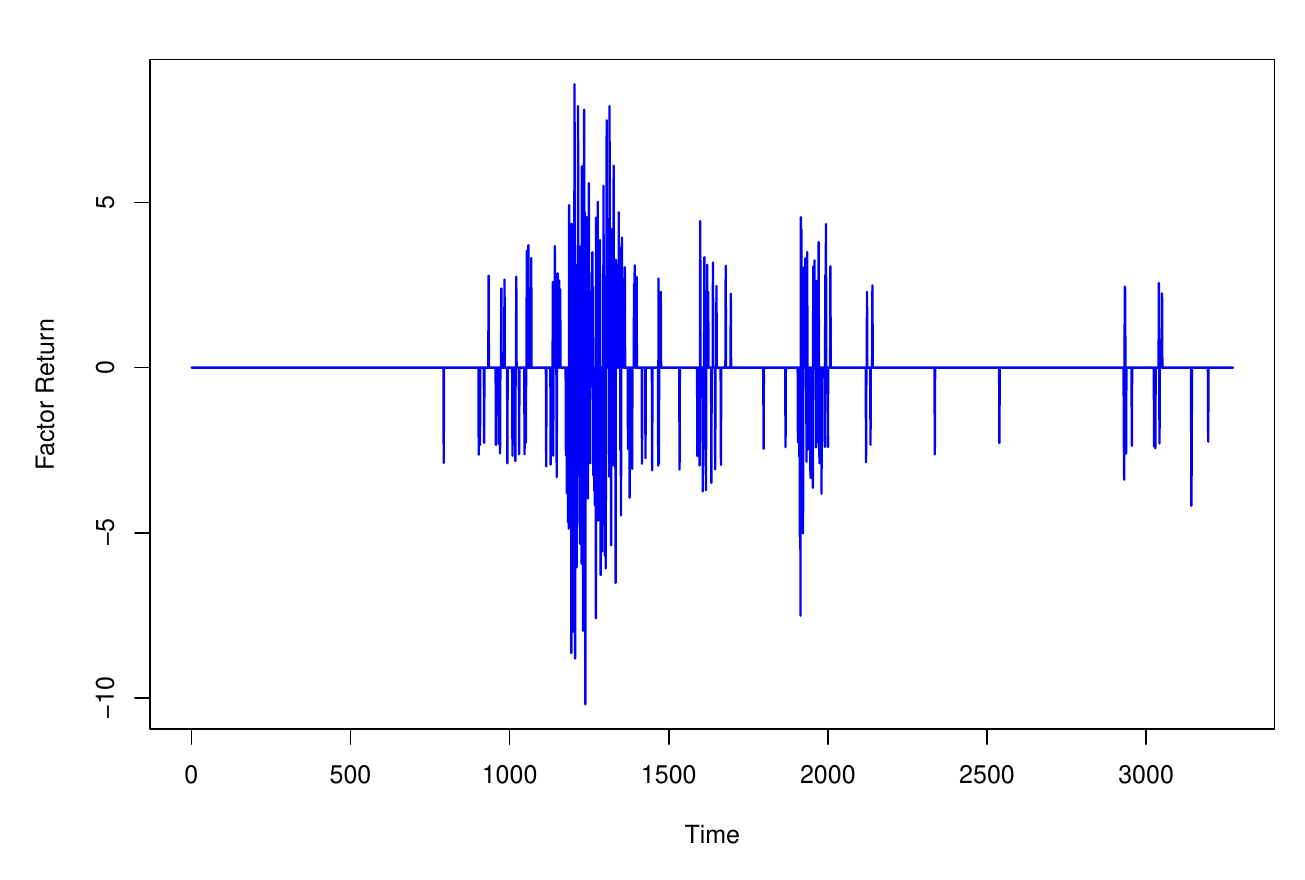}}
\caption{Time plots of the estimated sparse risk factors across the time horizon. The data consists of the daily stock returns for the balanced panel of S\&P 500 stocks from January 1st 2004 to December 31st 2016.}\label{fig-3}
\end{center}
\end{figure}

We then plot the estimated sparse factor returns in Figure~\ref{fig-3}.
Similar to the example in the {\it Introduction}, we find that  ${\sum_{t=1}^{3273}(f_t^s)^2}/{\sum_{t=1}^{3273} f_t^2} = 62.5\%$,  indicating that the 211 nonzero factor values—representing only 6.4\% of the total 3273 time points—account for 62.5\% of the variance in the  non-sparse factor process 
$\bff_t$.
Moreover, the plot reveals approximately three distinct clusters of time periods associated with significant systematic risks affecting the stock market. The first cluster is centered around 2008, driven by the 2007--2008 financial crisis.  The second and third clusters are closely related, both associated with the European sovereign debt crisis.  The most prominent cluster is centered around 2008, driven by the 2007--2008 financial crisis. This period saw severe disruptions in the financial markets, with significant declines in stock returns across the board. For the two closely related clusters associated with the European sovereign debt crisis, the first cluster corresponds to the initial phase of the crisis in 2010, while the second cluster relates to the intensification of the crisis in 2011 and 2012.

\begin{figure}[htp]
\begin{center}
{\includegraphics[width=17cm,height=8cm]{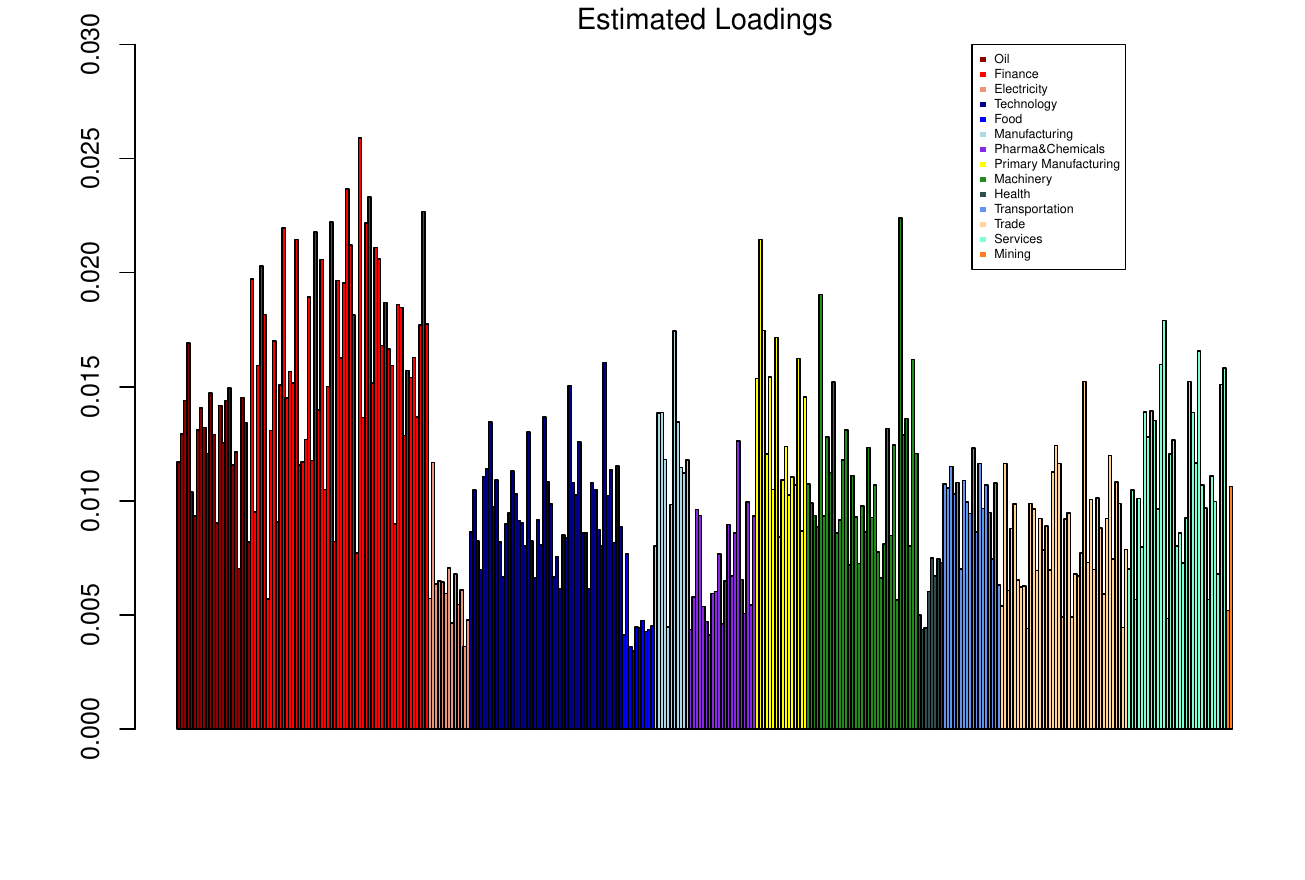}}
\caption{Barplots of the the estimated loadings, indicating the level of dependence of 14 different industry sectors on common risk factors. }\label{fig-20}
\end{center}
\end{figure}

Next, we estimate the loadings of the 332 stock returns given the sparse common factors. We present the estimated loadings, indicating the level of dependence of 14 different industry sectors on the common risk factor in Figure~\ref{fig-20}. The loadings quantify the extent to which each sector is influenced by these common factors. Here are the detailed observations based on the average dependence values: {\bf Oil Sector}: Exhibits moderate dependence on common risk factors, with an average loading of 0.0125.
{\bf Finance Sector}: Shows significant dependence, with an average loading of 0.0165.
 {\bf Electricity Sector}: Demonstrates relatively low dependence, with an average loading of 0.0062.
{\bf Technology Sector}: Reflects moderate dependence, with an average loading of 0.0097.
 {\bf Food Sector}: Indicates low dependence on common factors, with an average loading of 0.0046.
{\bf Manufacturing Sector}: Displays moderate dependence, with an average loading of 0.0116.
 {\bf Pharma \& Chemicals Sector}: Shows relatively low dependence compared to other sectors, with an average loading of 0.0070.
{\bf Primary Manufacturing Sector}: Exhibits moderate dependence, with an average loading of 0.0133.
{\bf Machinery Sector}: Demonstrates moderate dependence, with an average loading of 0.0109.
 {\bf Health Sector}: Indicates relatively low dependence, with an average loading of 0.0061.
{\bf Transportation Sector}: Shows moderate dependence, with an average loading of 0.0099.
 {\bf Trade Sector}: Reflects moderate dependence, with an average loading of 0.0084.
 {\bf Services Sector}: Exhibits moderate dependence, with an average loading of 0.0111.
 {\bf Mining Sector}: Indicates relatively low dependence on common risk factors, with an average loading of 0.0079.

Overall, the Finance sector shows the highest average dependence on common risk factors, while sectors like Food, Health, and Electricity exhibit lower average dependence. These variations highlight the differing levels of sensitivity across industry sectors to common economic and market risk factors.


\subsection{Bridging Time and Risk Factors}

To further identify the systematic factors over time, referred to as time factors, Table S.VII  in  Sec. D of the Supplement provides a comprehensive list of dates with significant systematic risk factors, the reasons for these factors, and their associated time factors.  Each entry explains the specific event and its impact on stock returns, summarized according to the descriptions of the time factors in Table~S.VI of the Supplement. These reasons are extracted from the daily reports on {\it CNN Money} (\url{www.money.cnn.com})  after the market closes on each trading day.  This website was shut down in late 2018 and its content was merged into the main CNN Business section.

\begin{figure}[htp]
\begin{center}
{\includegraphics[width=14cm,height=6cm]{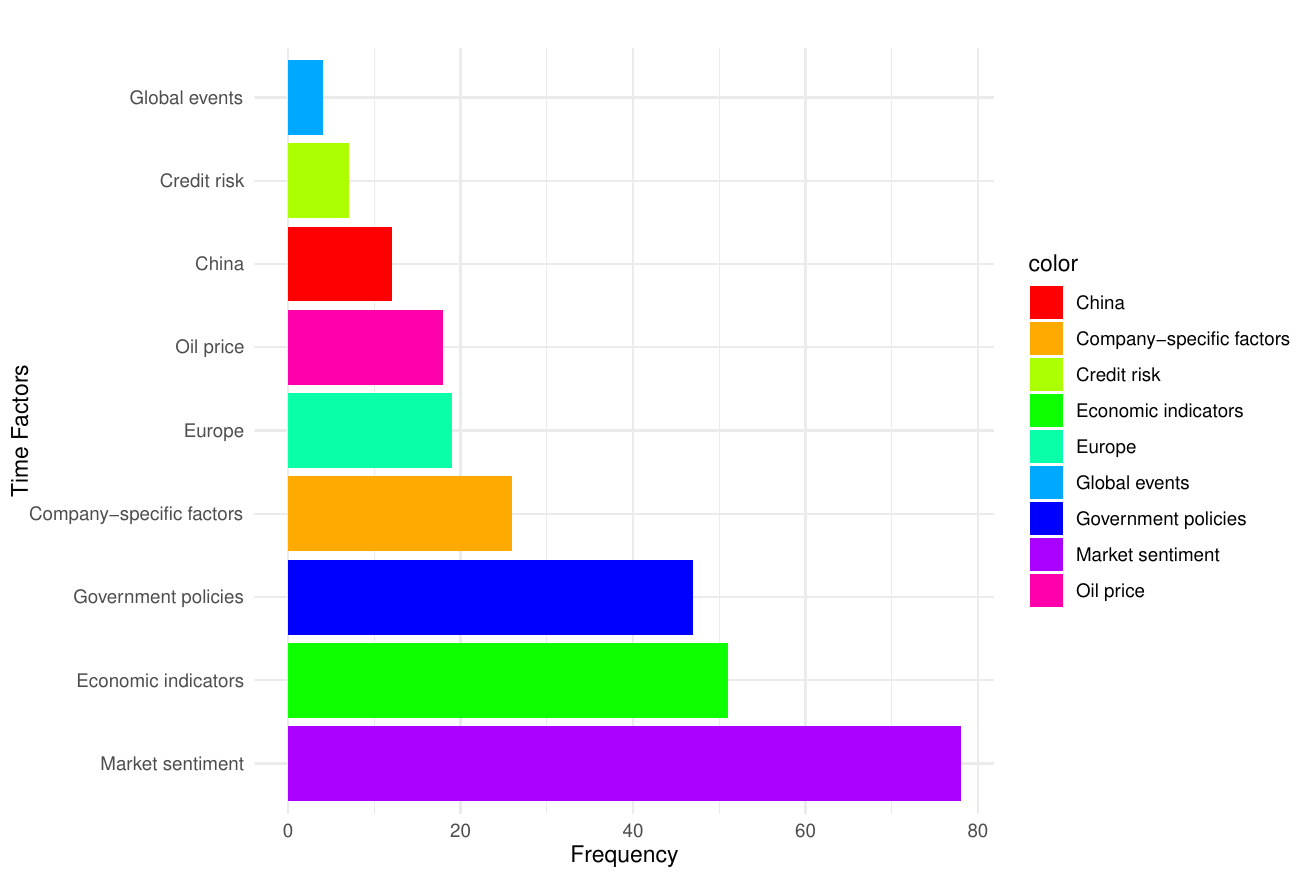}}
\caption{Frequency of Time Factors Influencing Stock Market Movements from January 2004 to December 2016. The chart visualizes the occurrence of various time factors, including market sentiment, economic indicators, government policies, company-specific factors, Europe, oil price, China, global events, and credit risk, based on their frequency of mention in Table~S.VII of the Supplement.}\label{fig-freq}
\end{center}
\end{figure}

We also plot the frequency charts of the nine factors over the time horizon in Figure~\ref{fig-freq}.
 The most frequently mentioned factor is market sentiment, highlighting its dominant role in driving stock price fluctuations. Economic indicators and government policies follow closely, indicating the significant impact of macroeconomic data and policy decisions on market behavior. Company-specific factors also play an important role, affecting individual stock performance and, by extension, broader market trends. Factors related to Europe, such as economic conditions and crises, have a moderate influence, reflecting the interconnectedness of global markets. Oil price fluctuations, China's economic activities, and global events are also notable contributors, underscoring the importance of global economic dynamics. Credit risk, while mentioned less frequently, remains a critical factor during periods of financial instability. Further implications of the results are discussed in Section B.1 of the Supplement for brevity.

		\section{Conclusion} \label{sec6}

Sparse factor modeling aims to extract meaningful insights by identifying a subset of relevant variables, enhancing interpretability while reducing computational complexity. This study presents a novel perspective by assuming that factor processes exhibit sparsity over time rather than imposing sparsity on loadings. This approach is particularly suited to scenarios where systematic co-movements are significant only during specific periods, such as financial crises or policy changes.

This paper introduces approximate factor models and a novel sparse asymptotic PCA (APCA) method, extending the asymptotic PCA framework of \cite{connor1986performance,connor1988risk} to high-dimensional settings. Theoretical properties of the proposed method, including the cross-sectional cross-validation approach for estimating sparsity, are rigorously established. Through simulation and empirical studies, the method is shown to be effective in delivering interpretable results and linking co-movements to specific events. This work not only advances the statistical methodology of sparse factor modeling but also establishes a meaningful connection between time-specific events and risk factors in economic and financial systems. 

\section*{Supplementary Material}
This supplementary material includes some examples of sparse factors over time horizons in finance and economics, additional tables and figures used in the numerical analyses, theoretical proofs of the main theorems, and a description of the text data extracted from {\it CNN Money} (\url{www.money.cnn.com}).
		
		
\section*{Disclosure Statement}
We declare that there are no relevant financial or non-financial competing interests to disclose.

		
	\end{onehalfspacing}

%
	
		\singlespacing
\ifx\undefined\BySame
\newcommand{\BySame}{\leavevmode\rule[.5ex]{3em}{.5pt}\ }
\fi
\ifx\undefined\textsc
\newcommand{\textsc}[1]{{\sc #1}}
\newcommand{\emph}[1]{{\em #1\/}}
\let\tmpsmall\small
\renewcommand{\small}{\tmpsmall\sc}
\fi

%

\newpage

\renewcommand{\thetheorem}{A.\arabic{theorem}}%
	\renewcommand{\theproposition}{A.\arabic{proposition}}%
	\renewcommand{\thelemma}{A.\arabic{lemma}}%
	\renewcommand{\theassumption}{A.\arabic{assumption}}%

	\setcounter{lemma}{0}
	\setcounter{proposition}{0}
	\setcounter{theorem}{0}
	\setcounter{assumption}{0}
  \setstretch{1.6}
	\begin{titlepage}
	
	\begin{center}
 {\huge Supplementary Material for ``Sparse Asymptotic PCA: Identifying Sparse Latent Factors Across  Time Horizon in High-Dimensional Time Series"}
	\end{center}

	 \thispagestyle{empty}
		\vspace{0.5cm}
		
		\begin{abstract}
The Online Appendix collects the mathematical proofs of the asymptotic results in Section 3 and the descriptions of the identified sparse factors in the empirical analysis that support the main text.
			\vspace{1cm}
			
			\noindent\textbf{Keywords:} Asymptotic Principal Components, Factor Analysis, Power Method, Sparsity, High-Dimension	
			
		\end{abstract}
	\end{titlepage}

	\setcounter{page}{1}
	
	\setcounter{section}{0}
	\setcounter{subsection}{0}

 	\renewcommand{\thesection}{\Alph{section}}

	\renewcommand{\thesubsection}{\thesection.\arabic{subsection}}
	
	\setcounter{equation}{0}
	\renewcommand{\theequation}{\thesection.\arabic{equation}}
	

	
	\renewcommand{\theequation}{S.\arabic{equation}}%
	\renewcommand{\thefigure}{S.\arabic{figure}} \setcounter{figure}{0}
	\renewcommand{\thetable}{S.\Roman{table}} \setcounter{table}{0}
	
	

	
	


		\section{Examples of Sparse Factors in Time Horizons}	

 Factor analysis is an important statistical tool for reducing the dimensions of large panels of economic and financial data. When the factors are latent and unobservable, the conventional PCA estimation procedure, though easy to implement and possessing good theoretical properties (\cite{bai2002determining} and \cite{bai2003inferential}), often results in loadings and factors that provide unclear economic insights and are difficult to interpret in practice. Under the proposed framework, sparse factors will improve the interpretability of the factors and offer a useful tool for understanding systematic risk over the timeline in economic and financial panel systems.
 

\begin{enumerate}[label=(\roman*)]
    \item {\it Asset Pricing Models}. In the Arbitrage Pricing Theory (APT) proposed by \cite{Ross1976}, a fundamental assumption is that a small number of factors can explain a large number of asset returns. In contrast to cases where the factors are observable, as in \cite{fama1993common}, the latent factors $\bff_t$'s
  extracted from the panel 
$\bx_t$'s are difficult to interpret. Financial economists widely recognize that time and risk are two critical factors that make finance challenging (see, for example, \cite{treynor1961market}). The finance subject would be incomplete without these two elements. The proposed framework provides an approach to bridge the systematic risk factor 
$\bff_t$ and time $t$ for financial returns. When the systematic/common factors are nonzero at certain time points, we may relate events such as macroeconomic conditions, news, government policies, and banking conditions to the nonzero systematic risk factors detected over that time period. Conversely, when the systematic factors are zero, we may conclude that the market was mainly driven by company-specific shocks during those corresponding time periods.

    \item {\it Disaggregate Business Cycle Analysis}.   \cite{gregory1999common} found that cross-country variations have common components, referred to as global shocks. In addition, each country also experiences country-specific shocks. It is clear that the global and country-specific shocks are reflected by the common factors 
$\bff_t$'s  and the idiosyncratic terms 
$\be_{t}$'s in Model (2.2), respectively. Under the proposed framework, sparse factors over a certain time period correspond to situations where global shocks do not play a dominant role in driving the cyclical variations in a country's economy. Conversely, nonzero factors will inspire researchers to discover the possible reasons for significant global shocks during the corresponding time period.


    \item {\it Consumption and Demand System}. Let $x_{i,t}$ be the budget share of good $i$ for agent $t$, where we have $N$ goods and $T$ agents in total. Consumer theory assumes that $x_{i,t}=\blambda_i'\bff(z_t)+e_{i,t}$ in a single price regime, where $\bff(z_t)$ is an $r$-dimensional unknown functional factors of $z_t$ with $z_t$ being agent $t$'s total expenditure. See Section 4 of \cite{lewbel1991rank} for details. For certain agents, their budget shares for good $i$ can be zero, implying that $\bff(z_t)$'s are zero for $t$ belonging to such agents. Therefore, the proposed sparse factor models may provide economists with one way to understand certain groups of households' preferences for specific goods.  See a similar argument in \cite{gabaix2014sparsity}.

    \item {\it Monitoring and Forecasting}. \cite{stock1998diffusion} and \cite{forni2000reference} have demonstrated that factor models provide an effective framework for monitoring economic activities, as large-dimensional economic panel data can often be summarized by a small number of factors. In this context, business cycles can be modeled through the co-movements of economic variables, which correspond to the common factors 
$\bff_t$ in Model (2.2) of the main text. When these factors exhibit sparsity, our approach allows us to track economic activities by identifying systematic risks over time. Furthermore, periods of heightened volatility in the economic panel may produce nonzero factors that are associated with specific events that influence the overall system. Additionally, sparse common factors (diffusion indices) extracted from a large panel of macroeconomic variables may also improve forecasting accuracy when combined with methods such as those proposed by \cite{gao2023supervised}.


\end{enumerate}

\section{Some Tables and Figures for Numerical Studies}

\begin{figure}[ht]
\begin{center}
{\includegraphics[width=13cm,height=7cm]{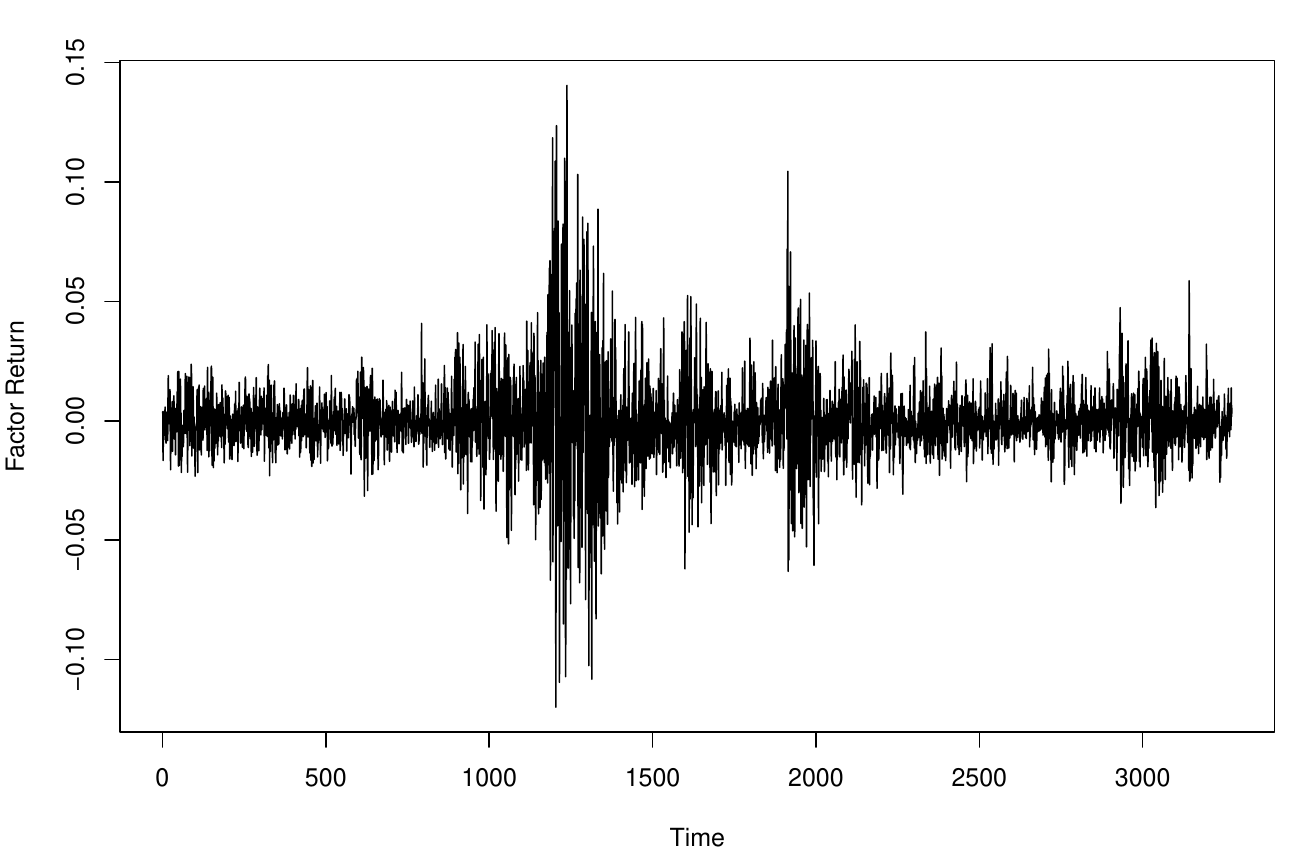}}
\caption{Time plots of the estimated risk factors across the time horizon using the Principal Component Analysis method. The data consists of the daily stock returns for the balanced panel of S\&P 500 stocks from January 1st 2004 to December 31st 2016.}\label{fig-00}
\end{center}
\end{figure}

\begin{figure}[ht]
\begin{center}
{\includegraphics[width=13cm,height=7cm]{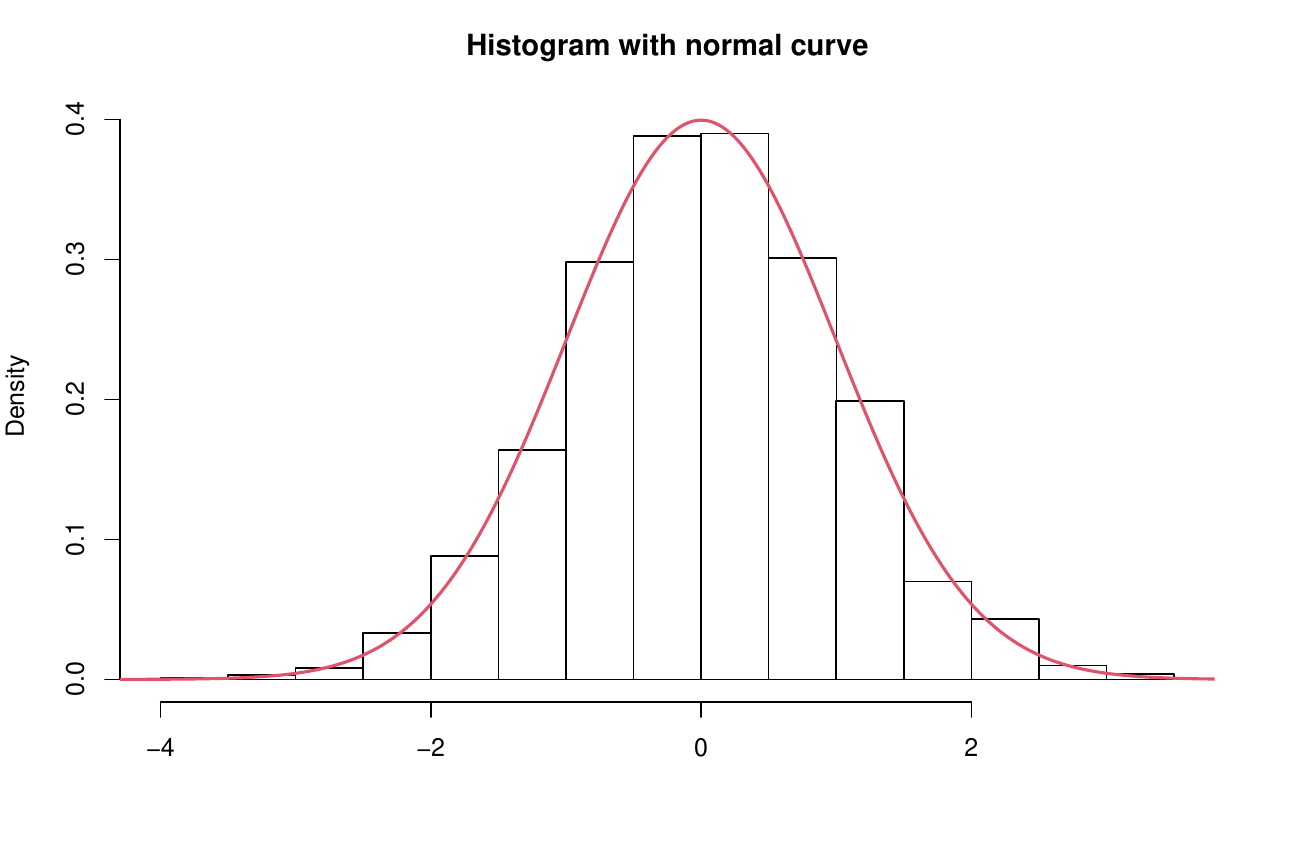}}
\caption{ Histograms of the estimated loading of $\sqrt{T}(\wh\blambda_{1}-\bH_s\blambda_1)$, where we set $(N,T)=(300,500)$ and $s=\left\lceil \sqrt{T} \right\rceil$.  The settings of the parameters are the same as those in Table~1 of the main text. The curve represents the density of normal distribution where the standard errors are estimated by the sample versions using the residuals. 2000 replications are used in the experiments.  }\label{fig-0}
\end{center}
\end{figure}

In this section, we provide additional tables and figures that support the numerical analysis in the main article. {Figure~\ref{fig-00} presents the time plots of the estimated risk factors over the time horizon of the dataset analyzed in Section 5, obtained using PCA.}
Figure~\ref{fig-0} plots the the empirical histograms of the estimated loading of $\sqrt{T}(\wh\blambda_{1}-\bH_s\blambda_1)$ for the example of one-factor case, where we set $(N,T)=(300,500)$ and $s=T/10$. Table~\ref{Table-a3} reports the estimation accuracy of factors when the number of factors $r=3$ for the example of the multi-factor case. Table~\ref{Table-a4} presents the empirical recovery rate of the sparsity in the factor processes when the number of factors $r=3$ for the multi-factor case. Table~\ref{Table-a5} reports the  empirical probabilities (EP) of determining the number of factors using the ratio-based method in (2.12) of the main article  when the number of factors $r=3$. Table~\ref{Table-a6} presents the empirical probabilities (EP) of determining the sparsity parameter using the information criterion in (3.9)  when the number of factors $r=1$ to verify the accuracy of the proposed criterion.

In Table~\ref{Table-Com}, we outline the categorization of 332 Stocks in the S\&P 500 Index by Industry Sector, which are used in the empirical analysis. Table~\ref{Table-ft-des} presents the  nine significant systematic risks over the time horizon from January 1 2004 to December 31, 2016, identified by the proposed sparse factor modeling approach. 

\begin{table}[ht]
\caption{The estimation accuracy of factors when the number of factors $r=3$. The measure is defined as (4.3) of the main text. $N$ and $T$ denote the dimension and the sample size, respectively. 500 replications are used in the experiments.} 
          \label{Table-a3}
{\begin{center}
\begin{tabular}{cccccccccccc}
\toprule
&\multicolumn{5}{c}{$\be_t\overset{i.i.d.}{\sim} N({\bf 0},\bI_N)$}&&\multicolumn{5}{c}{$\be_t=\Phi\be_{t-1}+\bve_t,\bve_t\overset{i.i.d.}{\sim} N({\bf 0},\bI_N)$}\\
\cline{2-6}\cline{8-12}
&\multicolumn{5}{c}{$T$}&&\multicolumn{5}{c}{$T$}\\
\cline{2-6}\cline{8-12}
$N$&100&200&300&500&800&&100&200&300&500&800\\
\hline
50&0.090&0.096&0.097&0.099&0.010&&0.136&0.159&0.158&0.158&0.155\\
100&0.059&0.064&0.065&0.066&0.067&&0.099&0.105&0.105&0.105&0.112\\
150&0.047&0.050&0.052&0.053&0.053&&0.078&0.087&0.086&0.085&0.086\\
200&0.041&0.043&0.044&0.045&0.045&&0.068&0.070&0.073&0.073&0.072\\
300&0.032&0.034&0.034&0.035&0.036&&0.052&0.053&0.057&0.058&0.057\\
\bottomrule
\end{tabular}
  \end{center}}
\end{table}

\begin{table}[ht]
\caption{Empirical recovery rate of the sparsity in the factor processes when the number of factors $r=3$. The error in each replication is calculated as (4.4) of the main text. $N$ and $T$ denote the dimension and the sample size, respectively. 500 replications are used in the experiments.} 
          \label{Table-a4}
{\begin{center}
\begin{tabular}{cccccccccccc}
\toprule
&\multicolumn{5}{c}{$\be_t\overset{i.i.d.}{\sim} N({\bf 0},\bI_N)$}&&\multicolumn{5}{c}{$\be_t=\Phi\be_{t-1}+\bve_t,\bve_t\overset{i.i.d.}{\sim} N({\bf 0},\bI_N)$}\\
\cline{2-6}\cline{8-12}
&\multicolumn{5}{c}{$T$}&&\multicolumn{5}{c}{$T$}\\
\cline{2-6}\cline{8-12}
$N$&100&200&300&500&800&&100&200&300&500&800\\
\hline
50&0.949&0.946&0.945&0.942&0.941&&0.931&0.931&0.918&0.918&0.917\\
100&0.966&0.960&0.960&0.958&0.957&&0.947&0.941&0.941&0.940&0.937\\
150&0.971&0.967&0.966&0.964&0.964&&0.956&0.949&0.949&0.949&0.948\\
200&0.973&0.971&0.969&0.970&0.969&&0.959&0.958&0.953&0.955&0.955\\
300&0.980&0.974&0.977&0.975&0.974&&0.968&0.964&0.965&0.963&0.962\\
\bottomrule
\end{tabular}
  \end{center}}
\end{table}

\begin{table}[ht]
\caption{Empirical probabilities (EP) of determining the number of factors using the ratio-based method in (2.12) of the main text  when the number of factors $r=3$.  $N$ and $T$ denote the dimension and the sample size, respectively. 500 replications are used in the experiments.} 
          \label{Table-a5}
{\begin{center}
\begin{tabular}{cccccccccccc}
\toprule
&\multicolumn{5}{c}{$\be_t\overset{i.i.d.}{\sim} N({\bf 0},\bI_N)$}&&\multicolumn{5}{c}{$\be_t=\Phi\be_{t-1}+\bve_t,\bve_t\overset{i.i.d.}{\sim} N({\bf 0},\bI_N)$}\\
\cline{2-6}\cline{8-12}
&\multicolumn{5}{c}{$T$}&&\multicolumn{5}{c}{$T$}\\
\cline{2-6}\cline{8-12}
$N$&100&200&300&500&800&&100&200&300&500&800\\
\hline
50&1&1&1&1&1&&1&0.986&1&1&1\\
100&1&1&1&1&1&&0.998&1&1&1&1\\
150&1&1&1&1&1&&0.998&1&1&1&1\\
200&1&1&1&1&1&&1&1&1&1&1\\
300&1&1&1&1&1&&1&1&1&1&1\\
\bottomrule
\end{tabular}
  \end{center}}
\end{table}

\begin{table}[ht]
\caption{Empirical probabilities (EP) of determining the sparsity parameter using the information criterion in (3.9)  when the number of factors $r=1$.  $N$ and $T$ denote the dimension and the sample size, respectively. 500 replications are used in the experiments.} 
          \label{Table-a6}
{\begin{center}
\begin{tabular}{cccccccccccc}
\toprule
&\multicolumn{5}{c}{$\be_t\overset{i.i.d.}{\sim} N({\bf 0},\bI_N)$}&&\multicolumn{5}{c}{$\be_t=\Phi\be_{t-1}+\bve_t,\bve_t\overset{i.i.d.}{\sim} N({\bf 0},\bI_N)$}\\
\cline{2-6}\cline{8-12}
&\multicolumn{5}{c}{$T$}&&\multicolumn{5}{c}{$T$}\\
\cline{2-6}\cline{8-12}
$N$&100&200&300&500&800&&100&200&300&500&800\\
\hline
50&1&1&1&1&1&&0.882&0.998&0.958&0.942&0.878\\
150&1&1&1&1&1&&1&1&0.996&1&0.988\\
200&1&1&1&1&1&&1&1&1&0.998&0.998\\
300&1&1&1&1&1&&1&1&1&1&1\\
\bottomrule
\end{tabular}
  \end{center}}
\end{table}

{\begin{center}
\begin{table}[hpt]\footnotesize
\caption{Categorization of 332 Stocks in the S\&P 500 Index by Industry Sector. The data spans from January 1, 2004, to December 31, 2016.} 
          \label{Table-Com}
          \begin{tabular}{l|p{11cm}}
\toprule
Category & Stock Ticker\\
\hline
Oil&APA, APC, CHK, CNX, COP, CVX, HAL, HP, MRO, MUR, NBR, NE, NFG, NFX, OXY, PXD, RDC, RIG, SLB, SRE, TSO, VLO, XOM\\
\hline
Finance&ACAS, AET, AFL, AIG, AIV, AJG, ALL, AMG, AMT, AXP, BAC, BEN, BK, BXP, C, CI, CINF, CME, COF, FII, FITB, GS, HBAN, HCN, HCP, HIG, HRB, JNS, JPM, KEY, LNC, MBI, MET, MMC, MTG, NTRS, PFG, PLD, PNC, PRU, RF, RJF, SLM, SNV, SPG, SPY, STI, STT, TMK, TROW, UDR, UNM, USB, WFC, XL, ZION\\
\hline
Electricity&AEP, AES, CMS, CNP, FE, LNT, NI, PCG, PEG, PNW, PPL, SO, XEL\\
\hline
Technology&AAPL, ADBE, ADI, ADP, ADSK, AKAM, AMD, AMZN, APH, CA, CERN, CSC, CSCO, CTSH, CTXS, CVG, FFIV, FISV, HAR, HPQ, IBM, INTC, INTU, JBL, JNPR, KLAC, LLL, LLTC, LMT, MCHP, MSFT, MU, NCR, NTAP, NVDA, ORCL, QCOM, SNPS, STX, SWKS, SYMC, TXN, UIS, VRSN, WDC, XLNX, XRX, YHOO\\
\hline
Food&CAG, CCE, CPB, GIS, HSY, K, KO, MKC, MO, PEP\\
\hline
Manufacturing& CTAS, GCI, IP, JCI, KMB, LEG, LPX, MAS, NYT, RL, WY\\
\hline
Pharma \& Chemicals& ABT, AMGN, APD, AVP, BMY, CL, CLX, ENDP, GILD, IFF, JNJ, LLY, MON, MRK, MYL, OLN, PFE, PG, PPG, PRGO, PX\\
\hline
Primary manufacturing& AA, AKS, ATI, COH, CTB, GLW, GT, NKE, NWL, OI, SNA, SWK, TUP, USG, UTX, WOR\\
\hline
Machinery&A, AMAT, AME, BA, BC, BGG, BHI, CAT, CMI, COL, FLIR, FOSL, FTI, GD, GE, GRMN, HAS, HON, HRS, IR, ITT, LRCX, MAT, NOC, PBI, PCAR, PKI, ROK, RTN, TEN, TER, TKR, TMO, TXT, WHR\\
\hline
Health&BAX, BCR, BDX, MDT, MMM, SYK, VAR, XRAY\\
\hline
Transportation&ALK, CCI, CCL, CMCSA, CSX, CTL, DISH, EXPD, JBHT, KSU, LUV, LVLT, NSC, TDW, UNP, UPS, VIAB, VZ\\
\hline
Trade&ABC, ANF, AZO, BBBY, BBY, CAH, COST, CVS, DLTR, FAST, FL, GPC, GPS, GWW, HD, HSIC, JCP, JWN, KMX, KR, KSS, LOW, MCD, MCK, ORLY, PDCO, RAD, ROST, SBUX, SHW, SPLS, SVU, SYY, TGT, TIF, TSCO, URBN, WEN, WMT, YUM\\
\hline
Services&ACN, ADS, APOL, EBAY, ESRX, FLR, IGT, INCY, IPG, IT, KBH, LEN, LH, MAR, MCO, NFLX, OMC, PAYX, PCLN, PHM, PWR, R, RCL, RHI, SEE, SRCL, THC, TWX, UHS, URI, WYNN\\
\hline
Mining&NEM, VMC\\
\bottomrule
\end{tabular}
\end{table}
 \end{center}}

\begin{figure}[htp]
\begin{center}
{\includegraphics[width=14cm,height=6cm]{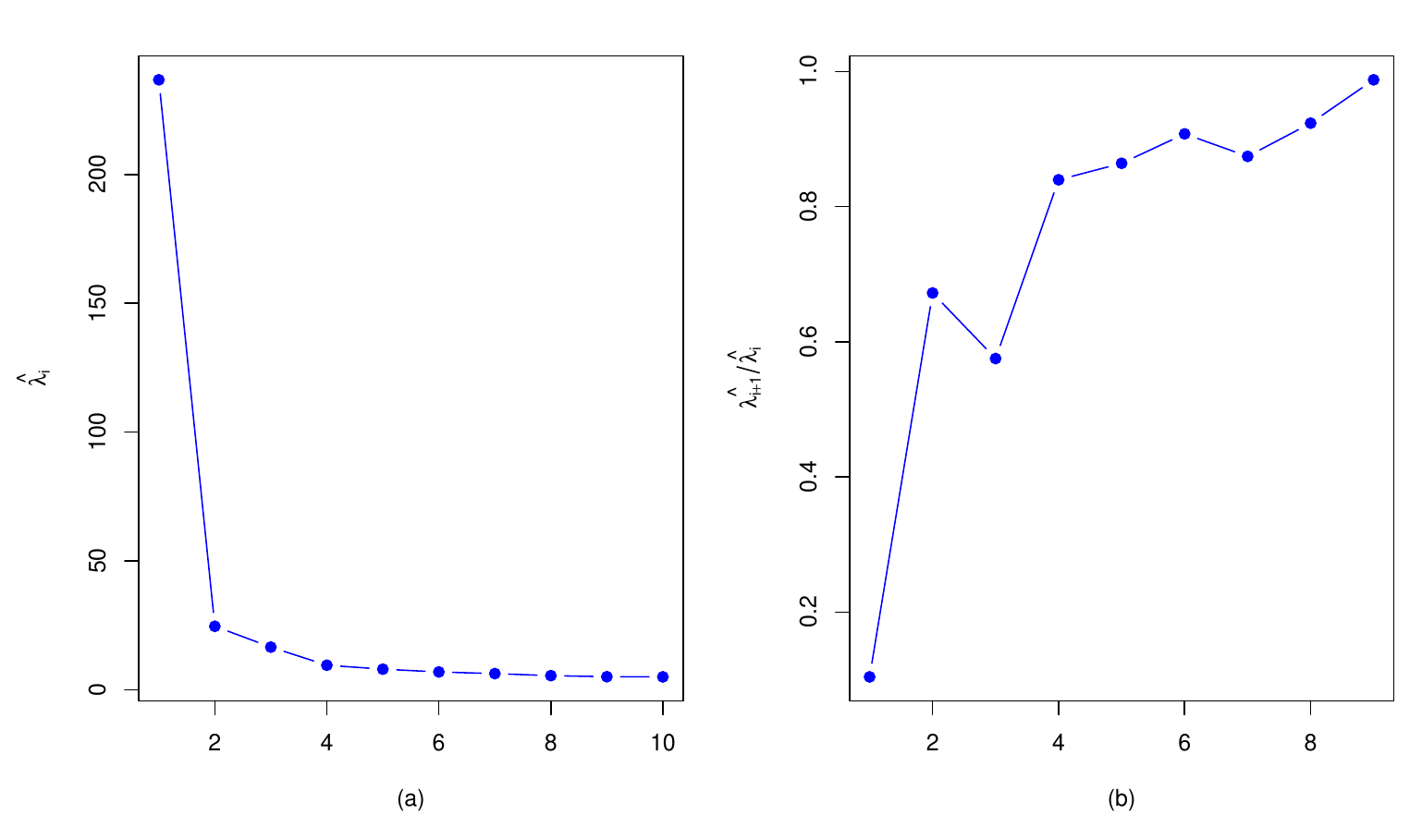}}
\caption{(a) The first $10$ eigenvalues of $\bX\bX'$;  (b) The plots of the ratios for the eigenvalues $\wh\lambda_i$ of the matrix $\wh \bS$.}\label{fig-2}
\end{center}
\end{figure}


{\begin{center}
\begin{table}[htp]
\caption{The nine significant systematic risks over the time horizon from January 1 2004 to December 31, 2016, identified by the proposed sparse factor modeling approach.} 
          \label{Table-ft-des}
          \begin{tabular}{l|p{10cm}}
\toprule
Time Factor & Description\\
\hline
Economic Indicators&Economic activities such as GDP growth, inflation, interest rates, and consumer confidence, sector reports, and job reports, etc.\\
\hline
Government Policies&Government policies, such as interest rate policies, tax policies, trade policies, and monetary policies, etc.\\
\hline
Global Events&Global events such as political instability, terrorist attacks, natural disasters, and pandemics, etc.\\
\hline
Market Sentiment&Market sentiment, which refers to the overall mood of investors.\\
\hline
Company-specific Factors& Company-specific factors, such as financial performance, management quality, and competitive position, can also impact financial markets. \\
\hline
Credit Risk&Credit risk refers to the possibility of borrowers defaulting on their obligations, affecting lenders and investors.\\
\hline
Oil Price& A reference price for buyers and sellers of crude oil such as West Texas Intermediate (WTI), Brent Crude, Dubai Crude, OPEC Reference Basket, Tapis crude, Bonny Light, Urals oil, Isthmus, and Western Canadian Select (WCS).\\
\hline
China& All significant economic activities associated with China such as a big decline in its financial market or a slowdown in its economic growth.\\
\hline
Europe& All significant economic activities associated with Europe and major European countries such as a debt crisis and the Brexit.\\
\bottomrule
\end{tabular}
\end{table}
 \end{center}}


\subsection{Further Implications of the Empirical Analysis}

From Tables~\ref{Table-ft} and~\ref{Table-ft-des}, we can see that these factors are all important to the stability of the financial market. In general, they affect the financial market in the following ways:
\begin{enumerate}[label=(\arabic*)]
 \item  {\bf Economic Indicators}: This includes employment reports, GDP growth, inflation rates, interest rates, and consumer confidence indices. For example, the weaker-than-expected December jobs report on January 4, 2008, exacerbated worries about a potential recession, leading to significant negative returns.
\item {\bf Government Policies}: These involve fiscal and monetary policy decisions such as interest rate cuts, bailouts, and tax policies. For instance, the Federal Reserve’s emergency interest rate cut on January 23, 2008, initially had a mixed impact but ultimately contributed to a market bounce back.
\item {\bf Market Sentiment}: Refers to the overall mood and confidence of investors. Market sentiment was notably affected on October 28, 2008, when investors dove back into stocks despite the ongoing financial turmoil, reflecting a temporary boost in confidence.
\item {\bf Company-specific Factors}: Events specific to individual companies, such as financial performance, management changes, and competitive positioning, can significantly impact their stock prices and, by extension, the broader market.
\item {\bf Credit Risk}: This pertains to the risk of defaults and financial instability within the banking sector. Renewed credit market fears on November 7, 2007, following concerns about Bear Stearns, led to a significant drop in stock returns.
\item {\bf Oil Price}: Fluctuations in crude oil prices directly impact stock returns. For example, a spike in oil prices on June 6, 2008, contributed to significant negative returns, highlighting the sensitivity of the market to energy costs.
\item {\bf Global Events}: Includes political instability, terrorist attacks, natural disasters, and pandemics. The news of explosions at the Boston Marathon on April 15, 2013, resulted in a sharp market decline.
\item {\bf China}: Economic activities and market movements within China. The big decline in Chinese stocks on February 27, 2007 had a notable impact on global markets.
\item {\bf Europe}: Economic activities and crises within Europe, including the debt crisis and Brexit. The downgrade of Greece’s debt rating to junk status on April 27, 2010, sparked fears about the European debt crisis, leading to significant negative returns.
\end{enumerate}

 From Tables~\ref{Table-ft} and \ref{Table-ft-des}, we have the following findings and implications:\\

 \begin{enumerate}[label=(\arabic*)]
 \item {\bf Understanding Systematic Risks:} Identifying the key factors that drive systematic risks provides valuable insights for investors and policymakers. By recognizing these factors, they can make more informed decisions to mitigate risks and capitalize on potential opportunities.
     \item {\bf Temporal Patterns and Market Sensitivity:} The clustering around the 2008 financial crisis and the European debt crisis underscores how prolonged economic uncertainties impact market stability. The varying reactions to government policies and economic indicators suggest the critical role of investor confidence during crises. For example, the mixed market reaction to the Federal Reserve's interest rate cut on January 23, 2008, illustrates the complexity of market sentiment during turbulent periods.

\item {\bf Policy Implications:} Our findings emphasize the importance of timely and effective government interventions, such as interest rate cuts and bailouts, in stabilizing the markets. Clear communication from policymakers can significantly influence investor sentiment and market outcomes. The market’s positive response to the Federal Reserve’s emergency interest rate cut on January 23, 2008, and the subsequent bounce back in stock prices highlight the effectiveness of such interventions.

\item {\bf Sector-Specific Insights:} The financial sector's vulnerability to systemic risks, particularly credit market fears and bank-specific news, is evident. Additionally, fluctuations in oil prices impact not only the energy sector but also broader market sentiments. The significant impact of renewed credit market fears on November 7, 2007, following concerns about Bear Stearns, underscores the interconnectedness of the financial sector and overall market stability.

\item {\bf Global Interconnectedness:} The influence of events in China and Europe on US stock returns highlights the interconnected nature of global markets. This interconnectedness suggests that investors may benefit from diversifying their portfolios to mitigate risks associated with specific regions or events. The market’s reaction to the downgrade of Greece’s debt rating on April 27, 2010, and the subsequent fears about the European debt crisis illustrate the global ripple effects of regional economic events.
 \end{enumerate}

Our empirical analysis using sparse factor modeling provides valuable insights into the significant risk factors influencing stock returns over time. The identified clusters of risk periods and the associated factors offer a deeper understanding of market dynamics and the impact of various economic and political events. The findings underscore the importance of timely government interventions, clear communication from policymakers, and the need for diversification in investment portfolios to manage global risks effectively.

\section{Mathematical Proofs}

 We will use $C$ or $c$ to denote a generic constant the value of which may change at different places.
 We first introduce several notations used in the derivations. Let $\bV=\bF/\sqrt{T}$, and $\bv_i=\underline{\bff}_i/\sqrt{T}$ be the $i$-th columns of $\bV$. Define $\bS_c={\bF\bLambda'\bLambda\bF'}/{(NT)}$ and $\bS_e=\be\be'/(NT)$. It follows from Model (3) and Assumptions 2--4 of the main article that
\begin{equation}\label{deco:S}
\bS=\bS_c+\bS_R=\lambda_1\bv_1\bv_1'+...+\lambda_r\bv_r\bv_r'+\bS_e+\bSigma_{fe}+\bSigma_{fe}',
\end{equation}
where $\bS_R=\bS_e+\bSigma_{fe}+\bSigma_{fe}'$ and $\bSigma_{fe}=\bF\bLambda'\be'/(NT)$ .
We need a few lemmas first.

\begin{lemma}\label{lm0}
For the $i$-th iteration in Algorithm 2 of the main article, the matrix $\bB_i$ in step 7 of Algorithm 2 is symmetric for each $i=1,...,r+1$.
\end{lemma}

{\bf Proof.} We will prove the result by an induction method. For each iteration $i=1,...,r+1$, we will first show that
\begin{equation}\label{ind}
  \bB_i^2=\bB_i\quad\text{and}\quad\bB_i'\bB_i=\bB_i, 
\end{equation}
which is stronger than the result in Lemma~\ref{lm0} and implies that $\bB_i$ is symmetric.

Note that $\bB_0=\bI_T$ which obviously satisfies the identities in (\ref{ind}). From Algorithm~2, we see that $\bB_1=\bI_T-\wh\bv_1\wh\bv_1'$ which also satisfies (\ref{ind}) since $\wh\bv_1'\wh\bv_1=1$. Next, we assume that $\bB_j^2=\bB_j$ and $\bB_j'\bB_j=\bB_j$, then, by the fact that $\wh\bv_j'\bB_j\wh\bv_j=1$ and an elementary argument, we can show that
\[\bB_{j+1}^2=\bB_j^2-2\bB_j\bq_j\bq_j'+\bB_j\bq_j\wh\bv_j'\bB_j\bv_j\bv_j'\bB_j'=\bB_j(\bI_T-\bq_j\bq_j')=\bB_{j+1},\]
and
\[\bB_{j+1}'\bB_{j+1}=\bB_j-2\bB_j\bq_j\bq_j'+\bB_j\bq_j\wh\bv_j'\bB_j\wh\bv_j\bq_j'=\bB_{j+1},\]
where we used the assumption in (\ref{ind}) for $i=j$ and the fact that $\wh\bv_j'\bB_j\wh\bv_j=1$. Therefore, this proves (\ref{ind}) for all $i$. It follows from (\ref{ind}) that $\bB_i$ is a symmetric matrix for all $i=1,...,r+1$. This completes the proof. $\Box$

\begin{lemma}\label{lm1}
For any two unit vectors $\bx$, $\by\in R^{T}$, the singular values of $\bx\bx'-\by\by'$ are
\[\sqrt{1-(\bx'\by)^2},\sqrt{1-(\bx'\by)^2},0,...,0,\]
where there are $T-2$ zeros.
\end{lemma}

{\bf Proof.} It is a 1-dimensional case of Proposition 2.1 in \cite{vu2013minimax}. See also the descriptions for (3.3)--(3.4) in \cite{gao2021two}. Specifically, note that the only non-zero singular value of $\bx\bx'(\bI_T-\by\by')$ is 
\[\sqrt{\Tr(\bx\bx'(\bI_T-\by\by')^2\bx\bx')}=\sqrt{\Tr\{1-(\bx'\by)^2\}}=1-(\bx'\by)^2.\]
By Proposition 2.1 of \cite{vu2013minimax}, the singular values of $\bx\bx'-\by\by'$ are 
\[\sqrt{1-(\bx'\by)^2},\sqrt{1-(\bx'\by)^2},0,...,0,\]
where there are $T-2$ zeros. This completes the proof. $\Box$

\begin{lemma}\label{lm2}
For any two matrices $\bM,\bN\in R^{N}$, we have the following inequality,
\[\Tr(\bM'\bN)\leq \|\bM\|_2\|\bN\|_*,\]
where $\|\bN\|_*$ denotes the sum of the singular values of $\bN$.
\end{lemma}

{\bf Proof.} We perform a singular value decomposition to $\bN$ and obtain $\bN=\bU\bD\bV'$. Then,
\begin{align}
    \Tr(\bM'\bN)=\Tr(\bV'\bM'\bU\bD)=\sum_{i=1}^N(\bV'\bM'\bU)_{ii}\bD_{ii}
    \leq&\max_{i}|(\bV'\bM'\bU)_{ii}|\sum_{i=1}^N\bD_{ii}\notag\\
    \leq& \|\bV'\bM'\bU\|_2\|\bD\|_*\notag\\
    \leq &\|\bM\|_2\|\bN\|_*.
\end{align}
This completes the proof. $\Box$

\begin{lemma}\label{lm3}
Under Assumptions 1--7, the estimator $\wh\bv_1$ of $\bv_1$ satisfies
\[\sqrt{1-(\wh\bv_1'\bv_1)^2}\leq \frac{2}{\lambda_1-\lambda_2}\sup_{\bv\in S_1}|\bv'\bS_R\bv|,\]
where $\bS_R$ is defined as that in (\ref{deco:S}) and $S_1=\{\bv:\|\bv\|_2=1\,\,\text{and}\,\, \|\bv\|_0\leq 2s_1\}$.
\end{lemma}
{\bf Proof.}  We only focus on the case that $\wh\bv_1\neq\bv_1$ as Lemmas \ref{lm3} obviously holds when $\wh\bv_1=\bv_1$. For any two vectors $\bx$ and $\by$, we know the trace has the property that $\Tr(\bx\by')=\Tr(\by'\bx)$. Note that $\bS_c\bv_1=\lambda_1\bv_1$, then
\begin{equation}\label{eigen:1}
    \bS_c-\lambda_1\bv_1\bv_1'=(\bI_T-\bv_1\bv_1')\bS_c(\bI_T-\bv_1\bv_1').
\end{equation}
On the other hand, for any $\|\bv\|_2=1$,
\begin{align}\label{Eq:1}
\Tr(\bS_c(\bv_1\bv_1'-\bv\bv'))=&\Tr(\bv_1'\bS_c\bv_1)-\Tr(\bv'(\bS_c-\lambda_1\bv_1\bv_1')\bv)-\lambda_1\Tr(\bv'\bv_1\bv_1'\bv)\notag\\
=&\lambda_1-\lambda_1(\bv'\bv_1)^2-\bv'(\bI_T-\bv_1\bv_1')\bS_c(\bI_T-\bv_1\bv_1')\bv,
\end{align}
where we use Equation (\ref{eigen:1}) in the last step. 

Let $\bb=(\bI_T-\bv_1\bv_1')\bv/\|(\bI_T-\bv_1\bv_1')\bv\|_2$, it follows that $\bb'\bv_1=0$ and therefore, $\bb\in\text{span}\{\bv_2,...,\bv_T\}$. By (\ref{deco:S}), it follows that
$\bb'\bS_c\bb\leq \lambda_2$, which implies that
\begin{equation}\label{proj:2}
    \bv'(\bI_T-\bv_1\bv_1')\bS_c(\bI_T-\bv_1\bv_1')\bv\leq \lambda_2\|(\bI_T-\bv_1\bv_1')\bv\|_2^2=\lambda_2-\lambda_2(\bv'\bv_1)^2.
\end{equation}
By (\ref{Eq:1}) and (\ref{proj:2}), we replace $\bv$ by the estimator $\wh\bv_1$ of $\bv_1$ and obtain
\begin{equation}\label{Inq:1}
    \Tr(\bS_c(\bv_1\bv_1'-\wh\bv_1\wh\bv_1'))\geq (\lambda_1-\lambda_2)\{1-(\wh\bv_1'\bv_1)^2\}.
\end{equation}

Now turn to the original optimization problem. Note that $\|\wh\bv_1\|_0\leq s_1$ and $\|\bv_1\|_0\leq s_1$, and $\wh\bv_1$ is the one that maximizes the variance $\bv'\bS\bv$ with the constraint $\|\bv\|_0\leq s_1$, then
\begin{equation}\label{diff}
    \Tr(\bS(\wh\bv_1\wh\bv_1'-\bv_1\bv_1'))=\wh\bv_1'\bS\wh\bv_1-\bv_1'\bS\bv_1>0.
\end{equation}
Combining (\ref{Inq:1}) and (\ref{diff}) gives 
\begin{equation}\label{Eq:2}
    1-(\wh\bv_1'\bv_1)^2\leq \frac{1}{\lambda_1-\lambda_2}\Tr((\bS-\bS_c)(\wh\bv_1\wh\bv_1'-\bv_1\bv_1'))=\frac{1}{\lambda_1-\lambda_2}\Tr(\bS_R(\wh\bv_1\wh\bv_1'-\bv_1\bv_1')).
\end{equation}
Note that there are only $s_1$ nonzero elements in $\wh\bv_1$ and $\bv_1$, and the indexes of their non-zero elements are not necessarily identical, we define a diagonal matrix $\bD_1$ where $i$-th  diagonal entry is $1$ when the corresponding $i$th element in $\wh\bv_1$ or $\bv_1$ are nonzero. Then $\bD_1\wh\bv_1=\wh\bv_1$ and $\bD_1\bv_1=\bv_1$, and 
\[\Tr((\bS_R(\wh\bv_1\wh\bv_1'-\bv_1\bv_1')=\Tr(\bS_R\bD_1(\wh\bv_1\wh\bv_1'-\bv_1\bv_1')\bD_1)=\Tr(\bD_1\bS_R\bD_1(\wh\bv_1\wh\bv_1'-\bv_1\bv_1')).\]
By Lemma \ref{lm2} and Lemma \ref{lm1},
\[\Tr(\bD_1\bS_R\bD_1(\wh\bv_1\wh\bv_1'-\bv_1\bv_1'))\leq \|\bD_1\bS_R\bD_1\|_2\|\wh\bv_1\wh\bv_1'-\bv_1\bv_1'\|_*=\|\bD_1\bS_R\bD_1\|_22\sqrt{1-(\wh\bv_1'\bv_1)^2}.\]
It follows from (\ref{Eq:2}) and the above one that
\begin{equation}\label{Eq:3}
    \sqrt{1-(\wh\bv_1'\bv_1)^2}\leq \frac{2}{\lambda_1-\lambda_2}\|\bD_1\bS_R\bD_1\|_2\leq \frac{2}{\lambda_1-\lambda_2}\sup_{\|\bv\|_2=1}|\bv'\bD_1\bS_R\bD_1\bv|.
\end{equation} 
Note that $\bD_1\bv$ have at most $2s_1$ nonzero elements and $\|\bD_1\bv\|_2\leq  1$, we have
\[\sup_{\|\bv\|_2=1}|\bv'\bD_1\bS_R\bD_1\bv|\leq \sup_{\|\bv\|_2=1}\frac{\bv'\bD_1}{\|\bD_1\bv\|_2}\bS_R\frac{\bD_1\bv}{\|\bD_1\bv\|_2}\|\bD_1\bv\|_2^2\leq \sup_{\bv\in S_1}|\bv'\bS_R\bv|,\]
where the set $S_1$ is define in Lemma \ref{lm3}. Then (\ref{Eq:3}) reduces to
\[\sqrt{1-(\wh\bv_1'\bv_1)^2}\leq \frac{2}{\lambda_1-\lambda_2}\sup_{\bv\in S_1}|\bv'\bS_R\bv|.\]
This completes the proof. $\Box$

The following lemma is from Theorem 2.8.1 of \cite{vershynin2018high} or Lemma D.2 of \cite{vu2013minimax}.
\begin{lemma}\label{lm4}
(Bernstein's inequality). Let $Y_1,...,Y_N$ be independent random variables with zero mean. Then
\[P\left(\left|\sum_{i=1}^NY_i\right|>t\right)\leq 2\exp\left(-c\min\left(\frac{t^2}{\sum_{i=1}^N\|Y_i\|_{\psi_1}^2},\frac{t}{\max_{i\leq N}\|Y_i\|_{\psi_1}}\right)\right),\]
where 
\[\|Y_i\|_{\psi_1}=\inf_{K>0}\{K:E\exp(|Y_i|/K)\leq 2\}.\]
\end{lemma}

\begin{lemma}\label{lm5}
Let Assumptions 1--7 hold. For any $\bv\in S_1$, we have
\[P(|\bv'\be\be'\bv'-E(\bv'\be\be'\bv')|>t)\leq 2\exp\left(-C\min\left(\frac{t^2}{N},t\right)\right). \]
As a result,
\[P(|\bv'\bS_e\bv'-E(\bv'\bS_e\bv')|>t)\leq 2\exp\left(-C\min\left({NT^2t^2},NTt\right)\right). \]
\end{lemma}

{\bf Proof.} Without loss of generality, we assume the first $2s_1$ elements of $\bv$ are nonzero, i.e. $\bv=(v_1,...,v_{2s_1},0,...,0)'$. Then,
$\bv'\be=(\bv'\underline{\be}_1,...,\bv'\underline{\be}_N)$,
where 
\[\bv'\underline{\be}_i=\sum_{j=1}^{2s_1}v_je_{i,j}.\]
Define $Z_i=\bv'\underline{\be}_i$, it follows from Assumption~7 that $Z_i$ is a sub-Gaussian random variable. By Lemma 2.7.6 of \cite{vershynin2018high}, $Z_i^2$ is a sub-exponential random variable and $\|Z_i^2\|_{\psi_1}\leq C$. 
Note that $\bv'\be\be'\bv=\sum_{i=1}^NZ_i^2$, by Lemma \ref{lm4}, we have
\begin{align}
    P\left(\left|\sum_{i=1}^N(Z_i^2-E(Z_i^2))\right|>t\right)\leq& 2\exp\left(-c\min\left(\frac{t^2}{\sum_{i=1}^N\|Z_i^2\|_{\psi_1}^2},\frac{t}{\max_{i\leq N}\|Z_i^2\|_{\psi_1}}\right)\right),\notag\\
    \leq &2\exp\left(-C\min\left(\frac{t^2}{N},t\right)\right),
\end{align}
for some constant $C>0$. This completes the proof. $\Box$

\begin{lemma}\label{lm7}
Let Assumptions 1--7 hold. For any vector $\bv\in S_1$, we have
\[P(|\bv'\bSigma_{fe}\bv|>x)\leq N\exp(-C(N\sqrt{T}x)^\gamma)+\exp(-CNTx^2)+\exp(-CNTx^2\exp(\frac{(N\sqrt{T}x)^{\gamma(1-\gamma)}}{(\log(N\sqrt{T}x))^\gamma})),\]
where $\gamma=(1/\gamma_1+1)^{-1}$ for any $\gamma_1>0$.
\end{lemma}

{\bf Proof.} Let $W_j=\bv'\frac{\bF}{\sqrt{T}}\blambda_j$, by definition, we have
\[\bv'\bSigma_{fe}\bv=\frac{1}{NT}\bv'\bF\bLambda'\be'\bv=\frac{1}{NT}\sum_{j=1}^N\bv'\bF\blambda_j\underline{\be}_j'\bv=\frac{1}{N\sqrt{T}}\sum_{j=1}^NW_jZ_j,\]
where $Z_j$ is defined as that in the proof of Lemma~\ref{lm5}. Note that $W_j$ and $Z_j$ are independent of each other and they all have finite variances. In addition, Assumption~5 implies that $Z_j$ are independent across $j=1,...,N$. By Assumption~7, $W_jZ_j$ satisfies condition (2.7) in \cite{merlevede2011bernstein} with $\gamma_2=1$. Since $Z_j$ are independent over $j=1,...,N$, then condition (2.6) in \cite{merlevede2011bernstein} is satisfied for any $\gamma_1>0$. By Theorem 1 of \cite{merlevede2011bernstein}, we have
\begin{equation}\label{max:e}
    P(|\sum_{j=1}^NW_jZ_j|>x)\leq N\exp(-Cx^\gamma)+\exp(-Cx^2/N)+\exp(-C\frac{x^2}{N}\exp(\frac{x^{\gamma(1-\gamma)}}{(\log(x))^\gamma})),
\end{equation}
where $\gamma=(1/\gamma_1+1)^{-1}$. Then Lemma~\ref{lm7} follows from (\ref{max:e}). $\Box$

\vskip 0.5cm
{\bf Proof of Theorem 1.} Note that
\begin{equation}
    \sup_{v\in S_1}|\bv'\bS_e\bv|\leq \sup_{v\in S_1}|\bv'(\bS_e-E\bS_e)\bv|+\sup_{v\in S_1}|\bv'(E\bS_e)\bv|=\Pi_1+\Pi_2.
\end{equation}
By Assumption~5, we have
$E\bS_e=\sigma\bI_T$, where
\[\sigma=\frac{\sigma_1^2+...+\sigma_N^2}{NT}.\]
As $\sigma\leq C/T$ for some constant $C>0$, we have
\begin{equation}\label{2term}
\Pi_2=\sup_{v\in S_1}|\bv'(E\bS_e)\bv|\leq C/T.
\end{equation}
Now turn to $\Pi_1$. Note that $S_1=\{\bv:\|\bv\|_2=1\,\,\text{and}\,\,\|\bv\|_0\leq 2s_1\}$. Let $\mathcal{S}_2^{T-1}$ be the set of $T$-dimensional unit vector and  $\mathcal{B}_0^d$ be the set of vectors satisfiying $\|\bv\|_0\leq d$. Then $S_1=\mathcal{S}_2^{T-1}\cap\mathcal{B}_0^{2s_1}$.

For every possible subset $I\subset\{1,...,T\}$ of size $2s_1$, it is well known that the minimal $\delta$-covering of $\mathcal{S}_2^{2s_1-1}$ in the Euclidean metric has cardinality at most $(1+2/\delta)^{2s_1}$.  See, for example, Proposition D.2 in \cite{vu2013minimax}. 
Let $\mathcal{N}$ be the minimal $\delta$-covering set of $\mathcal{S}_2^{2s_1-1}$ with all possible subsets of $\{1,...,T\}$ with size $2s_1$, it is not hard to see that
\[|\mathcal{N}|\leq\left(\begin{array}{c}
     T \\
     2s_1
\end{array}\right)(1+\frac{2}{\delta})^{2s_1}\leq \left(\frac{eT}{2s_1}\right)^{2s_1}\left(1+\frac{2}{\delta}\right)^{2s_1},\]
where we use the binomial coefficient bound for the first term.
Note that for every $\bx\in \mathcal{S}_2^{T-1}\cap \mathcal{B}_0^{2s_1}$, there exists $\by\in \mathcal{N}$ satisfying $\|\bx-\by\|_2\leq \delta$ and $\bx-\by\in\mathcal{B}_0^{2s_1}$. Therefore,
\[ \sup_{v\in S_1}|\bv'(\bS_e-E\bS_e)\bv|\leq (1-2\delta)^{-1}\sup_{\bv\in\mathcal{N}}|\bv'(\bS_e-E\bS_e)\bv|,\]
because the left-hand-side (LHS) satisfies $LHS\leq 2\delta* LHS+\sup_{\bv\in\mathcal{N}}|\bv'(\bS_e-E\bS_e)\bv|$.
We choose $\delta=1/4$, by the Bonferroni-type inequality and Lemma \ref{lm5},
\begin{align}\label{Eq:rt}
    P\left(\sup_{v\in S_1}|\bv'(\bS_e-E\bS_e)\bv|>t\right)\leq &P\left((1-2\delta)^{-1}\sup_{\bv\in\mathcal{N}}|\bv'(\bS_e-E\bS_e)\bv|>t\right)\notag\\
    \leq &P\left(\sup_{\bv\in\mathcal{N}}|\bv'(\bS_e-E\bS_e)\bv|>t/2\right)\notag\\
    \leq &|\mathcal{N}|P\left(|\bv'(\bS_e-E\bS_e)\bv|>t/2\right)\notag\\
    \leq &C\left(\frac{eT}{2s_1}\right)^{2s_1} 9^{2s_1}\exp\left(-C\min\left({NT^2t^2},NTt\right)\right).
\end{align}
Therefore, we can obtain from (\ref{Eq:rt}) that
\[\sup_{v\in S_1}|\bv'(\bS_e-E\bS_e)\bv|\leq C\max\{\sqrt{\frac{s_1\log(T)}{NT^2}},\frac{s_1\log(T)}{NT}\}\leq C\sqrt{\frac{s_1\log(T)}{NT^2}}+C\frac{s_1\log(T)}{NT}.\]
Combining (\ref{2term}) and the above inequality, we have
\begin{equation}\label{vsev}
 \sup_{v\in S_1}|\bv'\bS_e\bv|\leq C\left(\sqrt{\frac{s_1\log(T)}{NT^2}}+\frac{s_1\log(T)}{NT}+\frac{1}{T}\right).   
\end{equation}
By a similar argument and Lemma~\ref{lm7}, we can also show that

\begin{equation}\label{vsev1}
 \sup_{v\in S_1}|\bv'\bSigma_{fe}\bv|\leq C\left(\sqrt{\frac{s_1\log(T)}{NT}}\right).   
\end{equation}
By Assumption~4 that $\lambda_1-\lambda_2>c>0$, Theorem 1 follows from Lemma \ref{lm3}, (\ref{deco:S}), and the above two inequalities. This completes the proof. $\Box$
\vskip 0.5cm
{\bf Proof of Theorem 2.} According to Algorithms 1--3, we can repeat the proofs of Theorem 1 for the eigenvectors obtained in the subsequent iterations. So long as the previous eigenvector is consistent, we can similarly obtain the consistencies of the next few estimated eigenvectors. On the other hand, because the number of factors $r$ is finite, by (3.5), we have 
\[\rho(\wh\bV_1,\wh\bV_1)^2=2\Tr\left(\bI_r-\wh\bV'\bV\bV'\wh\bV\right).\]
Note that the columns in the true factor matrix $\bF$ are orthogonal, then the cross terms in $\wh\bV\bV$ are of the same order as that in Theorem~1 of the main text.  We can apply the previous results to each component of the one on the right-hand side, and the result in Theorem 2 holds since $r$ is finite. See also the descriptions for different distances between two matrices in Equations (3.2)--(3.5) of the main text. This completes the proof. $\Box$

To show the consistency of the cross-validation method, we first introduce some notation. We denote $\bF^0$ as the true factors with sparsity $s_0$, which is also the true one. $\bF^s$ is the true factor matrix which keeps  the largest $s$ elements in each column of $\bF^{0}$ if $s\leq s_0$ and it is equal to $\bF^0$ if $s\geq s_0$. $\wt\bF_1^s$ is the estimated factor matrix with sparsity $s$ using the training sample. Define
\[C_{NT}=\sqrt{\frac{s_0\log(T)}{NT}}+\frac{s_0\log(T)}{NT}+\frac{1}{T},\]
which is the upper bound for $\|\wh\bF-\bF\|_F/\sqrt{T}$ as shown in Theorem 2.
\vskip 0.5cm

{\bf Proof of Theorem~3.} Let $\wh\bF$ be the estimated sparse factors by the proposed algorithms. We apply the least-squares method to (2.3) of the main text and obtain
\begin{equation}\label{lbd:ht}
    \wh\bLambda'=(\wh\bF'\wh\bF)^{-1}\wh\bF'\bX=(\wh\bF'\wh\bF)^{-1}\wh\bF'\bF\bLambda'+(\wh\bF'\wh\bF)^{-1}\wh\bF'\be.
\end{equation}
It follows from (\ref{lbd:ht}) that
\begin{equation}
    \wh\blambda_i-\bH_s\blambda_i=(\frac{\wh\bF'\wh\bF}{T})^{-1}\frac{\wh\bF'\be_i}{T},
\end{equation}
where $\bH_s=(\wh\bF'\wh\bF)^{-1}\wh\bF'\bF$.  {Note that
\begin{align*}
   \max_{1 \leq i \leq N} \left| \frac{\wh{\bF}' \be_i}{T} \right| 
   &\leq \max_{1 \leq i \leq N} \left| \frac{1}{T} (\wh{\bF} - \bF)' \be_i \right| 
   + \max_{1 \leq i \leq N} \left| \frac{1}{T} \bF' \be_i \right|.
\end{align*}
We now consider the first term. Let \( \wt{\bF} \) denote the PCA factors obtained without thresholding. Since (\ref{lbd:ht}) is a least-squares method, we construct the sparse factor estimates \( \wh{\bF} \) by retaining only the largest \( \bs \) elements (in absolute value) of \( \wt{\bF} \) columnwise, using our algorithm without normalization.

Without loss of generality, suppose
\[
\bF = \begin{pmatrix} \bF_1 \\ \bF_2 \end{pmatrix}, \quad
\wt{\bF} = \begin{pmatrix} \wt{\bF}_1 \\ \wt{\bF}_2 \end{pmatrix}, \quad
\wh{\bF} = \begin{pmatrix} \wh{\bF}_1 \\ \wh{\bF}_2 \end{pmatrix},\quad
\be_i = \begin{pmatrix} \be_{i,1} \\ \be_{i,2} \end{pmatrix},
\]
where \( \bF_1 \) contains the non-sparse components and \( \bF_2 = \mathbf{0} \). Under the thresholding procedure, \( \wh{\bF}_1 = \wt{\bF}_1 \) and \( \wh{\bF}_2 = \mathbf{0} \).
Then,
\begin{align*}
\max_{1 \leq i \leq N} \left| \frac{1}{T} (\wh{\bF} - \bF)' \be_i \right|
&\leq \max_{1 \leq i \leq N} \left| \frac{1}{T} (\wh{\bF} - \wt{\bF})' \be_i \right|
+ \max_{1 \leq i \leq N} \left| \frac{1}{T} (\wt{\bF} - \bF)' \be_i \right| \\
&\leq \max_{1 \leq i \leq N} \left| \frac{1}{T} \wt{\bF}_2' \be_{i,2} \right|
+ \max_{1 \leq i \leq N} \left| \frac{1}{T} (\wt{\bF} - \bF)' \be_i \right|.
\end{align*}
By the argument in Lemma B.1 of \cite{bai2003inferential}, the second term is at most of order \( \delta_{NT}^{-2} = O_p(1/N + 1/T^2) \) for each \( i \), where we have used the improved rate \( \delta_{NT}^{-1} = O_p(\sqrt{1/N} + 1/T) \) as established in Proposition 1 of \cite{bai2023approximate}. The first term can be handled similarly by partitioning the true factors and the PCA-estimated factors, following the proof of Lemma B.1 in the same reference. Therefore, the first term is also at most of order \( O_p(1/N + 1/T^2) \) for each \( i \). Consequently,
\begin{align*}
   \max_{1 \leq i \leq N} \left| \frac{\wh{\bF}' \be_i}{T} \right|
   &= O_p\left( \left( \frac{1}{N} + \frac{1}{T^2} \right) \sqrt{\log N} \right) + O_p\left( \sqrt{ \frac{\log N}{T} } \right)=O_p(\frac{\sqrt{\log N}}{N}+\sqrt{\frac{\log N}{T}}).
\end{align*}
If \( \sqrt{T}/N = o(1) \), then the bound simplifies to
\[
\max_{1 \leq i \leq N} \left| \frac{\wh{\bF}' \be_i}{T} \right| = O_p\left( \sqrt{ \frac{\log N}{T} } \right).
\]
This completes the verification of Theorem~3(i).
}

Let $\frac{\wh\bF'\wh\bF}{T}\rightarrow_p\bQ$ and $\bGamma_i=\lim_{T\rightarrow\infty}\Var(\frac{1}{\sqrt{T}}\sum_{t=1}^T\bff_te_{i,t})$, it is straightforward that 
\[   \sqrt{T}(\wh\blambda_i-\bH_s\blambda_i)\longrightarrow_d N(0,\bQ^{-1}\bGamma_i\bQ^{-1}).\]
Note that $\frac{\wh\bF'\wh\bF}{T}\rightarrow_p\bI_r$ due to the average consistency of each estimated factor process and Assumption~2, by Assumptions~5--6 and the results in Theorem~2, we can show that
  \begin{equation}\label{dis:lbd}
       \sqrt{T}(\wh\blambda_i-\bH_s\blambda_i)\longrightarrow_d N(0,\sigma_i^2\bI_r),
    \end{equation}
    where $\sigma_i$ is defined in Assumption~5.
This completes the proof. $\Box$
\vskip 0.5 cm

{
{\bf Proof of Corollary 1.}
We first show the consistency of the information criterion in (2.11). Since the upper bound of the estimated factors is $C_{NT} = \sqrt{\log(T)/N} + 1/T$, it suffices to verify the conditions in Theorem 2 of \cite{bai2002determining}:
\begin{equation}\label{verify}
\frac{N+T}{NT}\log(T)\log\left( \frac{NT}{N+T} \right) \rightarrow 0, \quad \text{and} \quad C_{NT}^{-2} \frac{N+T}{NT}\log(T)\log\left( \frac{NT}{N+T} \right) \rightarrow \infty.
\end{equation}
These hold straightforwardly under the given $C_{NT}$, ensuring that the conditions in \cite{bai2002determining} are met. Hence, the information criterion in (2.11) consistently estimates $r$.

Next, we establish the consistency of the eigenvalue ratio method in (2.12). According to Equations (2)–(3) and the subsequent discussion in \cite{Ahn2013}, the conditions stated in Corollary~1, along with Assumptions~1–7, satisfy all the requirements of Theorem~1 in \cite{Ahn2013}. Therefore, the ratio-based estimator in (2.12) also consistently estimates $r$.
This completes the proof of the corollary.  $\Box$

}

To prove Theorem 4, we need the following lemmas.
\vskip 0.5 cm
\begin{lemma}\label{lm8}
    Let Assumptions 1--7 hold. For any $s$ with $1\leq s\leq s_0$, we have
    \[R(s,\wt\bF^{s})-R(s,\bF^{s})=O_p(C_{N_1T}),\]
    where we ignore the subscript in (2.13) and denote $\wt\bF^{s}$ as the estimated factor matrix from $\wt\bX_1$ with sparsity $s$, and $\bF^{s}$ as the true factor matrix retaining only keeps the top $s$ elements, both of which are normalized.
\end{lemma}

{\bf Proof.} We only show it for $r=1$ since the proof is similar for $r>1$. Suppose the true factor $\bF^s$ is the normalized true factor matrix such that $\bF^s{'}\bF^s/T=\bI_r$, which is similarly assumed in \cite{bai2013principal}. By (2.13), we have
\begin{equation}
    R(s,\wt\bF^{s})-R(s,\bF^s)=\frac{1}{N_2T}\|\wt\bX_2-\frac{1}{T}\wt\bF^s\wt\bF^s{'}\wt\bX_2\|_F^2-\frac{1}{N_2T}\|\wt\bX_2-\frac{1}{T}\bF^s\bF^s{'}\wt\bX_2\|_F^2.
\end{equation}
Note that
\[\|\wt\bX_2-\frac{1}{T}\wt\bF^s\wt\bF^s{'}\wt\bX_2\|_F^2=\Tr\{(\wt\bX_2-\frac{1}{T}\wt\bF^s\wt\bF^s{'}\wt\bX_2)(\wt\bX_2-\frac{1}{T}\wt\bF^s\wt\bF^s{'}\wt\bX_2)'\}\] and
\[\|\wt\bX_2-\frac{1}{T}\bF^s\bF^s{'}\wt\bX_2\|_F^2=\Tr\{(\wt\bX_2-\frac{1}{T}\bF^s\bF^s{'}\wt\bX_2)(\wt\bX_2-\frac{1}{T}\bF^s\bF^s{'}\wt\bX_2)'\}.\]
By an elementary argument, we can show that
\begin{equation}\label{er:df}
    R(s,\wt\bF^{s})-R(s,\bF^s)=\frac{1}{N_2T}\Tr\{(\frac{1}{T}\bF^s\bF^s{'}-\frac{1}{T}\wt\bF^s\wt\bF^s{'})\wt\bX_2\wt\bX_2'\}.
\end{equation}
Since
\begin{equation}\label{ffhat}
 \| \frac{1}{T}\bF^s\bF^s{'}-\frac{1}{T}\wt\bF^s\wt\bF^s{'}\|_2=\|\frac{1}{\sqrt{T}}(\bF^s-\wt\bF^s)\frac{\bF^s{'}}{\sqrt{T}}+\frac{\wt\bF^{s}}{\sqrt{T}}\frac{(\bF^s-\wt\bF^s)'}{\sqrt{T}}\|_2=O_p(C_{N_1T}),  
\end{equation}
and
\[\|\wt\bX_2\wt\bX_2'\|_2=O_p(N_2T),\]
it follows from (\ref{er:df}) that
\[ R(s,\wt\bF^{s})-R(s,\bF^s)\leq \frac{r}{N_2T}\|(\frac{1}{T}\bF^s\bF^s{'}-\frac{1}{T}\wt\bF^s\wt\bF^s{'})\wt\bX_2\wt\bX_2'\|_2=O_p(C_{N_1T}).\]
This completes the proof. $\Box$

\begin{lemma}\label{lm9}
   Let Assumptions 1--7 hold. For any $s$ with $ s< s_0$, there exists a constant $\tau_s>0$ such that
       \[R(s,\bF^{s})-R(s_0,\bF^{0})\geq \tau_s,\]
       with probability tending to one as $N,T\rightarrow \infty$. $\bF^s$ is the same as that in Lemma~\ref{lm8} and $\bF^{0}$ is the true factor matrix with true sparsity $s_0$.
\end{lemma}
{\bf Proof.} By a similar argument as that in (\ref{er:df}), we can obtain that 
\begin{equation}\label{full:d}
    R(s,\bF^s)-R(s_0,\bF^0)=\frac{1}{N_2T}\Tr\{[\bF^0(\bF^0{'}\bF^0)^{-1}\bF^0{'}-\bF^s(\bF^s{'}\bF^s)^{-1}\bF^{s}{'}]\wt\bX_2\wt\bX_2'\},
\end{equation}
where we may only impose one normalization condition on $\bF^s$ or $\bF^{0}$. For example, we may assume $\bF^{s}{'}\bF^s/T=\bI_r$ and $\bF^0=\bF^s+\bG$, where $G$ is a nonzero vector when $s<s_0$. Note that
\begin{equation}\label{x2x2}
\wt\bX_2\wt\bX_2'=\bF^0\bLambda_2'\bLambda_2\bF^0{'}+\bF^0\bLambda_2{'}\bE_2'+\bE_2\bLambda_2\bF^0{'}+\bE_2\bE_2',
\end{equation}
where $\bLambda_2$ and $\bE_2$ are the loading matrix and idiosyncratic term of $\wt\bX_2$, respectively.
Then,
\begin{align}\label{trace:d}
  \Tr\{[\bF^0(\bF^0{'}\bF^0)^{-1}\bF^0{'}-&\bF^s(\bF^s{'}\bF^s)^{-1}\bF^{s}{'}]\wt\bX_2\wt\bX_2'\}\notag\\
  =& \Tr\{ \bF^0\bLambda_2'\bLambda_2\bF^0{'}-\bF^s\bLambda_2'\bLambda_2\bF^0{'}-\bF^s(\bF^s{'}\bF^s)^{-1}\bF^s{'}\bG\bLambda_2'\bLambda_2\bF^0{'}\notag\\
  &+\bF^0\bLambda_2'\bE_2'+\bF^0(\bF^0{'}\bF^0)^{-1}\bF^0{'}\bE_2\bLambda_2\bF^0{'}-\bF^s(\bF^s{'}\bF^s)^{-1}\bF^{s}{'}\bF^0\bLambda_2'\bE_2'\notag\\
  &-\bF^s(\bF^s{'}\bF^s)^{-1}\bF^{s}{'}\bE_2'\bLambda_2\bF^0{'}-\bF^s(\bF^s{'}\bF^s)^{-1}\bF^{s}{'}\bE_2\bE_2'\}\notag\\
  =&I+II,
\end{align}
where
\[I=\Tr\{ \bF^0\bLambda_2'\bLambda_2\bF^0{'}-\bF^s\bLambda_2'\bLambda_2\bF^0{'}-\bF^s(\bF^s{'}\bF^s)^{-1}\bF^s{'}\bG\bLambda_2'\bLambda_2\bF^0{'}\},\]
and $II$ is the remaining one. It is obvious that $II$ is of a smaller order than $I$, and we only consider $I$.  Note that $\bF^0=\bF^s+\bG$, we have
\[I=\Tr\{\bG\bLambda_2'\bLambda_2\bG'-\bF^s(\bF^s{'}\bF^s)^{-1}\bF^{s}{'}\bG\bLambda_2'\bLambda_2\bG'\}.\]
Since each column in $\bF^s(\bF^s{'}\bF^s)^{-1}\bF^{s}{'}$ is a hat matrix and $\bG\bLambda_2'\bLambda_2\bG'$ is a positive semi-definite matrix, then there exists a constant $\tau_s>0$ such that $I\geq \tau_sN_2T$. Then Lemma \ref{lm9} follows from this result and (\ref{full:d})--(\ref{trace:d}). This completes the proof. $\Box$

\begin{lemma}\label{lm10}
   Let Assumptions 1--7 hold. For any $s$ with $ s\geq s_0$, we have
       \[R(s,\wt\bF^{s})-R(s_0,\wt\bF^{0})=O_p(C_{N_1T}^2),\]
       where 
\end{lemma}
{\bf Proof.} First, we observe that
\begin{align}
 |R(s,\wt\bF^{s})-R(s_0,\wt\bF^{0})|\leq& |R(s,\wt\bF^{s})-R(s_0,\bF^{0})|+|R(s_0,\bF^{0})-R(s_0,\wt\bF^{0})|\notag\\
 \leq &2\max_{s_0\leq s\leq smax}|R(s,\wt\bF^{s})-R(s_0,\bF^{0})|.
\end{align}
Thus, it is sufficient to prove for each $s$ with $s\geq s_0$,
\begin{equation}\label{max:r}
 |R(s,\wt\bF^{s})-R(s_0,\bF^{0})|=O_p(C_{N_1T}^2).   
\end{equation}
Note that
\begin{align}\label{full:2}
    R(s,\wt\bF^s)-R(s_0,\bF^0)=&\frac{1}{N_2T}\Tr\{[\bF^0(\bF^0{'}\bF^0)^{-1}\bF^0{'}-\wt\bF^s(\wt\bF^s{'}\wt\bF^s)^{-1}\wt\bF^{s}{'}]\wt\bX_2\wt\bX_2'\}\notag\\
    =&\frac{1}{N_2T}\Tr(\bL).
\end{align}
By (\ref{x2x2}), we can show that
\begin{align}\label{L:d}
\bL=&\bF^0\bLambda_2'\bLambda_2\bF^0{'}+\bF^0\bLambda_2'\bE_2'+\bF^0(\bF^0{'}\bF^0)^{-1}\bF^0{'}\bE_2\bLambda_2\bF^0{'}+\bF^0(\bF^0{'}\bF^0)^{-1}\bF^0{'}\bE_2\bE_2'\notag\\
&-\wt\bF^s(\wt\bF^s{'}\wt\bF^s)^{-1}\wt\bF^{s}{'}\bF^0\bLambda_2'\bLambda_2\bF^0{'}-\wt\bF^s(\wt\bF^s{'}\wt\bF^s)^{-1}\wt\bF^{s}{'}\bF^0\bLambda_2\bF^0{'}-\wt\bF^s(\wt\bF^s{'}\wt\bF^s)^{-1}\wt\bF^{s}{'}\bE_2\bLambda_2\bF^0{'}\notag\\
&-\wt\bF^s(\wt\bF^s{'}\wt\bF^s)^{-1}\wt\bF^{s}{'}\bE_2\bE_2'.
\end{align}
We only show the upper bound for the dominant term since those for the rest are similar.
Note that
\begin{align}
   \Tr\{ \bF^0\bLambda_2'\bLambda_2\bF^0{'}-&\wt\bF^s(\wt\bF^s{'}\wt\bF^s)^{-1}\wt\bF^{s}{'}\bF^0\bLambda_2'\bLambda_2\bF^0{'}\}\notag\\
    =&\Tr\{\bF^0\bLambda_2'\bLambda_2\bF^0{'}-\wt\bF^s\bLambda_2'\bLambda_2\bF^0{'}-\wt\bF^s(\wt\bF^s{'}\wt\bF^s)^{-1}\wt\bF^{s}{'}(\bF^0-\wt\bF^s)\bLambda_2'\bLambda_2\bF^0{'}\}\notag\\
    =&\Tr\{(\bF^0-\wt\bF^s)\bLambda_2'\bLambda_2\bF^0{'}-\wt\bF^s(\wt\bF^s{'}\wt\bF^s)^{-1}\wt\bF^{s}{'}(\bF^0-\wt\bF^s)\bLambda_2'\bLambda_2\bF^0{'}\}\notag\\
    =&\Tr\{(\bF^0-\wt\bF^s)\bLambda_2'\bLambda_2\bF^0{'}-(\bF^0-\wt\bF^s)\bLambda_2'\bLambda_2\wt\bF^s{'}\notag\\
    &-\wt\bF^s(\wt\bF^s{'}\wt\bF^s)^{-1}\wt\bF^{s}{'}(\bF^0-\wt\bF^s)\bLambda_2'\bLambda_2(\bF^0-\wt\bF^s){'}\}\notag\\
    =&\Tr\{(\bF^0-\wt\bF^s)\bLambda_2'\bLambda_2(\bF^0-\wt\bF^s){'}-\wt\bF^s(\wt\bF^s{'}\wt\bF^s)^{-1}\wt\bF^{s}{'}(\bF^0-\wt\bF^s)\bLambda_2'\bLambda_2(\bF^0-\wt\bF^s){'}\}.
\end{align}
It is not hat to show that
\[\|(\bF^0-\wt\bF^s)\bLambda_2'\bLambda_2(\bF^0-\wt\bF^s){'}\|_2\leq N_2T\|\frac{(\bF^0-\wt\bF^s)}{\sqrt{T}}\frac{\bLambda_2'\bLambda_2}{N_2}\frac{(\bF^0-\wt\bF^s){'}}{\sqrt{T}}\|_2 =O_p(N_2T C_{N_1T}^2).\]
Then, we can show that
\begin{equation}\label{1st:t}
    \frac{1}{N_2T}\Tr\{ \bF^0\bLambda_2'\bLambda_2\bF^0{'}-\wt\bF^s(\wt\bF^s{'}\wt\bF^s)^{-1}\wt\bF^{s}{'}\bF^0\bLambda_2'\bLambda_2\bF^0{'}\}=O_p(C_{N_1T}^2).
\end{equation}
By a similar argument for the remaining terms in (\ref{L:d}), we can show that
\[\frac{1}{N_2T}\Tr(\bL)=O_p(C_{N_1T}^2),\]
which implies (\ref{max:r}). This completes the proof. $\Box$

\vskip 0.5cm
{\bf Proof of Theorem 4.} We focus on the case where the number of partitions $J = 1$, as the argument extends similarly to a general $J > 1$. Under Assumption 3, which states that $s \asymp T$, we will show that for sufficiently large $T$,
\[\lim_{N\rightarrow\infty}P(PC(s)<PC(s_0))=0,\]
for all $s \neq s_0$ with $s \leq s_{\max}$.  {Since $s_0/T = c_0 \in (0,1)$ by Assumption 3, for any $s$ such that $s/T = c_s \in (0,1)$, we adopt the theoretical convention in the proof that $s < s_0$ implies $c_s < c_0$, and similarly, $s > s_0$ implies $c_s > c_0$.}  We first consider the case when $s<s_0$. Note that
\[PC(s)-PC(s_0)=R(s,\wt\bF^s)-R(s_0,\wt\bF^0)-r\frac{(s_0-s)}{T}C_Tg(N_1,T),\]
and
\[R(s,\wt\bF^s)-R(s_0,\wt\bF^0)=[R(s,\wt\bF^s)-R(s,\bF^s)]+[R(s,\bF^s)-R(s_0,\bF^0)]+[R(s_0,\bF^0)-R(s_0,\wt\bF^0)].\]
It follows from Lemma~\ref{lm8} that
\[R(s,\wt\bF^s)-R(s,\bF^s)=O_p(C_{N_1T}),\]
and
\[R(s_0,\bF^0)-R(s_0,\wt\bF^0)=O_p(C_{N_1T}).\]
Lemma~\ref{lm9} implies that
\[R(s,\bF^s)-R(s_0,\bF^0)\geq\tau_s>0.\]
Then,
\[P(PC(s)-PC(s_0)<0)=P(\tau_s+O_p(C_{N_1T})<r\frac{(s_0-s)}{T}C_Tg(N_1,T))\rightarrow 0,\]
since $C_Tg(N_1,T)\rightarrow 0$ and $C_{N_1T}$ is small enough for large $T$.

Next, we consider the case when $s\geq s_0$. Similar,
\[P(PC(s)-PC(s_0)<0)=P(R(s_0,\wt\bF^0)-R(s,\wt\bF^s)>r\frac{(s-s_0)}{T}C_Tg(N_1,T)).\]
By Lemma~\ref{lm10}, we have 
\[R(s_0,\wt\bF^0)-R(s,\wt\bF^s)=O_p(C_{N_1T}^2).\]
Since $C_{N_1T}^{-2}C_Tg(N_1,T)\rightarrow\infty$, we have that
\[P(R(s_0,\wt\bF^0)-R(s,\wt\bF^s)>r\frac{(s-s_0)}{T}C_Tg(N_1,T))\rightarrow 0,\]
as $N\rightarrow\infty$ and $T$ is large enough. This completes the proof. $\Box$

\section{Description of the Identified Sparse Factors}
In this section, we report the identified significant factors of stock returns over the time horizon from January 1, 2004, to December 31, 2016, studied in the empirical analysis. A comprehensive list of dates with significant systematic risk factors, the reasons for these factors, and their associated time points are provided in Table~\ref{Table-ft}. 

  
{\begin{center}\tiny
\begin{longtable}{lcp{7cm}p{5cm}}
\caption{Significant sparse factors over the time horizon from January 1, 2004, to December 31, 2016. Dates are formatted as {\it yyyymmdd}. The signs in the Factor Return column indicate whether the significant time factor resulted in a positive ($+$) or negative ($-$) return on that day. Reasons in the Reason column are extracted from daily reports on  {\it CNN Money} (\url{www.money.cnn.com}) after the market closes each trading day. The Time Factor column summarizes the corresponding factor for each trading day with significant common risk.  This website was shut down in late 2018 and its content was merged into the main CNN Business section.} 
          \label{Table-ft}\\
\toprule
Date	&	Factor Return	&	{Reason}	&	Time Factor	\\
\hline
20070227	&	$-$	&	A big decline in Chinese stocks, weakness in some key readings on the US economy and news that Vice President Dick Cheney was the apparent target in a Taliban suicide bombing attack in Afghanistan	&	China, Economic indicators, Global events (terrorist attack)	\\
20070803	&	$-$	&	Credit market fears, sparked by Wall Street bank Bear Stearns	&	Credit risk, Market sentiment	\\
20070809	&$-$&	As new credit concerns surfaced, sparking a steep selloff	&Credit risk, Market sentiment\\
20070828	&$-$&	Credit card defaults keep rising	&Credit risk\\
20070918	&	$+$	&	After the Federal Reserve cut a key short-term interest rate by a half-percentage point	&	Government policies	\\
20071019	&$-$&As Wachovia revives worries about bank sector	&Market sentiment\\
20071101&	$-$	&On credit fears	&Credit risk, Market sentiment\\
20071107	&	$-$	&	Renewed credit market fears	&	Credit risk, Market sentiment	\\
20071113	&$+$&	After Wal-Mart's earnings report	&Company-specific factors\\
20071128	&	$+$	&	Expectations the Federal Reserve will continue its rate cutting campaign and on strength from the embattled financials	&	Government policies, Market sentiment	\\
20071211	&	$-$	&	After the Federal Reserve cut the fed funds rate by a quarter-percentage point, as expected, but disappointed some investors looking for a bigger cut	&	Government policies, Market sentiment	\\
20080104	&	$-$	&	After a weaker-than-expected December jobs report exacerbated worries that the economy may be falling into recession	&	Economic indicators, Market sentiment	\\
20080115	&F$-$&	After Citigroup's weak earnings and a big drop in retail sales revived worries about the threat of recession	&Company-specific factors, Market sentiment\\
20080117	&	$-$	&	Recession worries following comments from Federal Reserve Chairman Ben Bernanke, Merrill Lynch's big quarterly loss and weak readings on the housing and manufacturing sectors	&	Market sentiment, Company-specific factors, Economic indicators	\\
20080123	&	$+$	&	The Federal Reserve  stepped in Tuesday and announced an emergency intermeeting interest rate cut, a decision that initially had a mixed impact on stocks, but helped pave the way for Wednesday's bounce back	&	Government policies	\\
20080205	&	$-$	&	After a report showing a big slowdown in the services sector of the economy and cautionary comments from a Fed official amplified fears that a recession is underway or imminent	&	Company-specific factors, Government policies	\\
20080229	&	$-$	&	After AIG's record loss added to worries about the financial sector and more weak economic news intensified fears about a recession	&	Company-specific factors	\\
20080306&	$-$	&As investors eyed the latest wave of credit market woes	&Credit risk, Market sentiment\\
20080311	&	$+$	&	Announcement: the Federal Reserve will lend up to  \$$200$ billion to banks and lenders as a means of loosening up tight credit markets	&	Government policies	\\
20080318	&	$+$	&	After the Federal Reserve cut the fed funds rate by three-quarters of a percentage point, surprising investors looking for a larger cut	&	Government policies	\\
20080401	&	$+$	&	Investors cheered signs that the companies hit hardest by the credit market crisis seem to be working through the problems	&	Market sentiment	\\
20080606	&	$-$	&	After oil prices spiked more than $\$$ $11$ a barrel and the May jobs report showed a big jump in the unemployment rate	&	Oil price, Economic indicators	\\
20080626	&	$-$	&	Selling accelerated following a record surge in oil prices	&	Oil price	\\
20080708	&	$+$	&	Falling oil prices and a stronger dollar	&	Oil price	\\
20080709	&	$-$	&	More worries about Freddie Mac and Fannie Mae's ability to raise capital exacerbated credit market and corporate profit jitters	&	Market sentiment	\\
20080716	&	$+$	&	Encouraging news from the banking and airline sectors Falling oil prices also helped spark a strong stock market rally	&	Economic indicators, Oil price	\\
20080717	&	$+$	&	A string of large oil price declines and more encouraging earnings results from the financial sector	&	Oil price, Economic indicators	\\
20080724	&	$-$	&	Renewed fears about the battered housing market and rising unemployment	&	Market sentiment	\\
20080729	&	$+$	&	Strong financial reports, rising consumer confidence and falling oil prices	&	Economic indicators, Oil price	\\
20080805	&	$+$	&	Oil prices fell sharply and investors appeared to take solace in the Federal Reserve's assessment of the nation's economy	&	Oil price, Government policies	\\
20080808&	$+$	&Oil prices tumbled a 3-month low near $\$$ $115$ a barrel	&Oil price\\
20080904	&	$-$	&	Mixed retail sales, lower oil prices and dour labor market readings amplified worries about a global economic slowdown	&	Economic indicators, Oil price	\\
20080909	&	$-$	&	Worries about Lehman Brothers' ability to raise capital, and about the extent of AIG's mortgage-related losses, exacerbated broad recession fears	&	Company-specific factors	\\
20080915	&	$-$	&	After one of the most calamitous days in US financial services history resulted in Bank of America's \$$50$ billion deal to buy Merrill Lynch and the bankruptcy filing of Lehman Brothers	&	Company-specific factors	\\
20080917	&	$-$	&	The government's emergency rescue of AIG amplified fears about the stability of financial markets	&	Government policies	\\
20080918	&	$+$	&	On a CNBC report that the government is working on a more permanent solution to absorbing bad debt	&	Government policies	\\
20080919	&	$+$	&	The government's plan to help rescue banks from toxic mortgage debt soothed investors at the end of a gut-churning week on Wall Street	&	Government policies	\\
20080922	&	$-$	&	As investors worried about the specifics of the government's \$$700$ billion bailout plan and rocketing oil prices - which saw its biggest one-day dollar gain ever	&	Market sentiment	\\
20080929	&	$-$	&	After the House rejected the government's \$$700$ billion bank bailout plan	&	Government policies	\\
20080930	&	$+$	&	On bets that Congress will pass a version of the government's \$$700$ billion package, following Monday's crushing defeat	&	Market sentiment, Government policies	\\
20081002	&	$-$	&	As frozen credit markets and weak economic reports amplified jitters ahead of the House vote on the \$$700$ billion bank rescue plan	&	Market sentiment, Government policies	\\
20081006	&	$-$	&	As the \$$700$ billion bank bailout plan and European government attempts to prop up faltering banks failed to comfort panicky investors	&	Market sentiment, Government policies	\\
20081007	&	$-$	&	As the Federal Reserve's plan to loosen credit markets failed to temper investor pessimism	&	Market sentiment, Government policies	\\
20081009	&	$-$	&	As panicked investors dumped stocks across the board	&	Market sentiment	\\
20081013	&	$+$	&	As investors bet that the worst of the credit crisis is over, following a series of global initiatives announced over the last few days	&	Market sentiment	\\
20081015	&	$-$	&	A weak retail sales report and dour forecasts from the Federal Reserve, coupled with sober comments from Fed Chairman Ben Bernanke, sent stocks tumbling	&	Economic indicators, Government policies	\\
20081016	&	$+$	&	As the lowest oil prices in more than a year gave investors a reason to scoop up shares battered in the recent market selloff	&	Oil price	\\
20081020	&	$+$	&	As investors welcomed talk of a second economic stimulus plan and an improvement in key lending rates	&	Market sentiment, Government policies	\\
20081021	&	$-$	&	As mixed corporate earnings reports gave investors a reason to retreat after the previous session's big rally	&	Economic indicators	\\
20081022	&	$-$	&	As weak earnings and slumping oil prices amplified fears of a global recession	&	Oil price	\\
20081024	&	$-$	&	As Wall Street joined a worldwide market slump on bets that a recession is imminent - if not already under way	&	Market sentiment	\\
20081027	&	$-$	&	As recession jitters outweighed relief that the government's programs to shore up the financial system have gotten underway	&	Market sentiment	\\
20081028	&	$+$	&	As investors dove back into stocks near the end of one of the worst months in Wall Street history	&	Market sentiment	\\
20081031	&	$+$	&	Capping off a strong week at the end of one of the worst months in Wall Street history	&	Market sentiment	\\
20081104	&	$+$	&	As millions of Americans battered by the weakened economy turned out to vote for the next President of the United States	&	Market sentiment	\\
20081105	&	$-$	&	As Barack Obama's historic victory gave way to renewed worries about the struggling economy	&	Market sentiment	\\
20081106	&	$-$	&	As fears of a prolonged recession sent investors running for the exits	&	Market sentiment	\\
20081110	&	$-$	&	As ongoing recession fears overshadowed any relief about China's \$$586$ billion stimulus plan and the government's revamping of its deal with AIG	&	China, Government policies	\\
20081111	&	$-$	&	As recession fears trumped a new government and mortgage industry plan to help troubled homeowners	&	Market sentiment, Government policies	\\
20081112	&	$-$	&	As investors bet that a long and deep recession is on the horizon	&	Market sentiment	\\
20081113	&	$+$	&	As the major stock gauges bounced back from levels not seen since 2003	&	Market sentiment	\\
20081114	&	$-$	&	As the worst retail sales on record ignited fears of a long recession	&	Economic indicators	\\
20081117	&	$-$	&	As investors eyed Citigroup's massive job losses and a weak manufacturing report, while awaiting the fate of a potential bailout for the automakers	&	Company-specific factors, Economic indicators	\\
20081119	&	$-$	&	As ongoing anxiety about the economy and uncertainty about the future of the auto industry weighed on the market	&	Market sentiment	\\
20081120	&	$-$	&	As fears of a prolonged recession sparked a massive selloff	&	Market sentiment	\\
20081121	&	$+$	&	After reports surfaced that President-elect Barack Obama will nominate New York Federal Bank President Timothy Geithner as his new Treasury secretary	&	Government policies	\\
20081124	&	$+$	&	As Citigroup's massive rescue package and President-elect Obama's picks for his economic team pushed investors off the sidelines	&	Company-specific factors, Government policies	\\
20081126	&	$+$	&	As investors scooped up stocks hit in the recent selloff, ahead of the Thanksgiving holiday	&	Market sentiment	\\
20081201	&	$-$	&	As investors bailed out following confirmation that the US is mired in a recession and indications that it's likely to continue for some time	&	Market sentiment	\\
20081202	&	$+$	&	As investors welcomed signs that the automakers might get a bailout after all	&	Government policies	\\
20081203&	$+$	&Oil prices fall after government inventory report	&Oil price\\
20081204	&	$-$	&	As a rash of job cuts at major companies added to jitters ahead of the November jobs report	&	Economic indicators	\\
20081205	&	$+$	&	After a brutal November employment report, as investors extended the recent trend of buying despite the bad news	&	Economic indicators	\\
20081208	&	$+$	&	As investors welcomed President-elect Barack Obama's plan to create jobs and revive the economy, and reports that government help for the automakers is on the way	&	Government policies	\\
20081211	&	$-$	&	On worries that the $\$$ $14$ billion auto rescue bill won't pass in the Senate due to Republican opposition	&	Government policies	\\
20081216	&	$+$	&	After the Federal Reserve cut a key short-term interest rate to the lowest level on record, and signaled it had more tools available to help the economy as the recession stretches on	&	Government policies, Market sentiment	\\
20081222	&	$-$	&	Amid concerns about fourth-quarter corporate earnings, falling oil prices and ongoing woes in the auto industry	&	Market sentiment, Oil price, Economic indicators	\\
20081230	&	$+$	&	As investors scooped up a variety of shares hit hard in the 2008 stock market battering	&	Market sentiment	\\
20090102	&	$+$	&	With investors starting off a new year on the right foot, after an abysmal 2008	&	Market sentiment	\\
20090106	&	$+$	&	As investors looked beyond the Federal Reserve's dour outlook on the economy and instead scooped up shares hit in last year's big selloff	&	Market sentiment	\\
20090107	&	$-$	&	After weak labor market reports and dour forecasts from Alcoa and Intel gave investors reasons to retreat after the recent rally	&	Economic indicators, Company-specific factors	\\
20090109	&	$-$	&	After a government report showed another big monthly drop in payrolls, resulting in the biggest annual job loss since just after World War II	&	Economic indicators	\\
20090112	&	$-$	&	By concerns about Citigroup's potential deal with Morgan Stanley - and the start of the fourth-quarter earnings reporting period	&	Company-specific factors	\\
20090114	&	$-$	&	As a bleak retail sales report and more dour news from the banking sector amplified fears of a prolonged recession	&	Economic indicators	\\
20090120	&	$-$	&	As investors looked beyond President Barack Obama's historic inauguration to the battered economy he inherits	&	Market sentiment	\\
20090121	&	$+$	&	As investors welcomed IBM's earnings and scooped up bank shares hit hard in the recent retreat	&	Company-specific factors	\\
20090122	&	$-$	&	As a management shakeup at Bank of America and Microsoft's earnings disappointment weighed on investor sentiment	&	Company-specific factors	\\
20090128	&	$+$	&	As investors took comfort in reports that the Obama administration and the Federal Reserve are taking steps to get credit flowing again and help staunch the economic slowdown	&	Government policies	\\
20090129	&	$-$	&	Following more dire news on earnings, housing and employment	&	Economic indicators	\\
20090130	&	$-$	&	As investors eyed abysmal reports on economic growth and quarterly earnings	&	Economic indicators	\\
20090206	&	$+$	&	As optimism about the government's economic stimulus bill and the new version of the bank bailout plan countered unease following the brutal January jobs report	&	Government policies, Economic indicators	\\
20090210	&	$-$	&	As the government's bank rescue plan failed to reassure investors burned by the 14-month old recession	&	Government policies	\\
20090217	&	$-$	&	On fears that the government's efforts to slow the recession won't be sufficient	&	Market sentiment	\\
20090219	&	$-$	&	As fears of a prolonged recession sent stock investors heading for the exits	&	Market sentiment	\\
20090223	&	$-$	&	As investors continue to worry that the government's efforts to slow the recession won't be sufficient	&	Market sentiment	\\
20090224	&	$+$	&	After comments from Fed Chairman Ben Bernanke that downplayed bank takeover fears helped to spark a big rally	&	Government policies, Market sentiment	\\
20090227	&	$-$	&	On worries about the government taking a bigger chunk of Citigroup and a bleak reading on the economy, again touching 12-year lows	&	Government policies, Economic indicators	\\
20090302	&	$-$	&	After insurance company American International Group's huge quarterly loss added to worries about the financial sector and the economy	&	Company-specific factors	\\
20090304	&	$+$	&	Following reports that China's economy may be improving and as government officials unveiled details of the \$$75$ billion foreclosure fix	&	China	\\
20090305	&	$-$	&	As investors waded through more grim news: GM said its survival is in doubt, bank shares took a beating, and Citigroup fell below a buck	&	Company-specific factors	\\
20090310	&	$+$	&	After Citigroup cooled some worries about its future and regulators said they may reinstate a key trading rule	&	Company-specific factors, Government policies	\\
20090312	&	$+$	&	As investors scooped up banks and other shares hit in a selloff that left the Dow at 12-year lows	&	Market sentiment	\\
20090317	&	$+$	&	As investors continued to dig out from 12-year lows	&	Market sentiment	\\
20090318	&	$+$	&	After the Federal Reserve said it would buy up to \$300 billion in long-term government bonds	&	Government policies	\\
20090320	&	$-$	&	As investors pulled back after the recent run	&	Market sentiment	\\
20090323	&	$+$	&	After Treasury's plan to buy up billions in bad bank assets and a better-than-expected existing home sales report raised hopes that the economy is stabilizing	&	Government policies, Economic indicators	\\
20090326	&	$+$	&	As the March market run shows its legs	&	Market sentiment	\\
20090330	&	$-$	&	As auto and bank woes spark a selloff after the rally	&	Economic indicators	\\
20090402	&	$+$	&	After key accounting rule that has impact on banks is changed G-20 also in focus	&	Government policies	\\
20090407	&	$-$	&	After a four-week advance, on worries about banks and autos and the start of the quarterly reporting period	&	Market sentiment	\\
20090409	&	$+$	&	After Wells Fargo forecast a nearly \$$3$ billion quarterly profit, adding to hopes that the banking sector is stabilizing	&	Company-specific factors	\\
20090414	&	$-$	&	After a weaker-than-expected retail sales report gave investors a reason to retreat following a five-week run	&	Company-specific factors, Market sentiment	\\
20090416	&	$+$	&	The major stock gauges touch the best levels in months on JPMorgan Chase earnings	&	Company-specific factors	\\
20090420	&	$-$	&	On worries about financial sector earnings, despite Bank of America's better-than-expected quarterly results	&	Market sentiment	\\
20090421	&	$+$	&	As worries about corporate results were countered by renewed hopes that the financial sector is closer to stabilizing	&	Market sentiment	\\
20090424	&	$+$	&	After Ford, Microsoft and American Express reported results that met or topped analysts' expectations	&	Company-specific factors	\\
20090429	&	$+$	&	After the Federal Reserve held interest rates steady, as expected, but issued a slightly more upbeat economic outlook	&	Government policies	\\
20090504	&	$+$	&	As a better-than-expected housing market report intensified hopes that the economy is closer to stabilizing	&	Economic indicators	\\
20090508	&	$+$	&	After a government report showed employers cut fewer jobs than expected last month	&	Economic indicators	\\
20090511	&	$-$	&	As investors took a step back after propelling the major stock gauges by more than $30\%$ each in just two months	&	Market sentiment	\\
20090513	&	$-$	&	After a weaker-than-expected retail sales report gave investors a reason to retreat	&	Economic indicators	\\
20090518	&	$+$	&	After positive news about the US housing market, including an upbeat profit forecast from Lowes, as well as an upgrade of Bank of America	&	Economic indicators, Company-specific factors	\\
20090526	&	$+$	&	After a report showing consumer confidence hit an eight-month high offset dismal housing news	&	Market sentiment	\\
20090601	&	$+$	&	As better-than-expected readings on manufacturing activity raised hopes that a global economic recovery is brewing	&	Economic indicators	\\
20090615	&	$-$	&	As weaker oil prices and more geopolitical unrest raised worries that the recession may not be waning as soon as some had hoped	&	Oil price, Global events (Geopolitical events)	\\
20090622	&	$-$	&	As the World Bank's weak outlook on global growth and a selloff in commodity prices sent investors heading for the exits	&	Market sentiment	\\
20090702	&	$-$	&	After a worse-than-expected jobs report hammered hopes that the economy is close to stabilizing	&	Economic indicators	\\
20090713	&	$+$	&	As investors welcomed an analysts' improved outlook on Goldman Sachs one day ahead of its quarterly report	&	Market sentiment, Company-specific factors	\\
20090715	&	$+$	&	After Intel's forecast for a second-half pickup and the Federal Reserve's improved outlook reassured wary investors	&	Company-specific factors	\\
20090723	&	$+$	&	As investors welcomed better-than-expected quarterly results and home sales	&	Economic indicators	\\
20090817	&	$-$	&	As worries that nervous consumers will pressure a fragile recovery dragged stocks lower after a five-month advance	&	Market sentiment	\\
20090901	&	$-$	&	On worries that the market gains have raced ahead of any economic recovery	&	Market sentiment	\\
20091001	&	$-$	&	After a bigger-than-expected rise in weekly jobless claims and a weaker-than-expected reading on manufacturing sparked worries about the pace of the economic recovery	&	Economic indicators	\\
20091028	&	$-$	&	As a weaker-than-expected new home sales report added to questions about the strength of the economic recovery	&	Economic indicators	\\
20091029	&	$+$	&	As a strong report on economic growth in the third quarter reassured investors that the recovery is on track	&	Economic indicators	\\
20091030	&	$-$	&	As investors dumped a variety of shares at the end of a rough week and choppy month on Wall Street	&	Market sentiment	\\
20091109	&$+$	&As investor optimism gained momentum	&Market sentiment\\
20100204	&	$-$	&	A growing debt crisis in Europe	&	Europe (Debt crisis)	\\
20100427	&	$-$	&	After Standard \& Poors cut Greece's debt rating to junk and lowered Portugal's debt rating,	&	Europe (Debt crisis)	\\
20100504	&	$-$	&	On worries that the global recovery could suffer if Europe's efforts to contain Greece's debt problems don't succeed, and if China's efforts to slow its booming economy go too far	&	Europe (Debt crisis), China	\\
20100506	&	$-$	&	Fears about the spread of the European debt crisis	&	Europe (Debt crisis)	\\
20100510	&	$+$	&	After European officials approved a nearly $1$ trillion rescue plan to contain the debt crisis in troubled nations and stabilize the euro	&	Government policies	\\
20100520	&	$-$	&	Worries about how the European debt crisis and slump in the euro will impact the global recovery fueled the selling	&	Europe (Debt crisis)	\\
20100527	&	$+$	&	After Chinese officials dismissed reports that they're reviewing their nation's investment in European bonds amid concerns about the continent's debt problems	&	China	\\
20100601	&	$-$	&	As worries about the global economic outlook overshadowed better-than-expected readings on the US economy	&	Market sentiment	\\
20100602&	$+$	&Fueled by a rebounding energy sector	&Company-specific factors\\
20100604	&	$-$	&	After a government report showed employers added fewer jobs than expected last month and the euro plunged to a new 4-year low, reviving worries about the health of the European economyEconomic Indicators	&	Economic indicators, Europe (Debt crisis)	\\
20100610	&	$+$	&	As concerns over Europe's debt crisis and its impact on the global recovery were calmed by a sharp boost in Chinese exports and a strengthening euro	&	Europe (Debt crisis), China	\\
20100615	&$+$&As worries about Europe's debt woes hurting U.S. growth eased	&Europe (Debt crisis)\\
20100629	&	$-$	&	After a big drop in consumer confidence and signs of a bigger slowdown in the global economy	&	Market sentiment	\\
20100707	&	$+$	&	As investors came back after the recent bloodletting, spurred on by State Street's improved earnings forecast	&	Economic indicators	\\
20100716	&	$-$	&	After financial firms Bank of America and Citigroup reported weaker quarterly revenue and a plunge in consumer sentiment revived concerns about the economic outlook	&	Company-specific factors	\\
20100722	&	$+$	&	After better-than-expected earnings and forecasts from 3M, Caterpillar, AT\&T and UPS helped reassure investors about the pace of the economic recovery	&	Economic indicators, Company-specific factors	\\
20100811	&	$-$	&	After a report showed the US trade gap widened, and foreign data cast doubt on overseas demand for American goods	&	Economic indicators	\\
20100901	&	$+$	&	As investors cheered signs of strength in the manufacturing sector	&	Economic indicators	\\
20100924&	$+$	&As a report on durable goods edged slightly ahead of forecasts	&Economic indicators\\
20110222	&	$-$	&	Libya's escalating political crisis sparked a sharp sell-off in US stocks, as oil prices continued to skyrocket	&	 Global events (geopolitical events)	\\
20110601	&	$-$	&	Weak economic data has started to snowball	&	Economic indicators	\\
20110727&	$-$&	As Congress remained stalled on resolving the debt ceiling	&Government policies\\
20110802	&	$-$	&	As fears about a weak US economy were enflamed after investors got another disappointing economic report - this time on consumer spending	&	Economic indicators	\\
20110804	&	$-$	&	As fear about the global economy spooked investors	&	Market sentiment	\\
20110808	&	$-$	&	As the debt crisis in Europe, lackluster economic news and a downgrade to the US credit rating	&	Europe (Debt crisis), Credit risk	\\
20110809	&	$+$	&	After the Federal Reserve said it will keep interest rates exceptionally low until 2013	&	Government policies	\\
20110810	&	$-$	&	As investors were confronted with mounting fears about Europe's ongoing debt crisis, this time in France	&	Europe (Debt crisis)	\\
20110811	&	$+$	&	On positive earnings and labor market news	&	Economic indicators	\\
20110818	&	$-$	&	As renewed concerns about the US and global economies	&	Market sentiment	\\
20110823	&	$+$	&	Following a report from the FDIC that showed the number of failing banks shrank for the first time in nearly five years	&	Economic indicators	\\
20110829	&	$+$	&	A Greek bank deal, a solid US consumer spending report and relief that Hurricane Irene caused less damage than expected	&	Government policies	\\
20110902	&	$-$	&	After a government report showing no job growth in August stoked fears that the US may be headed into another recession	&	Economic indicators	\\
20110907	&	$+$	&	The gains came as concerns over Europe's debt crisis eased and investors geared up for President Obama's highly anticipated jobs speech Thursday evening	&	Europe (Debt crisis)	\\
20110909	&	$-$	&	As bad news out of Europe kept piling up	&	Europe (Debt crisis)	\\
20110921	&	$-$	&	Fed disappoints	&	Government policies	\\
20110922	&	$-$	&	Fear factor	&	Market sentiment	\\
20110928	&	$-$	&	Worries over stalled global growth prospects and concerns that European leaders may not be moving fast enough to solve the region's debt problems	&	Market sentiment	\\
20110930	&	$-$	&	As investors remain worried about the debt crisis in Europe and the outlook for global economic growth	&	Market sentiment	\\
20111003	&	$-$	&	With worries about Greece's solvency still in the spotlight	&	Market sentiment	\\
20111004	&	$+$	&	Bear market bounce	&	Market sentiment	\\
20111006	&$+$	&On Euro optimism	&Europe\\
20111010	&	$+$	&	As investors cheered a pledge from European leaders to unveil a plan for solving the eurozone's debt crisis by the end of the month	&	Government policies	\\
20111017	&	$-$	&	As worries about Europe's debt crisis dominated	&	Market sentiment, Europe (Debt crisis)	\\
20111018	&	$+$	&	Following a report suggesting that Europe's bailout fund may get a big boost	&	Market sentiment, Government policies	\\
20111025	&$-$	&Consumer confidence fell sharply to 39.8 in October	&Market sentiment\\
20111027	&	$+$	&	After European Union leaders agreed to expand Europe's bailout fund and take major losses on Greek bonds	&	Government policies	\\
20111031	&	$-$	&	As investors continued to scrutinize the eurozone debt deal	&	Europe (Debt crisis)	\\
20111101	&	$-$	&	New fears about the fate of the European rescue plan	&	Market sentiment	\\
20111109	&	$-$	&	Italy fears	&	Europe (Debt crisis)	\\
20111123	&	$-$	&	As eurozone fears rumbled on and a preliminary report showed that Chinese manufacturing slowed sharply	&	Europe (Debt crisis), China	\\
20111128	&	$+$	&	On robust Black Friday sales	&	Economic indicators	\\
20111130	&	$+$	&	After the Federal Reserve said it will work with other central banks to support the global economy	&	Government policies	\\
20111208	&$-$	&After European Central Bank President Mario Draghi refused to commit to offering broad assistance to troubled eurozone	&Government policies\\
20111220	&	$+$	&	As concerns about the European debt crisis eased and investors welcomed signs of strength in the US housing market	&	Market sentiment, Economic indicators	\\
20120601	&	$-$	&	As Europe's debt crisis remains unresolved and the US economy is showing new signs of distress	&	Europe (Debt crisis), Market sentiment	\\
20120606	&$+$&	As a report on eurozone economic growth revealed no further weakness	&Economic indicators\\
20120621&	$-$	&China's manufacturing activity continued to drop off in June	&China\\
20120629	&	$+$	&	A deal among European leaders to help struggling eurozone banks buoyed global markets	&	Government policies	\\
20130415	&	$-$	&	Following the news of explosions at the Boston Marathon	&	Global events (terrorist attack)	\\
20140203	&$-$&	After the Institute for Supply Management's monthly index showed that manufacturing activity last month expanded at its weakest pace since May &Economic indicators\\
20150821&	$-$	&As global growth concerns accelerated selling pressure to push the Dow and Nasdaq into correction territory&	Market sentiment\\
20150824	&	$-$	&	Deep fears about China's economic slowdown	&	China	\\
20150826	&	$+$	&	The huge rally represents a rebound following six days of dramatic selling that was driven by serious concerns about how China's slowing economy will impact the rest of the world	&	China	\\
20150827	&$+$	&The U.S. economy grew 3.7\% in the second quarter, a very big upward revision than the first official estimate, 2.3\%&	Economic indicators\\
20150901	&	$-$	&	Following more fears about a slowdown in China's economy	&	China	\\
20150928&	$-	$&The Consumer Confidence Index dropped, indicating a potential weakening of the US economy&	Economic indicators\\
20160107&	$-$&	As China's stock market crashed 7\% overnight and crude oil plummeted to the lowest level in more than 12 years	&China, Oil price\\
20160202&	$-$	&As renewed declines in oil prices weighed amid mixed reaction to some key earnings reports	&Oil price\\
20160212&	$+$	&As European and American bank stocks — as well as oil prices — bounced sharply	&Company-specific factors, Oil price\\
20160113	&	$-$	&	Low oil prices, as concerns about global economic slowdown	&	Oil price	\\
20160129	&	$+$	&	A realization that if the US avoids a recession -- as most economists think it will -- beaten-down stocks could be a good buy	&	Market sentiment	\\
20160624	&	$-$	&	British voters chose to leave the European Union	&	Europe (Brexit)	\\
20160627	&	$-$	&	British voters chose to leave the European Union	&	Europe (Brexit)	\\
20160909&	$-	$ &As concerns the Federal Reserve might raise interest rates this month loomed following comments made by key Fed officials	&Market sentiment, Government policies\\
\bottomrule
 
\end{longtable}
 \end{center}}


@book{bai2010spectral,
  title={Spectral analysis of large dimensional random matrices},
  author={Bai, Zhidong and Silverstein, Jack W},
  volume={20},
  year={2010},
  publisher={Springer}
}

@article{onatski2012asymptotics,
  title={Asymptotics of the principal components estimator of large factor models with weakly influential factors},
  author={Onatski, Alexei},
  journal={Journal of Econometrics},
  volume={168},
  number={2},
  pages={244--258},
  year={2012},
  publisher={Elsevier}
}

@article{su2017time,
  title={On time-varying factor models: Estimation and testing},
  author={Su, Liangjun and Wang, Xia},
  journal={Journal of Econometrics},
  volume={198},
  number={1},
  pages={84--101},
  year={2017},
  publisher={Elsevier}
}

@article{jiang2023revisiting,
  title={Revisiting asymptotic theory for principal component estimators of approximate factor models},
  author={Jiang, Peiyun and Uematsu, Yoshimasa and Yamagata, Takashi},
  journal={arXiv preprint arXiv:2311.00625},
  year={2023}
}

@article{johnstone2009consistency,
  title={On consistency and sparsity for principal components analysis in high dimensions},
  author={Johnstone, Iain M and Lu, Arthur Yu},
  journal={Journal of the American Statistical Association},
  volume={104},
  number={486},
  pages={682--693},
  year={2009},
  publisher={Taylor \& Francis}
}

@article{connor1988risk,
  title={Risk and return in an equilibrium APT: Application of a new test methodology},
  author={Connor, Gregory and Korajczyk, Robert A},
  journal={Journal of Financial Economics},
  volume={21},
  number={2},
  pages={255--289},
  year={1988},
  publisher={Elsevier}
}

@article{connor1986performance,
  title={Performance measurement with the arbitrage pricing theory: A new framework for analysis},
  author={Connor, Gregory and Korajczyk, Robert A},
  journal={Journal of Financial Economics},
  volume={15},
  number={3},
  pages={373--394},
  year={1986},
  publisher={Elsevier}
}

@book{campbell1997econometrics,
  title={The Econometrics of Financial Markets},
  author={Campbell, John Y and Lo, Andrew W and MacKinlay, A Craig},
  year={1997},
  publisher={Princeton University Press}
}

@article{davis1970rotation,
  title={The rotation of eigenvectors by a perturbation. III},
  author={Davis, Chandler and Kahan, William Morton},
  journal={SIAM Journal on Numerical Analysis},
  volume={7},
  number={1},
  pages={1--46},
  year={1970},
  publisher={SIAM}
}

@article{ma2013sparse,
  title={Sparse principal component analysis and iterative thresholding},
  author={Ma, Zongming},
 journal={The Annals of Statistics},
  volume={41},
  number={2},
  pages={772--801},
  year={2013}
}

@inproceedings{moghaddam2006generalized,
  title={Generalized spectral bounds for sparse LDA},
  author={Moghaddam, Baback and Weiss, Yair and Avidan, Shai},
  booktitle={Proceedings of the 23rd international conference on Machine learning},
  pages={641--648},
  year={2006}
}

@article{bai2023approximate,
  title={Approximate factor models with weaker loadings},
  author={Bai, Jushan and Ng, Serena},
  journal={Journal of Econometrics},
  volume={235},
  number={2},
  pages={1893--1916},
  year={2023},
  publisher={Elsevier}
}

@article{campbell2001have,
  title={Have individual stocks become more volatile? An empirical exploration of idiosyncratic risk},
  author={Campbell, John Y and Lettau, Martin and Malkiel, Burton G and Xu, Yexiao},
  journal={The Journal of Finance},
  volume={56},
  number={1},
  pages={1--43},
  year={2001},
  publisher={Wiley Online Library}
}

@article{goyal2003idiosyncratic,
  title={Idiosyncratic risk matters!},
  author={Goyal, Amit and Santa-Clara, Pedro},
  journal={The Journal of Finance},
  volume={58},
  number={3},
  pages={975--1007},
  year={2003},
  publisher={Wiley Online Library}
}

@article{andrews1991heteroskedasticity,
  title={Heteroskedasticity and autocorrelation consistent covariance matrix estimation},
  author={Andrews, Donald WK},
  journal={Econometrica},
  volume={59},
  number={3},
  pages={817--858},
  year={1991},
  publisher={JSTOR}
}

@article{uematsu2022estimation,
  title={Estimation of sparsity-induced weak factor models},
  author={Uematsu, Yoshimasa and Yamagata, Takashi},
  journal={Journal of Business \& Economic Statistics},
  volume={41},
  number={1},
  pages={213--227},
  year={2022},
  publisher={Taylor \& Francis}
}

@article{gabaix2014sparsity,
  title={A sparsity-based model of bounded rationality},
  author={Gabaix, Xavier},
  journal={The Quarterly Journal of Economics},
  volume={129},
  number={4},
  pages={1661--1710},
  year={2014},
  publisher={MIT Press}
}

@article{gao2023supervised,
  title={Supervised Dynamic PCA: Linear Dynamic Forecasting with Many Predictors},
  author={Gao, Zhaoxing and Tsay, Ruey S},
  journal={Journal of the American Statistical Association},
volume={Forthcoming},
  year={2024}
}

@article{merlevede2011bernstein,
  title={A Bernstein type inequality and moderate deviations for weakly dependent sequences},
  author={Merlev{\`e}de, Florence and Peligrad, Magda and Rio, Emmanuel},
  journal={Probability Theory and Related Fields},
  volume={151},
  number={3-4},
  pages={435--474},
  year={2011},
  publisher={Springer}
}

@article{treynor1961market,
  title={Market value, time, and risk},
  author={Treynor, Jack L},
  journal={Unpublished manuscript dated
August 8, 1961, No. 95--209.},
  year={1961}
}

@article{stock1998diffusion,
  title={Diffusion indexes},
  author={Stock, James H and Watson, Mark W},
  journal={NBER Working Paper 6702},
 year={1998},
publisher={MIT press}
}

@article{fama1993common,
  title={Common risk factors in the returns on stocks and bonds},
  author={Fama, Eugene F and French, Kenneth R},
  journal={Journal of Financial Economics},
  volume={33},
  number={1},
  pages={3--56},
  year={1993},
  publisher={Elsevier}
}

@article{pan2008modelling,
  title={Modelling multiple time series via common factors},
  author={Pan, Jiazhu and Yao, Qiwei},
  journal={Biometrika},
  volume={95},
  number={2},
  pages={365--379},
  year={2008},
  publisher={Oxford University Press}
}

@article{stewart1990matrix,
  title={Matrix perturbation theory},
  author={Stewart, Gilbert W and Sun, Ji-guang},
  journal={Academic Press},
  year={1990}
}

@article{huang2022scaled,
  title={Scaled PCA: A new approach to dimension reduction},
  author={Huang, Dashan and Jiang, Fuwei and Li, Kunpeng and Tong, Guoshi and Zhou, Guofu},
  journal={Management Science},
  volume={68},
  number={3},
  pages={1678--1695},
  year={2022},
  publisher={INFORMS}
}

@article{gao2019banded,
  title={Banded spatio-temporal autoregressions},
  author={Gao, Zhaoxing and Ma, Yingying and Wang, Hansheng and Yao, Qiwei},
  journal={Journal of Econometrics},
  volume={208},
  number={1},
  pages={211--230},
  year={2019},
  publisher={Elsevier}
}

@article{bai2013principal,
  title={Principal components estimation and identification of static factors},
  author={Bai, Jushan and Ng, Serena},
  journal={Journal of Econometrics},
  volume={176},
  number={1},
  pages={18--29},
  year={2013},
  publisher={Elsevier}
}

@article{pelger2022interpretable,
  title={Interpretable sparse proximate factors for large dimensions},
  author={Pelger, Markus and Xiong, Ruoxuan},
  journal={Journal of Business \& Economic Statistics},
  volume={40},
  number={4},
  pages={1642--1664},
  year={2022},
  publisher={Taylor \& Francis}
}

@article{jolliffe2003modified,
  title={A modified principal component technique based on the LASSO},
  author={Jolliffe, Ian T and Trendafilov, Nickolay T and Uddin, Mudassir},
  journal={Journal of Computational and Graphical Statistics},
  volume={12},
  number={3},
  pages={531--547},
  year={2003},
  publisher={Taylor \& Francis}
}

@article{shen2008sparse,
  title={Sparse principal component analysis via regularized low rank matrix approximation},
  author={Shen, Haipeng and Huang, Jianhua Z},
  journal={Journal of Multivariate Analysis},
  volume={99},
  number={6},
  pages={1015--1034},
  year={2008},
  publisher={Elsevier}
}

@article{witten2009penalized,
  title={A penalized matrix decomposition, with applications to sparse principal components and canonical correlation analysis},
  author={Witten, Daniela M and Tibshirani, Robert and Hastie, Trevor},
  journal={Biostatistics},
  volume={10},
  number={3},
  pages={515--534},
  year={2009},
  publisher={Oxford University Press}
}

@article{kristensen2017diffusion,
  title={Diffusion indexes with sparse loadings},
  author={Kristensen, Johannes Tang},
  journal={Journal of Business \& Economic Statistics},
  volume={35},
  number={3},
  pages={434--451},
  year={2017},
  publisher={Taylor \& Francis}
}

@article{forni2000generalized,
  title={The generalized dynamic-factor model: Identification and estimation},
  author={Forni, Mario and Hallin, Marc and Lippi, Marco and Reichlin, Lucrezia},
  journal={Review of Economics and Statistics},
  volume={82},
  number={4},
  pages={540--554},
  year={2000},
  publisher={MIT Press 238 Main St., Suite 500, Cambridge, MA 02142-1046, USA journals~…}
}

@article{lam2012factor,
  title={Factor modeling for high-dimensional time series: inference for the number of factors},
  author={Lam, Clifford and Yao, Qiwei},
  journal={The Annals of Statistics},
  pages={694--726},
  year={2012},
  publisher={JSTOR}
}

@article{gao2022modeling,
  title={Modeling high-dimensional time series: A factor model with dynamically dependent factors and diverging eigenvalues},
  author={Gao, Zhaoxing and Tsay, Ruey S},
  journal={Journal of the American Statistical Association},
  volume={117},
  number={539},
  pages={1398--1414},
  year={2022},
  publisher={Taylor \& Francis}
}

@article{lewbel1991rank,
  title={The rank of demand systems: theory and nonparametric estimation},
  author={Lewbel, Arthur},
  journal={Econometrica},
  pages={711--730},
  year={1991},
  publisher={JSTOR}
}

@article{stock1989new,
  title={New indexes of coincident and leading economic indicators},
  author={Stock, James H and Watson, Mark W},
  journal={NBER Macroeconomics Annual},
  volume={4},
  pages={351--394},
  year={1989},
  publisher={MIT press}
}

@article{gregory1999common,
  title={Common and country-specific fluctuations in productivity, investment, and the current account},
  author={Gregory, Allan W and Head, Allen C},
  journal={Journal of Monetary Economics},
  volume={44},
  number={3},
  pages={423--451},
  year={1999},
  publisher={Elsevier}
}

@book{forni2000reference,
  title={Reference cycles: the NBER methodology revisited},
  author={Forni, Mario and Hallin, Marc and Lippi, Marco and Reichlin, Lucrezia and others},
  volume={2400},
  year={2000},
  publisher={Centre for Economic Policy Research}
}

@article{vu2013minimax,
  title={Minimax sparse principal subspace estimation in high dimensions},
  author={Vu, Vincent Q and Lei, Jing},
  journal={The Annals of Statistics},
  volume={41},
  number={6},
  pages={2905--2947},
  year={2013},
  publisher={Institute of Mathematical Statistics}
}

@book{vershynin2018high,
  title={High-dimensional probability: An introduction with applications in data science},
  author={Vershynin, Roman},
  volume={47},
  year={2018},
  publisher={Cambridge university press}
}

@article{gao2021two,
  title={A Two-Way Transformed Factor Model for Matrix-Variate Time Series},
  author={Gao, Zhaoxing and Tsay, Ruey S},
  journal={Econometrics and Statistics},
  volume={27},
  pages={83--101},
  year={2021},
  publisher={Elsevier}
}

@article{yuan2013,
	author = {Yuan, Xiao-Tong and Tong Zhang},
	journal = {Journal of Machine Learning Research},
	number = {4},
	pages = {899-925},
	title = {Truncated power method for sparse eigenvalue problems},
	volume = {14},
	year = {2013}}

@article{mackey2008,
	author = {Mackey, Lester},
	journal = {Advances in neural information processing systems},
	title = {Deflation methods for sparse PCA},
	volume = {21},
	year = {2008}}

@article{zou2006,
	author = {Zou, Hui and Trevor Hastie and Robert Tibshirani},
	journal = {Journal of Computational and Graphical Statistics},
	pages = {265--286},
	title = {Sparse principal component analysis},
	volume = {15(2)},
	year = {2006}}

@article{gaotsay2022,
  title={Divide-and-conquer: a distributed hierarchical factor approach to modeling large-scale time series data},
  author={Gao, Zhaoxing and Tsay, Ruey S},
  journal={Journal of the American Statistical Association},
  volume={118},
  number={544},
  pages={2698--2711},
  year={2023},
  publisher={Taylor \& Francis}
}

@article{fan2013large,
	author = {Fan, Jianqing and Liao, Yuan and Mincheva, Martina},
	journal = {Journal of the Royal Statistical Society: Series B (Statistical Methodology)},
	number = {4},
	pages = {603--680},
	publisher = {Wiley Online Library},
	title = {Large covariance estimation by thresholding principal orthogonal complements},
	volume = {75},
	year = {2013}}

@article{lettaupelger2018,
	author = {Lettau, M. and Pelger, M.},
	journal = {Review of Financial Studies},
	number = {5},
	pages = {2274-2325},
	title = {Factors that Fit the Time-Series and Cross-Section of Stock Returns},
	volume = {33},
	year = {2020}}

@article{pelger2019,
	author = {Pelger, M.},
	journal = {Journal of Finance},
	number = {4},
	pages = {2179-2220},
	title = {Understanding Systematic Risk: A High-Frequency Approach},
	volume = {75},
	year = {2020}}

@article{bai2002determining,
	author = {Bai, Jushan and Ng, Serena},
	journal = {Econometrica},
	number = {1},
	pages = {191--221},
	publisher = {Wiley Online Library},
	title = {Determining the number of factors in approximate factor models},
	volume = {70},
	year = {2002}}

@article{onatski2010determining,
	author = {Onatski, Alexei},
	journal = {The Review of Economics and Statistics},
	number = {4},
	pages = {1004--1016},
	publisher = {MIT Press},
	title = {Determining the number of factors from empirical distribution of eigenvalues},
	volume = {92},
	year = {2010}}

@article{bai2003inferential,
	author = {Bai, Jushan},
	journal = {Econometrica},
	number = {1},
	pages = {135--171},
	publisher = {Wiley Online Library},
	title = {Inferential theory for factor models of large dimensions},
	volume = {71},
	year = {2003}}

@article{Stock2002b,
	author = {Stock, James H and Watson, Mark W},
	journal = {Journal of the American Statistical Association},
	number = {460},
	pages = {1167--1179},
	publisher = {Taylor {\&} Francis},
	title = {{Forecasting Using Principal Components From a Large Number of Predictors}},
	volume = {97},
	year = {2002}}

@article{Chamberlain1983,
	author = {Chamberlain, Gary and Rothschild, Michael},
	journal = {Econometrica},
	number = {5},
	pages = {1281--1304},
	title = {Arbitrage, factor structure, and mean-variance analysis on large asset markets},
	volume = {51},
	year = {1983}}

@article{Ahn2013,
	author = {Ahn, Seung C and Horenstein, Alex R},
	journal = {Econometrica},
	number = {3},
	pages = {1203--1227},
	publisher = {Wiley Online Library},
	title = {Eigenvalue ratio test for the number of factors},
	volume = {81},
	year = {2013}}

@article{Ross1976,
	author = {Ross, Stephen A},
	journal = {Journal of Economic Theory},
	number = {3},
	pages = {341--360},
	title = {{The arbitrage theory of capital asset pricing}},
	volume = {13},
	year = {1976}}

@article{Stock2002a,
	author = {Stock, James H and Watson, Mark W},
	journal = {Journal of Business {\&} Economic Statistics},
	number = {2},
	pages = {147--162},
	title = {{Macroeconomic Forecasting Using Diffusion Indexes}},
	volume = {20},
	year = {2002}}
\end{document}